\documentclass[11pt,oneside]{book}

\usepackage{fancyhdr} 
\usepackage{lastpage} 
\usepackage{extramarks} 
\usepackage{graphicx} 
\usepackage{lipsum} 
\usepackage{hyperref}
\hypersetup{colorlinks=true,linkcolor=blue,citecolor=blue,filecolor=black}
\usepackage{caption}
\usepackage{setspace}
\usepackage{blindtext}
\usepackage{graphicx}
\usepackage{amsmath,amssymb}
\usepackage[usenames,dvipsnames]{xcolor}
\usepackage{hyperref}

\usepackage{url}
\usepackage{mathptmx}
\usepackage{helvet}
\usepackage{courier}
\usepackage{verbatim}
\usepackage{eucal}
\usepackage{bm,bbm}
\usepackage{marvosym}
\usepackage[makeroom]{cancel}
\usepackage{hyperref}

\usepackage{amsmath}
\usepackage{amssymb}

\newcommand{\be}{\begin{equation}}
\newcommand{\ee}{\end{equation}}

\newcommand{\bq}{\begin{equation}}
\newcommand{\eq}{\end{equation}}

\newcommand{\llambda}{x}

\usepackage{lipsum}

\let\OLDthebibliography\thebibliography
\renewcommand\thebibliography[1]{
  \OLDthebibliography{#1}
  \setlength{\parskip}{0pt}
  \setlength{\itemsep}{0pt plus 0.3ex}
}

\newcommand{\enterProblemHeader}[1]{
\nobreak\extramarks{#1}{#1 continued on next page\ldots}\nobreak
\nobreak\extramarks{#1 (continued)}{#1 continued on next page\ldots}\nobreak
}

\newcommand{\exitProblemHeader}[1]{
\nobreak\extramarks{#1 (continued)}{#1 continued on next page\ldots}\nobreak
\nobreak\extramarks{#1}{}\nobreak
}

\newcounter{homeworkProblemCounter} 

\newcommand{\homeworkProblemName}{}

\newcommand{\homeworkSectionName}{}

\newcommand{\hmwkTitle}{\Huge{Introduction to Random Matrices\\~\\Theory and Practice}} 
\newcommand{\hmwkClass}{} 
\newcommand{\hmwkClassTime}{} 
\newcommand{\hmwkClassInstructor}{} 
\newcommand{\hmwkAuthorName}{\Large{{\bf Giacomo Livan, Marcel Novaes, Pierpaolo Vivo}}} 

\def\Tr#1{\mbox{Tr} (#1) }

\title{
\vspace{2in}
\textmd{\textbf{\hmwkClass\ \hmwkTitle}}\\
\vspace{0.1in}\large{\textit{\hmwkClassInstructor\ \hmwkClassTime}}
\vspace{3in}
}

\author{\textbf{\hmwkAuthorName}}
\date{} 

\begin{document}

\maketitle
\tableofcontents





\chapter*{Preface}
\addcontentsline{toc}{chapter}{Preface}

This is a book for absolute beginners. If you have heard about random matrix theory, commonly denoted 
\textsf{RMT}, but you do not know what that is, then welcome!, this is the place for you. Our
aim is to provide a truly accessible introductory account of \textsf{RMT} for physicists and mathematicians at the beginning of their research career.
We tried to write the sort of text we would have loved to read when we were
beginning Ph.D. students ourselves.

Our book is structured with light and short chapters, and the style is informal. The calculations we found most instructive are spelt out in full. Particular attention is paid to the numerical verification of most analytical results. The reader will find the symbol [$\spadesuit$ \verb"test.m"] next to every calculation/procedure for which a numerical verification is provided in the associated file \verb"test.m" located at \\\boxed{\href{https://github.com/RMT-TheoryAndPractice/RMT}{https://github.com/RMT-TheoryAndPractice/RMT}}. We strongly believe that  theory without practice is of very little use: in this respect, our book differs from most available textbooks on this subject (not so many, after all).

Almost every chapter contains question boxes, where we try to anticipate and minimize possible points of confusion. Also, we include \textsf{To know more} sections at the end of most chapters, where we collect curiosities, material for extra readings and little gems - carefully (and arbitrarily!) cherrypicked from the gigantic literature on \textsf{RMT} out there.

Our book covers standard material - classical ensembles, orthogonal polynomial techniques, spectral densities and spacings - but also more advanced and modern topics - replica approach and free probability - that are not normally included in elementary accounts on \textsf{RMT}.

Due to space limitations, we have deliberately left out ensembles with complex eigenvalues, and many other interesting topics. Our book is not encyclopedic, nor is it meant as a surrogate or a summary of other excellent existing books. What we are sure about is that any seriously interested reader, who is willing to dedicate some of their time to read and understand this book till the end, will next be able to read and understand any other source (articles, books, reviews, tutorials) on \textsf{RMT}, without feeling overwhelmed or put off by incomprehensible jargon and endless series of ``It can be trivially shown that....".\\

So, what is a random matrix? Well, it is just a matrix whose elements are random variables.
No big deal. So why all the fuss about it? Because they are extremely useful! Just think
in how many ways random variables are useful: if someone throws a thousand (fair) coins,
you can make a rather confident prediction that the number of tails will not be too far
from $500$. Ok, maybe this is not really that useful, but it shows that sometimes it is far
more efficient to forego detailed analysis of individual situations and turn to
statistical descriptions.

This is what statistical mechanics does, after all: it abandons the deterministic (predictive) laws of
mechanics, and replaces them with a probability distribution on the space of possible microscopic states of your systems,
from which detailed statistical predictions at large scales can be made.

This is what \textsf{RMT} is about, but instead of replacing
deterministic numbers with random numbers, it replaces deterministic matrices with random
matrices. Any time you need a matrix which is too complicated to study, you can try
replacing it with a random matrix and calculate averages (and other statistical
properties). 

A number of possible applications come immediately to mind. For example, the
Hamiltonian of a quantum system, such as a heavy nucleus, is a (complicated) matrix. This was indeed one of the first
applications of \textsf{RMT}, developed by Wigner. Rotations are matrices; the metric of a
manifold is a matrix; the $S$-matrix describing the scattering of waves is a matrix; financial data
can be arranged in matrices; matrices are everywhere. In fact, there are many other
applications, some rather surprising, which do not come immediately to mind but which
have proved very fruitful. 

We do not provide a detailed historical account of how \textsf{RMT} developed, nor do we dwell too much on specific applications. The emphasis is on concepts, computations, tricks of the trade: all you needed to know (but were afraid to ask) to \emph{start} a hopefully long and satisfactory career as a researcher in this field.

It is a pleasure to thank here all the people who have somehow contributed to our
knowledge of \textsf{RMT}. We would like to mention in particular Gernot Akemann, Giulio Biroli, Eugene Bogomolny, Zdzis\l aw Burda, Giovanni Cicuta, Fabio D. Cunden, Paolo Facchi, Davide Facoetti, Giuseppe Florio, Yan V. Fyodorov, Olivier Giraud, Claude Godreche, Eytan
Katzav, Jon Keating, Reimer K\"uhn, Satya N. Majumdar, Anna Maltsev, Ricardo Marino,
Francesco Mezzadri, Maciej Nowak, Yasser Roudi,
Dmitry Savin, Antonello Scardicchio, Gregory Schehr, Nick Simm, Peter Sollich, Christophe Texier, Pierfrancesco Urbani, Dario
Villamaina, and many others.

This book is dedicated to the fond memory of Oriol Bohigas. \\

The final publication is available at Springer via \boxed{\href{http://dx.doi.org/10.1007/978-3-319-70885-0}{http://dx.doi.org/10.1007/978-3-319-70885-0}}.

\chapter{Getting Started}\label{chap:GettingStarted}

Let us start with a quick warm-up. We now produce a $N\times N$ matrix $H$ whose entries are independently sampled from a Gaussian probability density function (pdf)\footnote{\label{footnote1}You may already want to give up on this book. Alternatively, you can brush up your knowledge about random variables in Section \ref{sec_gettingstarted:1:secRandVar}.} with mean $0$ and variance $1$. One such matrix for $N=6$ might look like this:
\begin{equation}
H=
\left(
\begin{array}{cccccc}
 1.2448 & 0.0561 & -0.8778 & 1.1058 & 1.1759 & 0.7339 \\
 -0.1854 & 0.7819 & -1.3124 & 0.8786 & 0.3965 & -0.3138 \\
 -0.4925 & -0.6234 & 0.0307 & 0.8448 & -0.2629 & 0.7013 \\
 0.1933 & -1.5660 & 2.3387 & 0.4320 & -0.0535 & 0.2294 \\
 -1.0143 & -0.7578 & 0.3923 & 0.3935 & -0.4883 & -2.7609 \\
 -1.8839 & 0.4546 & -0.4495 & 0.0972 & -2.6562 & 1.3405 \\
\end{array}
\right)\ .\label{eq_gettingstarted:matrixH}
\end{equation}\\

Some of the entries are positive, some are negative, none is very far from $0$. There is no symmetry in the matrix at this stage, $H_{ij}\neq H_{ji}$.\\

Any time we try, we end up with a different matrix: we call all these matrices \emph{samples} or \emph{instances} of our \emph{ensemble}. The $N$ eigenvalues are in general complex numbers (try to compute them for $H$!).\\

To get \emph{real} eigenvalues, the first thing to do is to symmetrize our matrix. Recall that a real symmetric matrix has $N$ \emph{real} eigenvalues. We will not deal much with ensembles with complex eigenvalues in this book\footnote{...but we will deal a lot with matrices with complex \emph{entries} (and real eigenvalues).}.\\   

Try the following symmetrization $H_s =(H+H^T)/2$, where $(\cdot)^T$ denotes the transpose of the matrix. Now the symmetric sample $H_s$ looks like this: 
\begin{equation}
H_s=
\left(
\begin{array}{cccccc}
 1.2448 & -0.0646 & -0.6852 & 0.6496 & 0.0807 & -0.5750 \\
 -0.0646 & 0.7819 & -0.9679 & -0.3436 & -0.1806 & 0.0704 \\
 -0.6852 & -0.9679 & 0.0307 & 1.5917 & 0.0647 & 0.1258 \\
 0.6496 & -0.3436 & 1.5917 & 0.4320 & 0.1700 & 0.1633 \\
 0.0807 & -0.1806 & 0.0647 & 0.1700 & -0.4883 & -2.7085 \\
 -0.5750 & 0.0704 & 0.1258 & 0.1633 & -2.7085 & 1.3405 \\
\end{array}
\right)\ ,\label{eq_gettingstarted:matrixHs}
\end{equation}

whose six eigenvalues are now all real
\begin{equation}
\{-2.49316,-1.7534,0.33069,1.44593,2.38231,3.42944\}\ .
\end{equation}

Congratulations! You have produced your first random matrix drawn from the so-called GOE (Gaussian Orthogonal Ensemble)... a classic - more on this name later.\\

You can now do several things: for example, you can make the entries \emph{complex} or \emph{quaternionic} instead of real. In order to have real eigenvalues, the corresponding matrices need to be \emph{hermitian} and \emph{self-dual} respectively\footnote{Hermitian matrices have real elements on the diagonal, and complex conjugate off-diagonal entries. Quaternion self-dual matrices are $2N\times 2N$ constructed as \textsf{A=[X Y; -conj(Y) conj(X)]; A=(A+A')/2}, where \textsf{X} and \textsf{Y} are complex matrices, while \textsf{conj} denotes complex conjugation of all entries.} - better have a look at one example of the former, for $N$ as small as $N=2$
\begin{equation}
H_{her}=
\left(
\begin{array}{cccccc}
0.3252 &  0.3077 + 0.2803\mathrm{i}\\
   0.3077 - 0.2803\mathrm{i}  & -1.7115 \\
\end{array}
\right)\ .\label{eq_gettingstarted:matrixHhermitian}
\end{equation}\\

You have just met the Gaussian Unitary (GUE) and Gaussian Symplectic (GSE) ensembles, respectively - and are surely already wondering who invented these names.\\

We will deal with this jargon later. Just remember: the Gaussian Orthogonal Ensemble does \emph{not} contain orthogonal matrices - but real symmetric matrices instead (and similarly for the others).\\

\begin{figure}[ht]
\centering
\includegraphics[width=.75\columnwidth]{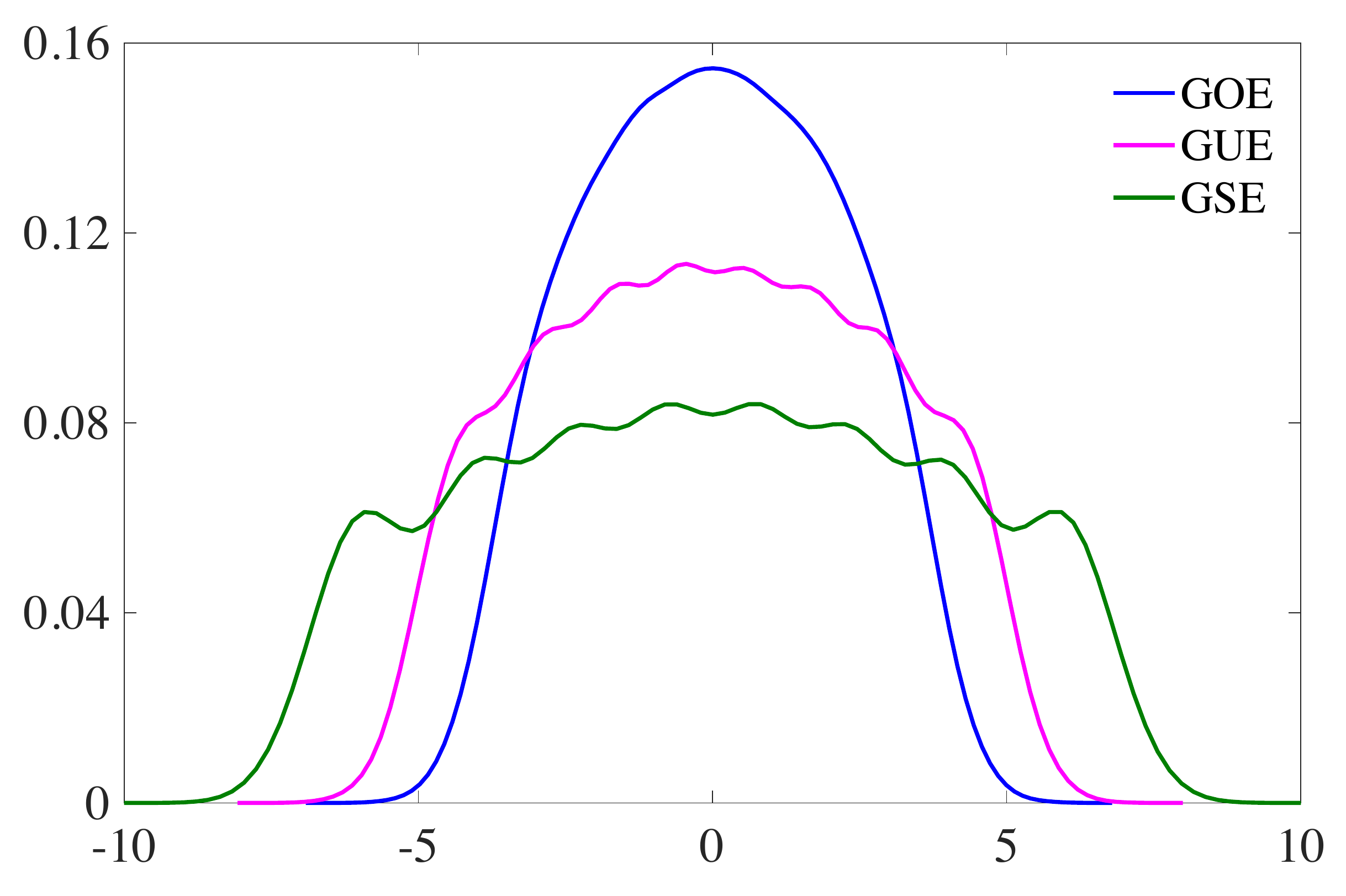}
\caption{Histograms of GOE, GUE and GSE eigenvalues ($N=8$ and $T=50000$ samples).}
\label{fig_gettingstarted:GXEfiniteN}
\end{figure}

Although single instances can sometimes be also useful, exploring the statistical properties of an ensemble typically requires collecting data from multiple samples. We can indeed now generate $T$ such matrices, collect the $N$ (real) eigenvalues for each of them, and then produce a \emph{normalized} histogram of the full set of $N\times T$ eigenvalues. With the code [$\spadesuit$ \verb"Gaussian_Ensembles_Density.m"], you may get a plot like Fig. \ref{fig_gettingstarted:GXEfiniteN} for $T=50000$ and $N=8$.
\\

Roughly half of the eigenvalues collected in total are positive, and half negative - this is evident from the symmetry of the histograms. These histograms are concentrated (significantly nonzero) over the region of the real axis enclosed by (for $N=8$)
\begin{itemize}
\item $\pm\sqrt{2N}\approx \pm 4$ (GOE),
\item $\pm\sqrt{4N}\approx \pm 5.65$ (GUE),
\item $\pm \sqrt{8N}\approx 8$ (GSE).
\end{itemize}

You can directly jump to the end of Chapter \ref{chap:SaddlePoint} to see what these histograms look like for big matrices.\\

\begin{center}
\fbox{\begin{minipage}{33em}
{\it Question.} Can I compute analytically the shape of these histograms? And what happens if $N$ becomes very large? \\ \\
$\blacktriangleright$ Yes, you can. In Chapters \ref{chap:FiniteN} and \ref{chap:FiniteNbeta14}, we will set up a formalism to compute exactly these shapes for any \emph{finite} $N$. In Chapter \ref{chap:SaddlePoint}, instead, we will see that for \emph{large} $N$ the histograms approach a limiting shape, called \emph{Wigner's semicircle law}. 
\end{minipage}}
\end{center}

\section{One-pager on random variables}\label{sec_gettingstarted:1:secRandVar}

Attributed to Giancarlo Rota is the statement that a random variable $X$ is neither random, nor is a variable\footnote{In the following we may use both upper and lower case to denote a random variable.}.\\ 

Whatever it is, it can take values in a discrete alphabet (like the outcome of tossing a die, $\{1,2,3,4,5,6\}$) or on an interval $\sigma$ (possibly unbounded) of the real line. For the latter case, we say that $\rho(\llambda)$ is the {\em
probability density function}\footnote{For example, for the GOE matrix \eqref{eq_gettingstarted:matrixHs} the diagonal entries were sampled from the Gaussian (or normal) pdf $\rho(\llambda)=\exp(-\llambda^2/2)/\sqrt{2\pi}$. We will denote the normal pdf with mean $\mu$ and variance $\sigma^2$ as $N(\mu,\sigma^2)$ in the following.} (pdf) of $X$ if
$\int_a^b d\llambda\rho(\llambda)$ is the probability that $X$ takes value in the
interval $(a,b)\subseteq\sigma$.\\ 

A die will not blow up and disintegrate in the air. One of the six numbers \emph{will} eventually come up. So the sum of probabilities of the outcomes should be $1$ ($=100\%$). People call this property \emph{normalization}, which for continuous variables just means $\int_\sigma d\llambda\rho(\llambda)=1$.\\

All this in theory.\\

In practice, sample your random variable many times and produce a normalized histogram of the outcomes. The pdf $\rho(\llambda)$ is nothing but the histogram profile as the number of samples gets sufficiently large. The \emph{average} of $X$ is $\langle X\rangle=\int d\llambda\rho(\llambda)\llambda$ and higher \emph{moments} are defined as $\langle X^n\rangle=\int
d\llambda\rho(\llambda)\llambda^n $. The \emph{variance} is $\mathrm{Var}(X)=\langle X^2\rangle-(\langle X \rangle)^2$, which is a measure of how broadly spread around the mean the pdf is.\\ 

The \emph{cumulative distribution function} $F(\llambda)$ is the probability that $X$ is smaller or equal to $\llambda$, $F(\llambda)=\int_{-\infty}^x dy\ \rho(y)$. Clearly, $F(\llambda)\to 0$ as $\llambda\to -\infty$ and $F(\llambda)\to 1$ as $\llambda\to +\infty$.\\

If we have two (continuous) random variables $X_1$ and $X_2$, they must be described by a {\em joint probability
density function} (jpdf) $\rho(\llambda_1,\llambda_2)$. Then, the quantity $
\int_a^bd\llambda_1\int_c^dd\llambda_2 \rho(\llambda_1,\llambda_2)$ gives the probability
that the first variable $X_1$ is in the interval $(a,b)$ and the other $X_2$ is,
simultaneously, in the interval $(c,d)$.\\

When the jpdf is \emph{factorized}, i.e. is the product of two density functions,
$\rho(\llambda_1,\llambda_2)=\rho_1(\llambda_1)\rho_2(\llambda_2)$, the variables are said
to be \emph{independent}, otherwise they are \emph{dependent}. When, in addition, we also have $\rho_1(x) = \rho_2(x)$, the random variables are called i.i.d. (independent and identically distributed). In
any case, $\rho(\llambda_1)=\int
 \rho(\llambda_1,\llambda_2)d\llambda_2$ is the \emph{marginal} pdf of $X_1$
when considered independently of $X_2$. \\

The above discussion can be generalized to an arbitrary number $N$ of random variables.
Given the jpdf $\rho(\llambda_1,\ldots,\llambda_N)$, the quantity $
\rho(\llambda_1,\ldots,\llambda_N)d\llambda_1\cdots d\llambda_N$ is the
probability that we find the first variable in the interval
$[\llambda_1,\llambda_1+d\llambda_1]$, the second in the interval
$[\llambda_2,\llambda_2+d\llambda_2]$, etc. The \emph{marginal} pdf $\rho(\llambda)$ that the first variable will be in the interval $[\llambda,\llambda+d\llambda]$ (ignoring the others) can be computed as \be \rho(\llambda)= \int
d\llambda_2\cdots\int d\llambda_N\rho(\llambda,\llambda_2,\ldots,\llambda_N).\ee

\begin{center}
\fbox{\begin{minipage}{33em}
{\it Question.} What is the jpdf $\rho[H]$ of the $N^2$ entries $\{H_{11},\ldots,H_{NN}\}$ of the matrix $H$ in \eqref{eq_gettingstarted:matrixH}? \\ \\
$\blacktriangleright$ The entries in $H$ are independent Gaussian variables, hence the jpdf is factorized as $\rho[H]\equiv\rho(H_{11},\ldots,H_{NN})=\prod_{i,j=1}^N \left[\exp\left(-H_{ij}^2/2\right)/\sqrt{2\pi}\right]$. 
\end{minipage}}
\end{center}

If a set of random variables is a function of another one, $\llambda_i=\llambda_i(\bm y)$, there is a relation between the jpdf of the two sets
\begin{equation}
\rho(\llambda_1,\ldots,\llambda_N)d\llambda_1\cdots d\llambda_N=\underbrace{\rho(\llambda_1(\bm y),\ldots,\llambda_N(\bm y))|J(\bm\llambda\to\bm y)|}_{\widehat{\rho}(y_1,\ldots,y_N)}d y_1\cdots d y_N\ ,
\end{equation}
where $J$ is the Jacobian of the
transformation, given by $J(\bm\llambda\to\bm y)=\det\left(\frac{\partial\llambda_i}{\partial y_j}\right)$. We will use this property in Chapter \ref{chap:ChangeVar}.

\begin{center}
\fbox{\begin{minipage}{33em}
{\it Question.} What is the jpdf of the $N(N-1)/2$ entries in the upper triangle of the symmetric matrix $H_s$ in \eqref{eq_gettingstarted:matrixHs}? \\ \\
$\blacktriangleright$ For $H_s$, you need to consider the diagonal and the off-diagonal entries separately: the diagonal entries are $(H_s)_{ii}=H_{ii}$, while the off-diagonal entries are $(H_s)_{ij}=(H_{ij}+H_{ji})/2$. As a result, 
\begin{equation}
\rho((H_s)_{11},\ldots,(H_s)_{NN})=\prod_{i=1}^N \left[\exp\left(-(H_s)_{ii}^2/2\right)/\sqrt{2\pi}\right]\prod_{i<j}\left[\exp\left(-(H_s)_{ij}^2\right)/\sqrt{\pi}\right]\ ,\label{eq_gettingstarted:1:jpdfentriesgaussian}
\end{equation}
i.e. the variance of off-diagonal entries is $1/2$ of the variance of diagonal entries. Make sure you understand why this is the case. This factor $2$ has very important consequences (see the last Question in Chapter \ref{chap:Classifiedmaterial}). From now on, for a real symmetric $H_s$ we will denote the jpdf of the $N(N-1)/2$ entries \emph{in the upper triangle} by $\rho[H]$ - dropping the subscript 's' when there is no risk of confusion.
\end{minipage}}
\end{center}

\chapter{Value the eigenvalue}\label{chap:valuetheeigenvalue}

In this Chapter, we start discussing the eigenvalues of random matrices.
\section{Appetizer: Wigner's surmise}
Consider a $2\times 2$ GOE matrix 
$
H_s=
\left(
\begin{array}{cc}
x_1 & x_3 \\
x_3 & x_2\\
\end{array}
\right)
$, with $x_1,x_2\sim N(0,1)$ and $x_3\sim N(0,1/2)$. What is the pdf $p(s)$ of the spacing $s=\lambda_2-\lambda_1$ between its two eigenvalues ($\lambda_2>\lambda_1$)?\\

The two eigenvalues are random variables, given in terms of the entries by the roots of the characteristic polynomial 
\begin{equation}
\lambda^2-\Tr {H_s}\lambda+\det(H_s)\ ,
\end{equation}
therefore $\lambda_{1,2}=\left(x_1+x_2\pm\sqrt{(x_1-x_2)^2+4 x_3^2}\right)/2$ and $s=\sqrt{(x_1-x_2)^2+4 x_3^2}$.\\

By definition, we have
\be
p(s)=\int_{-\infty}^\infty d x_1 dx_2 dx_3\frac{e^{-\frac{1}{2}x_1^2}}{\sqrt{2\pi}} \frac{e^{-\frac{1}{2}x_2^2}}{\sqrt{2\pi}} \frac{e^{-x_3^2}}{\sqrt{\pi}}\delta\left(s-\sqrt{(x_1-x_2)^2+4 x_3^2}\right)\ .
\ee

Changing variables as
\be
\begin{cases}
x_1-x_2  &= r \cos\theta\\
2 x_3 &= r \sin\theta\\
x_1+x_2 &= \psi\ 
\end{cases}
\qquad\Rightarrow\qquad
\begin{cases}
x_1  &= \frac{r \cos\theta +\psi}{2}\\
x_2 &= \frac{\psi-r \cos\theta }{2}\\\ 
x_3 &= \frac{r \sin\theta}{2}\ 
\end{cases} \ ,
\ee
and computing the corresponding Jacobian
\be
J=\det
\left(
\begin{array}{ccc}
\frac{\partial x_1}{\partial r} & \frac{\partial x_1}{\partial \theta} & \frac{\partial x_1}{\partial \psi}\\
\frac{\partial x_2}{\partial r} & \frac{\partial x_2}{\partial \theta} & \frac{\partial x_2}{\partial \psi}\\
\frac{\partial x_3}{\partial r} & \frac{\partial x_3}{\partial \theta} & \frac{\partial x_3}{\partial \psi}\\
\end{array}
\right)=\det
\left(
\begin{array}{ccc}
\cos\theta/2 & -r\sin\theta/2 & 1/2\\
-\cos\theta/2 & r\sin\theta/2 &1/2\\
\sin\theta/2 & r\cos\theta/2 & 0\\
\end{array}
\right)=-r/4\ ,
\ee
one obtains
\begin{align}
\nonumber p(s) &=\frac{1}{8\pi^{3/2}}\int_0^\infty dr\ r\delta(s-r)\int_0^{2\pi}d\theta \int_{-\infty}^\infty d\psi e^{-\frac{1}{2}\left[\left(\frac{r\cos\theta+\psi}{2}\right)^2+\left(\frac{-r\cos\theta+\psi}{2}\right)^2+\frac{r^2\sin^2\theta}{2}\right]}\\
&=\frac{\sqrt{4\pi}\ s}{8\pi^{3/2}}\int_0^{2\pi}d\theta e^{-\frac{1}{2}\left[\frac{s^2\cos^2\theta}{2}+\frac{s^2\sin^2\theta}{2}\right]}
=\boxed{\frac{s}{2}e^{-s^2/4}}\label{sec_valuetheeigenvalue:spacingpdffinal}\ .
\end{align}
Note that we used $\cos^2\theta+\sin^2\theta=1$ to achieve this very simple result: however, we could only enjoy this massive simplification because the variance of the off-diagonal elements was $1/2$ of the variance of diagonal elements - try to redo the calculation assuming a different ratio. Observe also that this pdf is correctly normalized, $\int_0^\infty ds\ p(s)=1$.\\

\begin{figure}[t]
\centering
\includegraphics[width=.75\columnwidth]{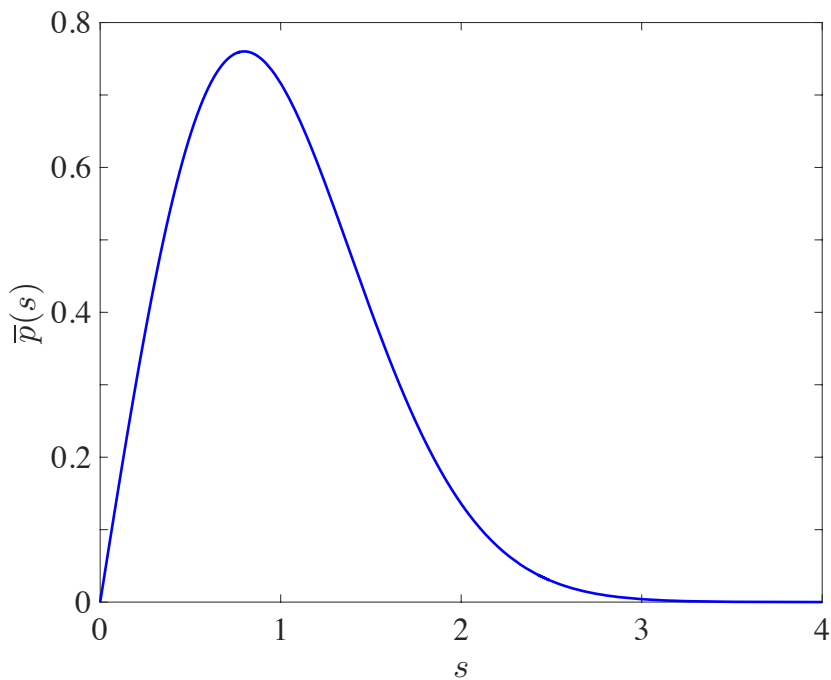}
\caption{Plot of Wigner's surmise.}
\label{fig_valuetheeigenvalue:wigner_surmise}
\end{figure}

It is often convenient to rescale this pdf and define $\bar{p}(s)=\langle s\rangle p\left(\langle s\rangle s\right)$, where $\langle s\rangle=\int_0^\infty ds p(s)s$ is the mean level spacing. Upon this rescaling, $\int_0^\infty \bar{p}(s)ds=\int_0^\infty s\bar{p}(s)ds=1$. For the GOE as above, show that $\bar{p}(s)=(\pi s/2)\exp(-\pi s^2/4)$, which is called \emph{Wigner's surmise}\footnote{Why is it defined a 'surmise'? After all, it is the result of an exact calculation! The story goes as follows: at a conference on Neutron Physics by Time-of-Flight, held at the Oak Ridge National Laboratory in 1956, people asked a question about the possible shape of the distribution of the spacings of energy levels in a heavy nucleus. E. P. Wigner, who was in the audience, walked up to the blackboard and guessed (= surmised) the answer given above.}, whose plot is shown in Fig. \ref{fig_valuetheeigenvalue:wigner_surmise}.\\

In spite of its simplicity, this is actually a quite deep result: it tells us that the probability of sampling two eigenvalues 'very close' to each other ($s\to 0$) is very small: it is as if each eigenvalue 'felt' the presence of the other and tried to avoid it (but not too much)! A bit like birds perching on an electric wire, or parked cars on a street: not too close, not too far apart. If this metaphor does not win you over, check this out \cite{ref_valuetheeigenvalue:Seba2}.

\section{Eigenvalues as correlated random variables}\label{sec_valuetheeigenvalue:secEigRandVar}

In the previous Chapter, we met the $N$ real eigenvalues $\{\llambda_1,\ldots,\llambda_N\}$ of a random matrix $H$. These eigenvalues are random variables described
by a jpdf\footnote{We will use the same symbol $\rho$ for both the jpdf of the entries in the upper triangle and of the eigenvalues.} $\rho(\llambda_1,\ldots,\llambda_N)$.

\begin{center}
\fbox{\begin{minipage}{33em}
{\it Question.} What does the jpdf of eigenvalues $\rho(\llambda_1,\ldots,\llambda_N)$ of a random matrix ensemble look like? \\ \\
$\blacktriangleright$ We will give it in Eq. \eqref{eq_valuetheeigenvalue:jpdgaussonehalf} for the Gaussian ensemble. Not for every ensemble the jpdf of eigenvalues is known.
\end{minipage}}
\end{center}

The important (generic) feature is that the $\{x_i\}$'s are {\bf not} independent: their jpdf does \emph{not} in general factorize. The most striking incarnation of this property is the so-called \emph{level repulsion} (as in Wigner's surmise): the eigenvalues of random matrices generically repel each other, while independent variables do not - as we show in the following section. 

\section{Compare with the spacings between i.i.d.'s}\label{sec_valuetheeigenvalue:iidspacing}

It is useful at this stage to consider the statistics of gaps between adjacent i.i.d. random variables. In this case, we will not see any repulsion.\\

Consider i.i.d. real random variables $\{X_1,\ldots,X_N\}$ drawn from a parent pdf $p_X(x)$ defined over a support $\sigma$. The corresponding cdf is $F(x)$. The labelling is purely conventional, and we do not assume that the variables are sorted in any order.\\

We wish to compute the \emph{conditional} probability density function $p_N(s|X_j=x)$ that, given that one of the random variables $X_j$ takes a value around $x$, there is another random variable $X_k$ ($k\neq j$) around the position $x+s$, and no other variables lie in between. In other word, a \emph{gap} of size $s$ exists between two random variables, one of which sits around $x$.\\

The claim is 
\be
p_N(s|X_j=x)=p_X(x+s)\left[1+F(x)-F(x+s)\right]^{N-2}\ .
\ee

The reasoning goes as follows: one of the variables sits around $x$ already, so we have $N-1$ variables left to play with. One of these should sit around $x+s$, and the pdf for this event is $p_X(x+s)$. The remaining $N-2$ variables need to sit either to the left of $x$ - and this happens with probability $F(x)$ - or to the right of $x+s$ - and this happens with probability $1-F(x+s)$.\\

Now, the probability of a gap $s$ between two adjacent particles, conditioned on the position $x$ of one variable, \emph{but irrespective of which variable this is} is obtained by the \emph{law of total probability} 
\be
p_N(s|\mathrm{any }\ X=x)= \sum_{j=1}^N p_N(s|X_j=x)\mathrm{Prob}(X_j=x)=N p_N(s|X_j=x)p_X(x)\ ,\label{eq_valuetheeigenvalue:pnsusinglawtotalprob}
\ee
where one uses the fact that the variables are i.i.d. and thus the probability that the particle $X_j$ lies around $x$ is the same for every particle, and given by $p_X(x)$.\\

To obtain the probability of a gap $s$ between any two adjacent random variables, no longer conditioned on the position of one of the variables, we should simply integrate over $x$ 

\be
p_N(s)=\int_\sigma dx\ p_N(s|\mathrm{any }\ X=x)=N \int_\sigma dx\ p_N(s|X_j=x)p_X(x)\ .\label{eq_valuetheeigenvalue:pnsusingiid}
\ee

As an exercise, let us verify that $p_N(s)$ is correctly normalized, namely $\int_0^\infty ds\ p_N(s)=1$. We have
\be
\int_0^\infty ds\ p_N(s)=N\int_0^\infty ds \int_\sigma dx\ p_X(x+s)\left[1+F(x)-F(x+s)\right]^{N-2}p_X(x)\ .
\ee
Changing variables $F(x+s)=u$ in the $s$-integral, and using $F(+\infty)=1$ and $du=F'(x+s)ds=p_X(x+s)ds$, we get
\be
\int_0^\infty ds\ p_N(s)=N\int_\sigma dx\ p_X(x)\underbrace{\int_{F(x)}^1 du [1+F(x)-u]^{N-2}}_{\frac{1-F(x)^{N-1}}{N-1 }}\ .
\ee
Setting now $F(x)=v$ and using $dv=F'(x)dx=p_X(x)dx$, we have
\be
\int_0^\infty ds\ p_N(s)=\frac{N}{N-1}\int_0^1 dv(1-v^{N-1})=1\ ,
\ee
as required.\\

As there are $N$ variables, it makes sense to perform the 'local' change of variables $s=\hat{s}/(N p_X(x))$ and consider the limit $N\to\infty$. The reason for choosing the scaling factor $N p_X(x)$ is that their typical spacing around the point $x$ will be precisely of order $\sim 1/ (N p_X(x))$: increasing $N$, more and more variables need to occupy roughly the same space, therefore their typical spacing goes down. The same happens locally around points $x$ where there is a higher chance to find variables, i.e. for a higher $p_X(x)$.\\

We thus have
\be
p_N\left(s=\frac{\hat{s}}{N p_X(x)}\Big|X_j=x\right)=p_X\left(x+\hat{s}/N p_X(x)\right)\left[1+F(x)-F\left(x+\hat{s}/N p_X(x)\right)\right]^{N-2}\ ,
\ee
which for large $N$ and $\hat{s}\sim\mathcal{O}(1)$, can be approximated as
\be
p_N\left(s=\frac{\hat{s}}{N p_X(x)}\Big|X_j=x\right)\approx p_X(x)e^{-\hat{s}}\ ,
\ee
therefore using \eqref{eq_valuetheeigenvalue:pnsusingiid}
\be
\lim_{N\to\infty}\hat{p}_N(\hat{s}):=\lim_{N\to\infty} p_N\left(s=\frac{\hat{s}}{N p_X(x)}\right)\frac{ds}{d\hat{s}}
=N\times \frac{1}{N}\int_\sigma dx p_X(x) e^{-\hat{s}}=e^{-\hat{s}}\ ,
\ee
the exponential law for the spacing of a Poisson process. From this, one deduces easily that i.i.d. variables do not repel, but rather attract: the probability of vanishing gaps, $\hat{s}\to 0$, does not vanish, as in the case of RMT eigenvalues!

\section{Jpdf of eigenvalues of Gaussian matrices}
The jpdf of eigenvalues of a $N\times N$ Gaussian matrix is given by\footnote{This jpdf goes back to the prehistory of RMT. It is an immediate consequence of Theorem 2 in \cite{ref_valuetheeigenvalue:Hsu2}, a 1939 statistics paper published in the journal \emph{Annals of Eugenics} (a rather scary title, isn't it?). In its full glory, it appeared explicitly for the first time in \cite{ref_valuetheeigenvalue:Rosenzweig2}.}
\begin{equation}
\rho(\llambda_1,\ldots,\llambda_N)=\frac{1}{\mathcal{Z}_{N,\beta}}
e^{-\frac{1}{2}\sum_{i=1}^N
\llambda_i^2}\prod_{j<k}|\llambda_j-\llambda_k|^\beta\ ,\label{eq_valuetheeigenvalue:jpdgaussonehalf}
\end{equation}
where 
\begin{equation}
\mathcal{Z}_{N,\beta}=(2\pi)^{N/2}\prod_{j=1}^N\frac{\Gamma(1+j\beta/2)}{\Gamma(1+\beta/2)}\label{eq_valuetheeigenvalue:partitionfunction}
\end{equation}
is a normalization constant\footnote{It can be computed via the so-called \emph{Mehta's integral}, a close relative of the celebrated \emph{Selberg's integral} \cite{ref_valuetheeigenvalue:Forrester_Warnaar}.}, enforcing $\int_{\mathbb{R}^N}d\bm\llambda\ \rho(\llambda_1,\ldots,\llambda_N)=1$, and $\beta=1,2,4$ is called the \emph{Dyson index}\footnote{The Dyson index is equal to the number of real variables needed to specify one entry of your matrix: $1$ for real, $2$ for complex and $4$ for quaternions. This is usually referred to as \emph{Dyson's threefold way}. For the Gaussian ensemble, then, GOE corresponds to $\beta=1$, GUE to $\beta=2$ and GSE to $\beta=4$.}. Henceforth, $d\bm x=\prod_{j=1}^N dx_j$. Note that the eigenvalues are considered to be \emph{unordered} here.\\

This jpdf corresponds \emph{exactly} to eigenvalues\footnote{For $\beta=4$, each matrix has $2N$ eigenvalues that are two-fold degenerate.} generated according to the algorithm in Chapter \ref{chap:GettingStarted}\footnote{Quite often, however, you find in the literature a Gaussian weight including extra factors, such as $\exp(-(\beta/2)\sum_i\llambda_i^2)$ or $\exp(-(N/2)\sum_i\llambda_i^2)$. One then needs to be very careful when comparing theoretical results (obtained with such conventions) to numerical simulations - in particular, a rescaling of the numerical eigenvalues by $\sqrt{\beta}$ or $\sqrt{N}$ before histogramming is essential in these two modified scenarios.}, and provided in the code  [$\spadesuit$ \verb"Gaussian_Ensembles_Density.m"].\\

Where does \eqref{eq_valuetheeigenvalue:jpdgaussonehalf} come from? Let us postpone the proof for a while and draw some conclusions by just staring at it for a few minutes.\\

The Gaussian factor $e^{-\frac{1}{2}\sum_{i=1}^N
\llambda_i^2}$ kills any configuration of eigenvalues $\{\bm\llambda\}$ where some $x_j$'s are ``big" (far from zero, in absolute value): the eigenvalues do not like to stay too far from the origin. On the other hand, the term $\prod_{j<k}|\llambda_j-\llambda_k|$ 
kills configurations where two eigenvalues get ``too close" to each other.\\ 

The ``repulsion" factor $\prod_{j<k}|\llambda_j-\llambda_k|$ has another effect: it makes the eigenvalues strongly non-independent! Every eigenvalue feels the presence of \emph{all} the others, and the jpdf \eqref{eq_valuetheeigenvalue:jpdgaussonehalf} does not factorize at all. Hence, the classical tools for independent random variables are of little use here. We will use \eqref{eq_valuetheeigenvalue:jpdgaussonehalf} in the next Chapter to deduce Wigner's semicircle law in a few simple steps.\\

This interplay between \emph{confinement} and \emph{repulsion} is the physical mechanism at the heart of many results in RMT.\\

As a final remark, go back to the spacing pdf in Eq. \eqref{sec_valuetheeigenvalue:spacingpdffinal}, which was obtained for $N=2$ and $\beta=1$ (a $2\times 2$ GOE matrix).
Armed with \eqref{eq_valuetheeigenvalue:jpdgaussonehalf} one may redo the calculation as
\begin{equation}
p(s)=\int_{-\infty}^\infty d\llambda_1 d\llambda_2 \rho(\llambda_1,\llambda_2)\delta(s-|\llambda_2-\llambda_1|)\ .
\end{equation}
Try to compute this integral, and recover Eq. \eqref{sec_valuetheeigenvalue:spacingpdffinal}.\\

\chapter{Classified Material}\label{chap:Classifiedmaterial}

In this Chapter, we continue setting up the formalism and provide a simple classification of matrix models.

\section{Count on Dirac}\label{sec_valuetheeigenvalue:secDirac}
\begin{center}
\fbox{\begin{minipage}{33em}
{\it Question} From the jpdf of eigenvalues $\rho(\llambda_1,\ldots,\llambda_N)$, how do I compute the shape of the histograms of the $N\times T$ eigenvalues as in Fig. \ref{fig_gettingstarted:GXEfiniteN}, for $T$ sufficiently large? \\ \\
$\blacktriangleright$ To cut a long story short, all you have to do is to take the marginal 
\begin{equation}
\rho(x)=\int\cdots\int d \llambda_2\cdots d\llambda_N\rho(\llambda,\llambda_2,\ldots,\llambda_N)\ ,\label{eq_valuetheeigenvalue:marginal1}
\end{equation}
and this function will reproduce the histogram profile you are after for any finite $N$. Note that $\rho(x)$ is correctly normalized to $1$, as your histogram is.\\
\end{minipage}}
\end{center}

Let us prove \eqref{eq_valuetheeigenvalue:marginal1}.\\

Take a single, fixed matrix $H$ with real eigenvalues - no randomness in here - and perform the following task: define a \emph{counting function} $n(x)$ such that $\int_a^b n(x^\prime)dx^\prime$ gives the fraction of eigenvalues $\llambda_i$ between $a$ and $b$.\\

The way to define it is to set\footnote{As we know, the Dirac delta function (or rather distribution) $\delta(x)$ is basically an extremely peaked function at the point $x=0$, like the limit of a Gaussian pdf as its variance goes to zero, $\delta(x)=\lim_{\epsilon\to 0^+} \frac{1}{2\sqrt{\pi\epsilon}}e^{-x^2/(4\epsilon)}$.}

\begin{equation}
n(\llambda)=\frac{1}{N}\sum_{i=1}^N\delta(\llambda-\llambda_i)\ ,\label{eq_valuetheeigenvalue:1:nx}
\end{equation}
the (normalized) sum of a set of ``spikes" at the location $\llambda_i$ of each eigenvalue. Using the following property of the delta function 
\begin{equation}
\int_{\mathcal{I}}d x\delta(x-x_0)f(x)=f(x_0)\qquad\mbox{ if }x_0\in\mathcal{I}\mbox{ and }0\mbox{ otherwise},
\end{equation}
we can show that indeed \eqref{eq_valuetheeigenvalue:1:nx} does the job properly\footnote{Compute 
\begin{equation}
N\int_a^b n(\llambda)d\llambda = \sum_{i=1}^N\int_a^b
\delta(\llambda-\llambda_i)d\llambda=\sum_{i=1}^N 
\chi_{[a,b]}(\llambda_i)\ ,
\end{equation}
where the indicator function $\chi_{[a,b]}(z)$ is equal to $1$ if $z\in (a,b)$ and $0$
otherwise. This is by definition the number of eigenvalues between $a$ and $b$, as
it should.}.\\

If $H$ is now a random matrix, the function $n(\llambda)$ becomes a \emph{random measure} on the real line - a function of $\llambda$ that changes from one realization of $H$ to another. The \emph{average} of it over the set of random eigenvalues $\{\llambda_1,\ldots,\llambda_N\}$ becomes interesting now\footnote{We use again the shorthand $d\bm{\llambda}=\prod_{j=1}^N d\llambda_j$.}
\begin{equation}
\langle n(\llambda)\rangle := \int\cdots\int d\bm\llambda \rho(\llambda_1,\ldots,\llambda_N)n(\llambda)=\frac{1}{N}\sum_{i=1}^N\int\cdots\int d\bm\llambda \rho(\llambda_1,\ldots,\llambda_N)\delta(\llambda-\llambda_i)=\rho(\llambda)\ ,\label{eq_valuetheeigenvalue:marginalcounting}
\end{equation}
where $\rho(\llambda)=\int\cdots\int d \llambda_2\cdots d\llambda_N\rho(\llambda,\llambda_2,\ldots,\llambda_N)$ is the marginal density of $\rho$. Try to prove the last equality in \eqref{eq_valuetheeigenvalue:marginalcounting} using the properties of delta function, and the fact that $\rho(\llambda_1,\ldots,\llambda_N)$ is symmetric upon the exchange $\llambda_i\to \llambda_j$. This is indeed the case for the Gaussian jpdf \eqref{eq_valuetheeigenvalue:jpdgaussonehalf} and will remain generally true.\\

\begin{figure}[t]
\centering
\includegraphics[width=.75\columnwidth]{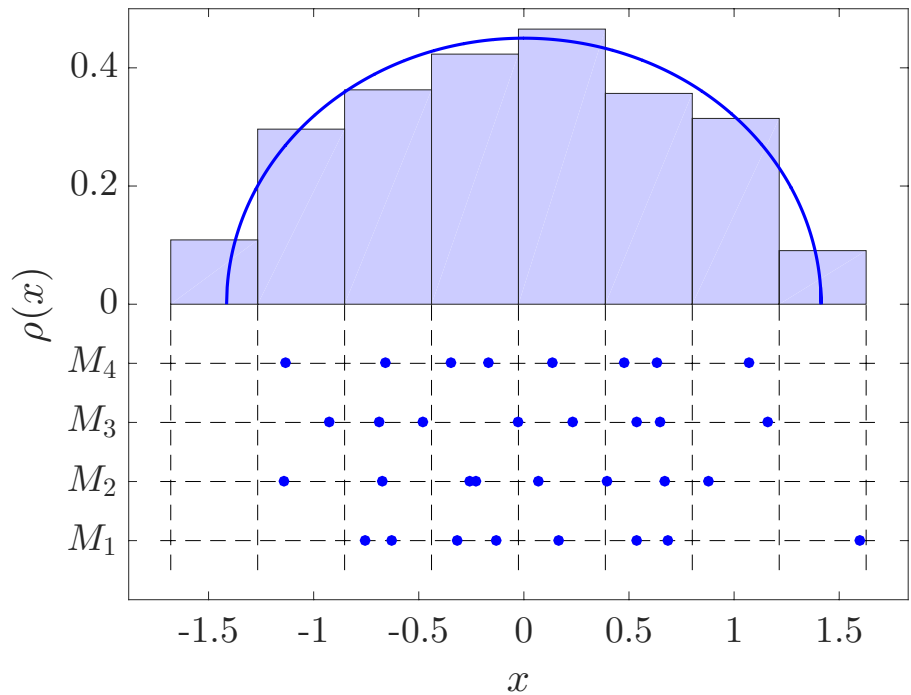}
\caption{Sets of $N=8$ randomly located ``spikes". A histogram of how many spikes occur around a given region of the real line is nothing but the average spectral density there.}
\label{fig_valuetheeigenvalue:spikes}
\end{figure}

The quantity $\langle n(\llambda)\rangle=\rho(\llambda)$ has many names: most often, it is called the \emph{(average) spectral density}. Fig. \ref{fig_valuetheeigenvalue:spikes} helps you visualize how $T=4$ sets of $N=8$ randomly located ``spikes" conspire to produce the continuous shape $\rho(\llambda)=\langle n(\llambda)\rangle$.\\

\begin{center}
\fbox{\begin{minipage}{33em}
{\it Question.} If $N$ becomes very large, what does the spectral density $\rho(\llambda)$ for the Gaussian ensemble look like? \\ \\
$\blacktriangleright$ For the jpdf $\rho(\llambda_1,\ldots,\llambda_N)$ given in \eqref{eq_valuetheeigenvalue:jpdgaussonehalf}, the precise statement for the spectral density $\rho(\llambda)=\int d\llambda_2\cdots d\llambda_N\rho(\llambda,\llambda_2,\ldots,\llambda_N)$ is
\begin{equation}
\lim_{N\to\infty}\sqrt{\beta N}\rho(\sqrt{\beta N}\llambda)=\rho_{\mathrm{SC}}(\llambda)\ ,\label{eq_classified:semicirclefirstdefinition}
\end{equation}
where $\rho_{\mathrm{SC}}(\llambda)=\frac{1}{\pi}\sqrt{2-\llambda^2}$ has a semicircular - or rather, semielliptical - shape. This is called \emph{Wigner's semicircle law}.\\
\end{minipage}}
\end{center}

\begin{figure}[h]
\centering
\includegraphics[width=.75\columnwidth]{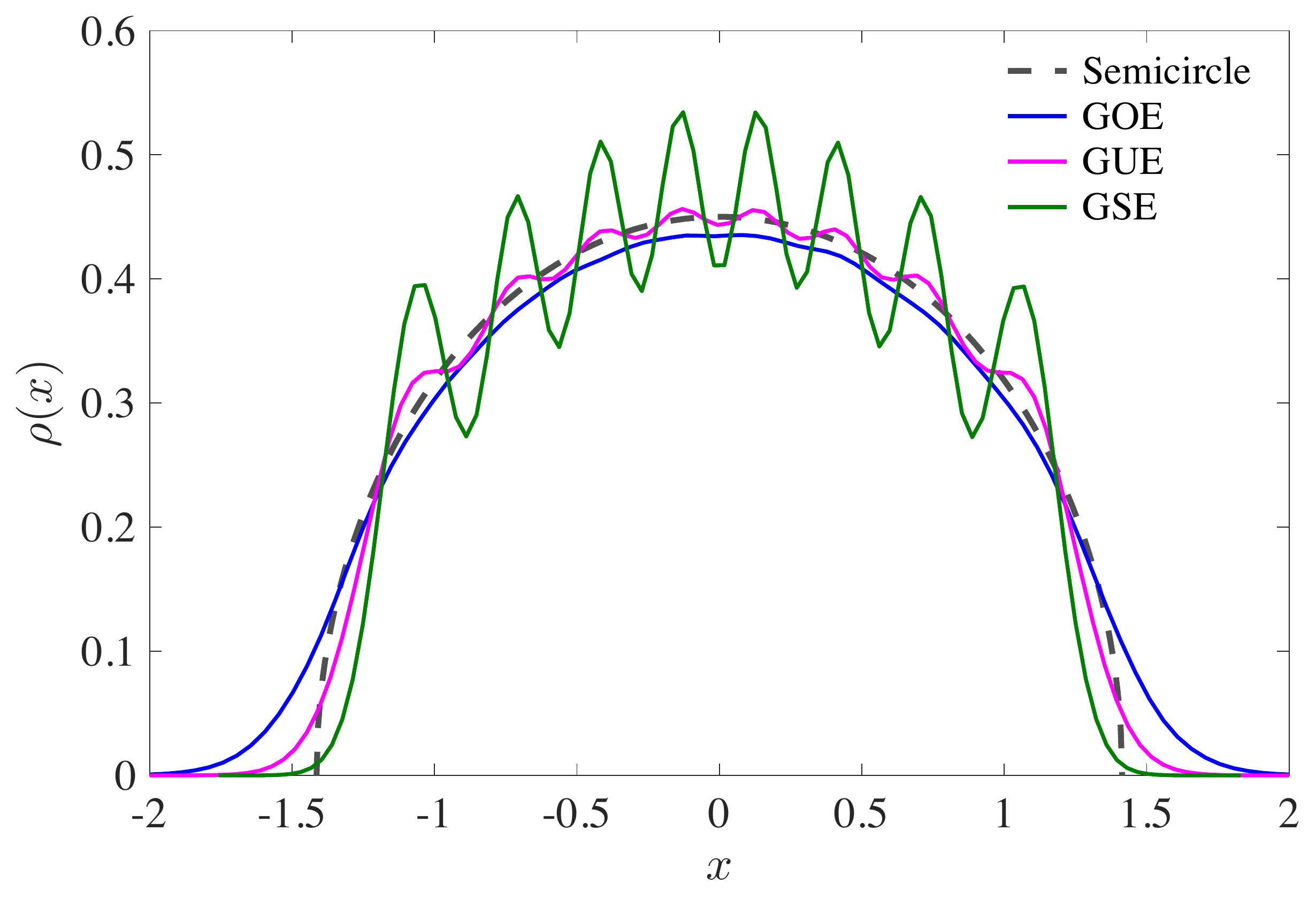}
\caption{Rescaled densities for $N = 8$ (GOE,GUE,GSE).}
\label{classifiedmaterial:finiteN}
\end{figure}

\clearpage

\begin{center}
\fbox{\begin{minipage}{33em}
{\it Question.} What is the meaning of the unexpected rescaling factor $\sqrt{\beta N}$?\\ \\

$\blacktriangleright$ This means that the histograms of eigenvalues for larger and larger $N$ become concentrated over the interval $[-\sqrt{2\beta N},\sqrt{2\beta N}]$, in agreement with our numerical findings in Fig. \ref{fig_gettingstarted:GXEfiniteN}. The points $\pm\sqrt{2\beta N}$ are called (spectral) \emph{edges}.\\

Note that:
\begin{enumerate}
\item The edges are \emph{growing} with $\sqrt{N}$ - bigger matrices have a wider range of eigenvalues, can you explain why? To get histograms that do \emph{not} become wider and wider with $N$, we need to divide each eigenvalue by $\sqrt{\beta N}$ before histogramming. This is what we do in Fig. \ref{classifiedmaterial:finiteN}, using the very same eigenvalues collected to produce Fig. \ref{fig_gettingstarted:GXEfiniteN}. You can see that the histograms for different $\beta$s nicely collapse on top of each other, reproducing an almost perfect semielliptical shape between $-\sqrt{2}$ and $\sqrt{2}$.
\item The edges are at $\pm\sqrt{2\beta N}$ for the jpdf $\rho(\llambda_1,\ldots,\llambda_N)$ given in \eqref{eq_valuetheeigenvalue:jpdgaussonehalf}. If you put \emph{ad hoc} extra factors in the exponential, like $\exp(-(\beta/2)\sum_i\llambda_i^2)$ or $\exp(-(N/2)\sum_i\llambda_i^2)$, as you sometimes find in the literature, this is tantamount to rescaling the eigenvalues by an appropriate factor. For example, for the choice $\exp(-(N/2)\sum_i\llambda_i^2)$, the edges are fixed - they do not grow with $N$ - at $\pm\sqrt{2\beta}$.
\item The edges of the semicircle are called \emph{soft}: for large but finite $N$, there is always a nonzero probability of sampling eigenvalues exceeding the edge points. For example, for a GOE matrix $10\times 10$, you have a tiny but nonzero probability to sample eigenvalues larger than $\sqrt{2\beta N}\approx 4.47...$. Other ensembles have spectral densities with \emph{hard} edges - this means impenetrable walls, which the eigenvalues can never cross.
\end{enumerate}
\end{minipage}}
\end{center}

\section{Layman's classification}
We deal here with ensembles of square matrices with real eigenvalues (the entries can be real, complex or quaternionic random variables). Can we classify these ensembles according to simple features? \\

A useful scheme (covering several scenarios encountered in real life) is the following (see Fig. \ref{fig_classifiedmaterial:diagram}):

\begin{enumerate}
\item {\bf Independent entries:} the first group on the left gathers matrix models whose entries are \emph{independent} random variables - modulo the symmetry requirements. Random matrices of this kind are usually called \emph{Wigner matrices}.\\ 

\emph{Examples:} in this category, you may find adjacency matrices of random graphs \cite{ref_valuetheeigenvalue:Kuehn}, or matrices with independent power-law entries (so-called \emph{L\'evy matrices} \cite{ref_valuetheeigenvalue:Cizeau2}), and power-law banded matrices \cite{ref_valuetheeigenvalue:Mirlin2} among others. Take a moment to download and read these papers - remember the following sentence, found on Richard Feynman's blackboard at the time of his death: ``Know how to solve every problem that has been solved".\\

\item {\bf Rotational invariance:} the second group on the right
is characterized by the so-called \emph{rotational invariance}. In essence, this property means that any two matrices that are related via a similarity transformation\footnote{$U$ is orthogonal/unitary/symplectic if $H$ is real symmetric/complex hermitian/quaternion self-dual, respectively. You surely have noticed that this is precisely the origin of the names given to the ensembles: Orthogonal, Unitary and Symplectic.} ${H}^\prime={U}{H}{U}^{-1}$ occur in the ensemble with the same probability
\begin{equation}
\rho[H] dH_{11}\cdots dH_{NN}=\rho[H'] dH'_{11}\cdots dH'_{NN}\ .
\end{equation}

This requires the following two conditions:
\begin{itemize}
\item $\rho[H] = \rho[{U}{H}{U}^{-1}]$. This means that the jpdf of the entries retains the same functional form before and after the transformation. This imposes a severe constraint on the allowable functional forms thanks to \emph{Weyl's lemma} \cite{ref_classified:weyl}, which states that $\rho[H]$ can only be a function of the traces of the first $N$ powers of $H$,
\begin{equation}
\rho[H]=\varphi\left(\mathrm{Tr}\ H,\mathrm{Tr}\ H^2,\ldots,\mathrm{Tr}\ H^N\right)\ .\label{eq_classified:Weyl}
\end{equation}
Since $\mathrm{Tr}\ H^n=\mathrm{Tr}\ ({U}{H}{U}^{-1})^n$ by the cyclic property of the trace, the $\Leftarrow$ implication is trivial.
\item $dH_{11}\cdots dH_{NN}=dH'_{11}\cdots dH'_{NN}$, i.e. the flat Lebesgue measure is invariant under conjugation by $U$. This is a classical result.
\end{itemize}

The rotational invariance property in essence means that the eigenvectors are not that important, as we can rotate our matrices as freely as we wish, and still leave their statistical weight unchanged.\\

\begin{figure}[t]
\centering
\includegraphics[width=.75\columnwidth]{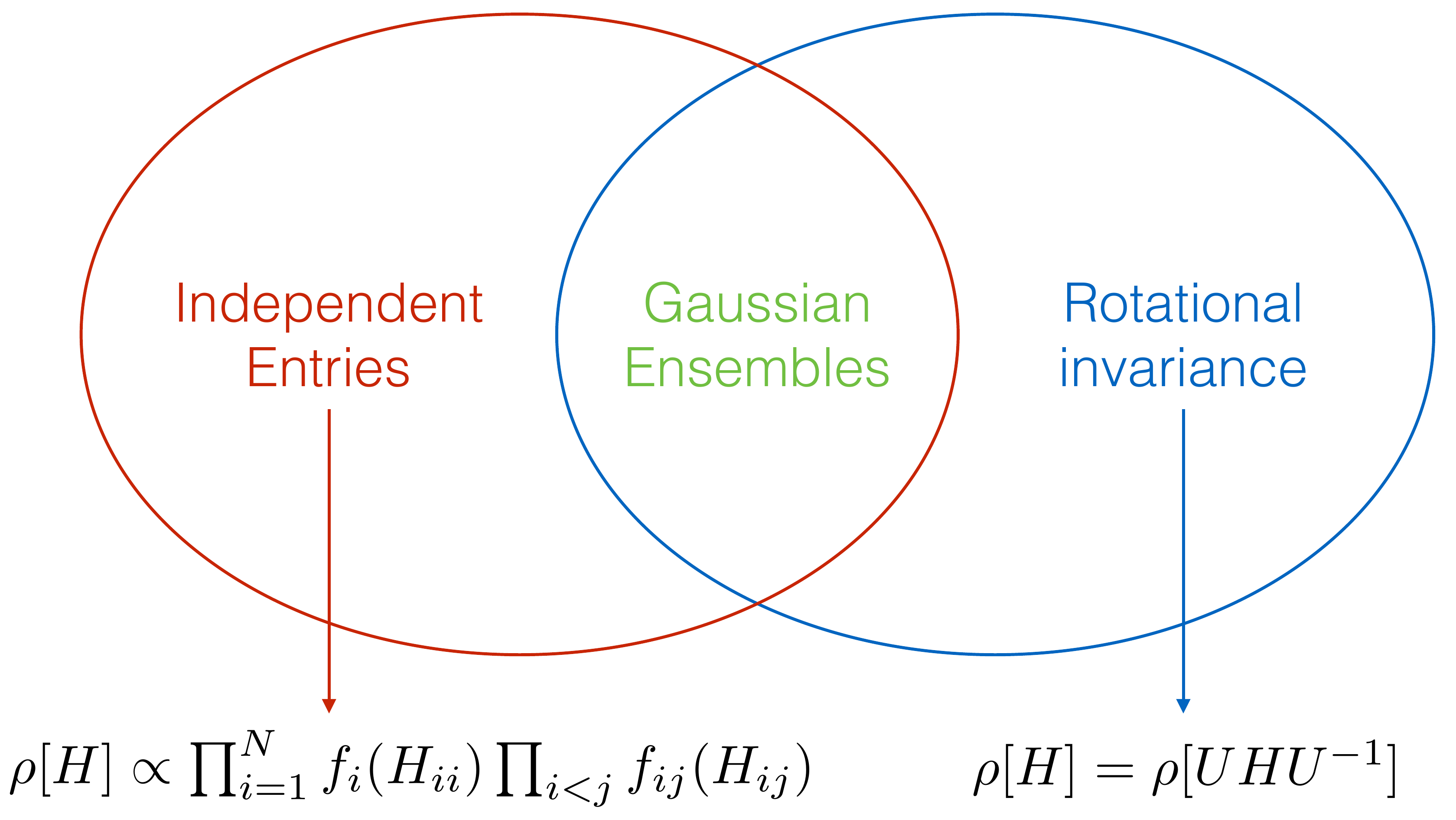}
\caption{Visualization of the layman's classification of random matrix ensembles.}
\label{fig_classifiedmaterial:diagram}
\end{figure}

\emph{Examples:} you may find in this category the Wishart-Laguerre (Chapter \ref{chap:Wishart}) and Jacobi classical ensembles, the so-called ``weakly-confined" ensembles \cite{ref_valuetheeigenvalue:Muttalib2} and many others. The same advice (``download-and-study") applies here.\\

\item What about the intersection between the two classes? It turns out that it contains \emph{only} the Gaussian ensemble\footnote{In its three incarnations: GOE, GUE and GSE.}.\\

This is a consequence of a theorem by Porter and Rosenzweig \cite{ref_valuetheeigenvalue:Rosenzweig2}. And is bad news, isn't it? We have to make a choice: 
if we insist that the ensemble has independent entries, then eigenvectors do matter. If we require a high level of rotational symmetry, then the entries get necessarily correlated. \emph{No free lunch} (beyond the Gaussian)!
\end{enumerate}

\begin{center}
\fbox{\begin{minipage}{33em}
{\it Question.} I can see that the Gaussian ensemble has independent entries. But I do not easily see that it has this ``rotational invariance". \\ \\
$\blacktriangleright$ This can be seen from the jpdf of entries in the upper triangle \eqref{eq_gettingstarted:1:jpdfentriesgaussian}. Show that you can rewrite this jpdf as
\begin{equation}
\rho[H_s]\propto \exp\left(-\frac{1}{2}\Tr {H_s^2}\right)\ ,\label{eq_valuetheeigenvalue:2:jpdfentriesgaussiantrace}
\end{equation}
where $\Tr\cdot$ is the matrix trace (the sum of diagonal element). For example, for the $2\times 2$ real symmetric matrix 
$
H_s=\left(\begin{array}{cc} a& b\\ b & c \\
\end{array}\right)
$, the trace of $H_s$ is $a+c$, and the trace of $H_s^2$ is $a^2+c^2+\mathbf{2}b^2$. You can actually rewrite \eqref{eq_gettingstarted:1:jpdfentriesgaussian} as \eqref{eq_valuetheeigenvalue:2:jpdfentriesgaussiantrace} \emph{only} thanks to that factor $2$...check this! Now, from \eqref{eq_valuetheeigenvalue:2:jpdfentriesgaussiantrace}, the rotational invariance property is much easier to see: for a similarity transformation ${H_s}^\prime={U}{H_s}{U}^{-1}$, one has $\Tr{{H_s}^{\prime 2}}=\Tr{H_s^2}$ (cyclic property of the trace).
\end{minipage}}
\end{center}

\section{To know more...}

\begin{enumerate}
\item Anything worth mentioning beyond the above classification? One important class is represented by the \emph{biorthogonal} ensembles: these are non-invariant, with non-independent entries, and yet their jpdf of eigenvalues is known in terms of the product of two determinants. Check these papers out \cite{ref_valuetheeigenvalue2:Borodin,ref_valuetheeigenvalue2:Desrosiers} for further information.
\item We suggest the following paper \cite{ref_valuetheeigenvalue:albrecht2} about ``histogramming without histogramming". Solid maths and an insightful and unconventional perspective on RMT spectra.
\item For a proof of the Porter-Rosenzweig theorem in the simplified $2\times 2$ case, as well as for a nice and pedagogical introduction to the Gaussian ensembles, we highly recommend the review \cite{ref_valuetheeigenvalue:fyodreview2}.
\item For the mathematically oriented reader, who is looking for more formal classifications of random matrix models, we recommend the mini-review \cite{ref_valuetheeigenvalue:Zirnbauer2} and references therein. 
\end{enumerate}

\chapter{The fluid semicircle}\label{chap:semicircle}

In this Chapter, we set up a statistical mechanics formalism to compute Wigner's semicircle law for Gaussian matrices. You will learn here the so-called ``Coulomb gas technique". 

\section{Coulomb gas}\label{sec_semicircle:secCoulomb}
The Coulomb gas (or fluid) technique is usually attributed to Dyson \cite{ref_semicircle:Dyson3}. Actually, a few years before, Wigner had already used it for the derivation of the semicircle law \cite{ref_semicircle:Wigner2}.\\

Take the jpdf for the Gaussian ensemble \eqref{eq_valuetheeigenvalue:jpdgaussonehalf} 
\begin{equation}
\rho(\llambda_1,\ldots,\llambda_N)=\frac{1}{\mathcal{Z}_{N,\beta}}
e^{-\frac{1}{2}\sum_{i=1}^N
\llambda_i^2}\prod_{j<k}|\llambda_j-\llambda_k|^\beta\ ,\label{eq_semicircle:jpdgaussonehalf}
\end{equation}
and rescale the eigenvalues as $x_i\to x_i\sqrt{\beta N}$.\\

The normalization constant now reads (set $C_{N,\beta}=(\sqrt{\beta N})^{N+\beta N(N-1)/2}$)
\begin{equation}
\mathcal{Z}_{N,\beta}=C_{N,\beta}\int_{\mathbb{R}^N} \prod_{j=1}^N d\llambda_j\ 
e^{-\frac{\beta}{2}N\sum_{i=1}^N
\llambda_i^2}\prod_{j<k}|\llambda_j-\llambda_k|^\beta=C_{N,\beta}\int_{\mathbb{R}^N} \prod_{j=1}^N d\llambda_j\ 
e^{-\beta N^2\mathcal{V}[\bm\llambda]}\ ,\label{eq_semicircle:jpdgaussboltz}
\end{equation}
where the \emph{energy} term in the exponent is
\begin{equation}
\mathcal{V}[\bm\llambda]=\frac{1}{2N}\sum_i \llambda_i^2-\frac{1}{2N^2}\sum_{i\neq j} \ln
|\llambda_i-\llambda_j|\ .\label{eq_semicircle:energyfirst}
\end{equation} The factor $1/2$ in front of the logarithmic term is due to the symmetrization from $i<j$ to $i\neq j$.\\
\vspace{6pt}\\
Stare at \eqref{eq_semicircle:jpdgaussboltz} intensely.\\

We have just exponentiated the product $\prod_{j<k}$, and obtained a canonical partition function\footnote{We are integrating the Gibbs-Boltzmann weight $e^{-\beta N^2\mathcal{V}[\bm\llambda]}$ over all possible positions of the particles.}!\\

\begin{figure}[h]
\centering
\includegraphics[width=.75\columnwidth]{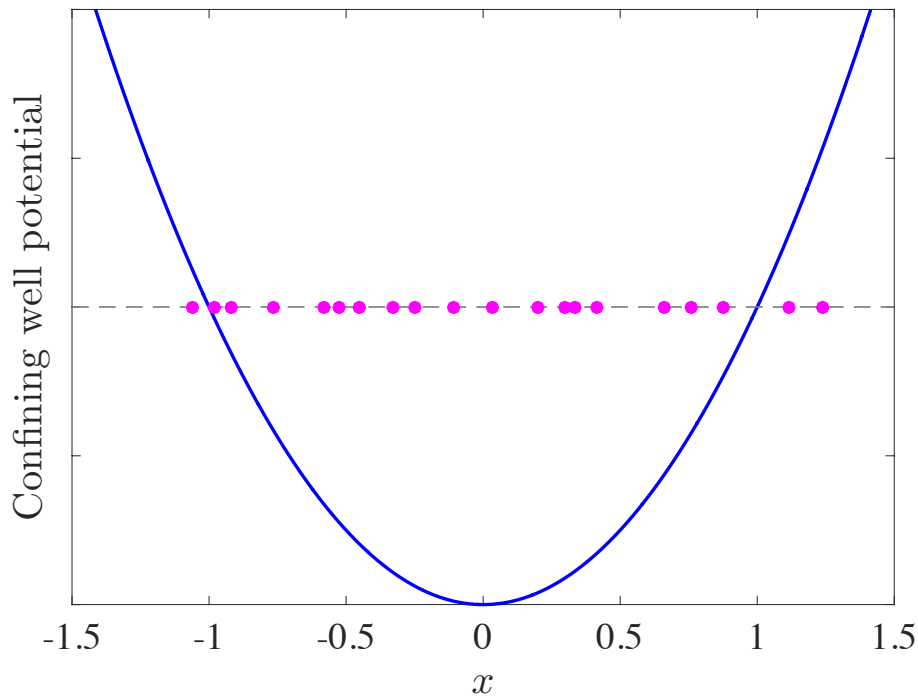}
\caption{Sketch of the quadratic confining potential, which prevents the particles from escaping towards $\pm\infty$.}
\label{fig:thefluidsemicircle_confining_well}
\end{figure}

The Gibbs-Boltzmann weight $e^{-\beta N^2\mathcal{V}[\bm\llambda]}$ corresponds to a thermodynamical fluid of particles with positions $\{x_1,\ldots,x_N\}$ on a line, in equilibrium at ``inverse temperature" $\beta$ under the effect of competing interactions: a quadratic (single-particle) potential (see fig. \ref{fig:thefluidsemicircle_confining_well}), and a repulsive (all-to-all) logarithmic term. The fluid is ``static", as there is no kinetic term in 
$\mathcal{V}[\bm\llambda]$.\\

The presence of the pre-factor $\beta N^2$ shows - at least formally - that the limit $N\to\infty$ is a simultaneous thermodynamic and zero-temperature limit.
A standard thermodynamic argument tells us how to find the equilibrium positions at zero temperature of the particles (eigenvalues) under such interactions: all we need to do is to minimize the \emph{free energy} $F=-(1/\beta)\ln \mathcal{Z}_{N,\beta}$ of this system. The calculation greatly simplifies in the limit $N\to\infty$.

\begin{center}
\fbox{\begin{minipage}{33em}
{\it Question.} Why is this called a ``Coulomb" gas? \\ \\
$\blacktriangleright$ Because we have a logarithmic interaction among charged particles. More precisely, we have a 2D ``fluid" of charges constrained to a line. We know that in 2D the electrostatic potential generated by a point charge is proportional to the \emph{logarithm} of the distance from it - while in 3D, this potential is inversely proportional to the distance, and in 1D is proportional to the distance. Therefore, a 2D charged fluid confined to a line is not quite the same as a 1D fluid!\\
 
A simple way to see this is by using Gauss's law, with a single charge $q$ sitting at the origin on a 2D plane. If we enclose the charge in a $1$-sphere $S$ (i.e. a circle), then we must have $\int_S \bm E \cdot \bm n\propto q$, where $\bm n$ is the normal vector to the circle. If you
assume that the electric field $\bm E$ is rotationally symmetric, i.e. $\bm E=E(r)\hat{\bm r}$, this turns into $E(r)2\pi r\propto q$, implying that $E(r)\propto q/r$. Integrating a field that goes like $1/r$ gives you a logarithmic potential. 
\end{minipage}}
\end{center}

\section{Do it yourself (before lunch)}
So, our goal is to find the free energy $F=-(1/\beta)\ln \mathcal{Z}_{N,\beta}$ for a large number of particles $N\to\infty$. As in many branches of physics, ``larger is easier".\\

We now provide a ``continuum" description of the fluid, based on the following steps.\\

{\bf 1. Introduce a counting function}\\

Define first a normalized one-point counting function
\begin{equation}
\label{eq_semicircle:2:counting} n(\llambda)=\frac{1}{N}\sum_{i=1}^N\delta(\llambda-\llambda_i)\ .
\end{equation}

This is a random function, satisfying $\int_{\mathbb{R}}d\llambda\ n(\llambda)=1$ and $n(\llambda)\geq 0$ everywhere. For finite $N$, this is just a collection of ``spikes" at the location of each eigenvalue. However, for large $N$, it is natural to assume that it will become a smooth function of $\llambda$. We will always work under this assumption\footnote{It may be helpful to think that $n(\llambda)$ is nothing but the limit for $\epsilon\to 0^+$ of a nascent delta function $n_{\epsilon}(\llambda)=\frac{1}{N}\sum_{i=1}^N\frac{e^{-(\llambda-\llambda_i)^2/4\epsilon}}{2\sqrt{\pi\epsilon}}$, where the limit $\epsilon\to 0^+$ is taken at the very end (after the limit $N\to\infty$).}\ .\\

{\bf 2. Coarse-graining procedure}\\

Instead of \emph{directly} summing - or rather integrating - over all configurations of eigenvalues $\{\llambda_1,\ldots,\llambda_N\}$, which in stat-mech we would call \emph{microstates} of our fluid, we first fix a certain one-point profile $n(\llambda)$ (non-negative, smooth and normalized).\\

Sketch your favorite function over $\mathbb{R}$ and call it $n(\llambda)$ - whatever you like, really, provided it is non-negative, smooth and normalized. Then, we sum over all microstates $\{\llambda_1,\ldots,\llambda_N\}$ \emph{compatible with} your sketch $n(\llambda)$ - in a sense to be made clearer. Finally, we sum over all possible (non-negative, smooth and normalized) $n(\llambda)$ you might have come up with in the first place.\\

This \emph{coarse-graining} procedure can be put on slightly cleaner grounds introducing the following representation of unity as a functional integral
\begin{equation}
 1 =\int \mathcal{D}[n(\llambda)]\delta\left[n(\llambda)-\frac{1}{N}\sum_{i=1}^N\delta(\llambda-\llambda_i)\right]\ ,\label{eq_semicircle:2:functint}\
\end{equation}
which enforces the definition \eqref{eq_semicircle:2:counting}. The functional integral runs (so to speak) over \emph{all possible} normalized, non-negative and smooth functions $n(\llambda)$. See \cite{ref_semicircle:MacKenzie} for more details on functional integrations. \\

Inserting this representation of unity inside the multiple integral \eqref{eq_semicircle:jpdgaussboltz} and exchanging the order of integrations, we end up with
\begin{equation}
\mathcal{Z}_{N,\beta} =C_{N,\beta}\int \mathcal{D}[n(\llambda)]\int_{\mathbb{R}^N} \prod_{j=1}^N d\llambda_j\ 
e^{-\beta N^2\mathcal{V}[\bm\llambda]}\delta\left[n(\llambda)-\frac{1}{N}\sum_{i=1}^N\delta(\llambda-\llambda_i)\right]\ .\label{eq_semicircle:jpdgaussboltzafterunity}
\end{equation}\\

{\bf 3. Convert sums into integrals }\\

Using the identities\footnote{Prove them inserting the definition of $n(\llambda)$ into the integrals and using properties of the delta function.}
\begin{align}
\sum_{i=1}^N f(\llambda_i) &= N\int_{\mathbb{R}} n(\llambda)f(\llambda)d\llambda\\
\sum_{i,j=1}^N g(\llambda_i,\llambda_j) &=N^2\iint_{\mathbb{R}^2} d\llambda d\llambda^\prime n(\llambda)n(\llambda^\prime)g(\llambda,\llambda^\prime)\ ,\label{eq_semicircle:2:energycontinuum:double}
\end{align}
we can rewrite the two terms in the energy \eqref{eq_semicircle:energyfirst} as
\begin{align}
&\frac{1}{2N}\sum_{i=1}^N \llambda_i^2 = \frac{1}{2N}\times N\int_{\mathbb{R}} n(\llambda)\llambda^2 d\llambda\label{eq_semicircle:2:sum1}\\
\nonumber & \frac{1}{2N^2}\sum_{i\neq j}\ln |\llambda_i-\llambda_j| =\frac{1}{2N^2}\left[\sum_{i,j}\ln |\llambda_i-\llambda_j| -\sum_i \ln \Delta(\llambda_i)\right]=\\
&\frac{1}{2N^2}\times N^2\iint_{\mathbb{R}^2} d\llambda d\llambda' n(\llambda)n(\llambda')\ln|\llambda-\llambda'|-\frac{1}{2N^2}\times N\int_{\mathbb{R}} d\llambda\ n(\llambda)\ln \Delta(\llambda)\ ,\label{eq_semicircle:2:sum2}
\end{align}
where $\Delta(x)$ is a position-dependent \emph{short-distance cutoff}. What does this mean?\\

Note that in the limit $\epsilon\to 0^+$, the double integral $\iint_{\mathbb{R}^2} d\llambda d\llambda' n_{\epsilon}(\llambda)n_{\epsilon}(\llambda')\ln|\llambda-\llambda'|$ is divergent. This physically corresponds to the infinite-energy contribution originated by two neighboring charges getting ``too close" to each other (the term $i=j$ in the sum $\sum_{i,j}\ln |\llambda_i-\llambda_j|$). The term $\int_{\mathbb{R}} d\llambda\ n_{\epsilon}(\llambda)\ln \Delta(\llambda)$ for $\epsilon\to 0^+$ ``renormalizes" the divergence and produces a finite result. More on how to plausibly fix $\Delta(\llambda)$ later.\\

{\bf 4. $\mathcal{V}[\bm \llambda]\to \mathcal{V}[n(\llambda)]$}\\

Note that in \eqref{eq_semicircle:2:sum1} and \eqref{eq_semicircle:2:sum2} the sums over eigenvalues $\{\llambda_1,\ldots,\llambda_N\}$ have been expressed through the counting function $n(\llambda)$, which - with a slight abuse of notation - will denote from now on its smooth limit as $N\to \infty$.\\

Therefore we can write
\begin{equation}
\mathcal{Z}_{N,\beta} =C_{N,\beta}\int \mathcal{D}[n(\llambda)]e^{-\beta N^2\mathcal{V}[n(\llambda)]}\underbrace{\int_{\mathbb{R}^N} \prod_{j=1}^N d\llambda_j\ 
\delta\left[n(\llambda)-\frac{1}{N}\sum_{i=1}^N\delta(\llambda-\llambda_i)\right]}_{I_{N}[n(\llambda)]}\ .\label{eq_semicircle:jpdgaussboltzafterunityBis}
\end{equation}
The functional $\mathcal{V}[n(\llambda)]$ reads
\begin{align}
\mathcal{V}[n(\llambda)] &=\frac{1}{2}\int_{\mathbb{R}} d\llambda\ \llambda^2 n(\llambda)-\frac{1}{2}\iint_{\mathbb{R}^2} d\llambda d\llambda^\prime
n(\llambda)n(\llambda^\prime)\ln |\llambda-\llambda^\prime |+\frac{1}{2N}\int_{\mathbb{R}}d\llambda\ n(\llambda)\ln\Delta(\llambda)\ .\label{eq_semicircle:2:energycontinuumscaled}
\end{align}\\

{\bf 5. Evaluate the integral $I_{N}[n(\llambda)]$ for large $N$}\\

We now have to evaluate
\begin{equation}
I_{N}[n(\llambda)]=\int_{\mathbb{R}^N} \prod_{j=1}^N d\llambda_j\ 
\delta\left[n(\llambda)-\frac{1}{N}\sum_{i=1}^N\delta(\llambda-\llambda_i)\right]
\end{equation}
in the limit $N\to\infty$.\\

It is quite easy to give a physical interpretation of this multiple integral. It is basically counting how many \emph{microstates} - microscopic configurations of the fluid charges - are compatible with a given \emph{macrostate} - the density profile $n(\llambda)$. We know from standard statistical mechanics arguments that the logarithm of this number should be proportional to the \emph{entropy} of the fluid. Let us see how.\\

Introducing a 'functional' analogue of the standard integral representation for the delta function \cite{ref_semicircle:Rammer}, we can write
\begin{align}
\nonumber  I_{N}[n(\llambda)] &=\int\mathcal{D}[\hat{n}(\llambda)]\int_{\mathbb{R}^N} \prod_{j=1}^N d\llambda_j\exp\left[\mathrm{i}N\int d\llambda\ n(\llambda)\hat{n}(\llambda)-\mathrm{i}\int d\llambda\ \hat{n}(\llambda)\sum_{i=1}^N\delta(\llambda-\llambda_i)\right]\\
\nonumber &=\int\mathcal{D}[\hat{n}(\llambda)]\exp\left[\mathrm{i}N\int d\llambda\ n(\llambda)\hat{n}(\llambda)\right]\left[\int_{\mathbb{R}}dy\ e^{-\mathrm{i}\int d\llambda\ \hat{n}(\llambda)\delta(\llambda-y)}\right]^N\\
&=\int\mathcal{D}[\hat{n}(\llambda)]e^{N S[\hat{n}(\llambda) | n(\llambda)]}\ ,\label{eq_semicircle:2:SPentropy}
\end{align}
where
\begin{equation}
S[\hat{n}(\llambda) | n(\llambda)]=\mathrm{i}\int d\llambda\ n(\llambda)\hat{n}(\llambda)+\mathrm{Log}\int_{\mathbb{R}}dy\ e^{-\mathrm{i}\hat{n}(y)}\ .\label{eq_semicircle:2:actionSPentropy}
\end{equation}\\

This type of integrals is music to the statistical physicist's ears! It is of the form $\int d(\cdot)\exp[\Lambda f(\cdot)]$, with $\Lambda\equiv N$ a very large parameter. Hence it can be evaluated with a Laplace (or saddle-point) approximation \cite{ref_semicircle: Wong}. \\

Finding the critical point of the action $S[\hat{n}(\llambda) | n(\llambda)]$
\begin{align}
0=\frac{\delta S}{\delta \hat{n}(\llambda)}=\mathrm{i}n(\llambda)-\mathrm{i}\frac{e^{-\mathrm{i}\hat{n}(\llambda)}}{\int_{\mathbb{R}}dy\ e^{-\mathrm{i}\hat{n}(y)}}\ ,
\end{align}
from which we obtain
\begin{equation}
e^{-\mathrm{i}\hat{n}(\llambda)}=n(\llambda)\int_{\mathbb{R}}dy\ e^{-\mathrm{i}\hat{n}(y)}\Rightarrow
\mathrm{i}\hat{n}(\llambda)=-\ln\ n(\llambda)-\mathrm{Log}\int_{\mathbb{R}}dy\ e^{-\mathrm{i}\hat{n}(y)}\ ,
\end{equation}
where we ignore spurious phases (recall that in the complex field $\mathrm{Log}\exp(z)$ may not just be equal to $z$!) that would make the action evaluated at the saddle-point complex. Substituting in \eqref{eq_semicircle:2:actionSPentropy}, we obtain 
\begin{equation}
I_{N}[n(\llambda)]\sim\exp\left[-N \int d\llambda\ n(\llambda)\ln\ n(\llambda) \right]\ ,\label{eq_semicircle:Shannon}
\end{equation}
to leading order in $N$. As expected, the term inside square brackets has precisely the form of the Shannon entropy of the density $n(\llambda)$.\\

{\bf 6. Evaluate $\Delta(\llambda)$}\\

Look back again at \eqref{eq_semicircle:2:energycontinuumscaled}. The short-distance cutoff $\Delta(\llambda)$ is yet to be fixed.\\

A standard, physically motivated argument - going back to Dyson for charges on a ring - posits that $\Delta(\llambda)$ - the so-called \emph{self-energy} term - should be taken of the form
\begin{equation}
\Delta(\llambda)\approx \frac{c}{N n(\llambda)}\ ,\label{eq_semicircle:2:Deltax}
\end{equation}
as the higher the density of particles around $\llambda$, the smaller the average distance between them\footnote{We have already met a similar argument in section \ref{sec_valuetheeigenvalue:iidspacing}.}. Also, $N$ charges spread over a distance of $\mathcal{O}(1)$ have a mean spacing $\sim \mathcal{O}(1/N)$, and this justifies the $1/N$ factor. This argument, however plausible, does not seem to have been made rigorous yet, though. Note, in particular, that the constant $c$ in \eqref{eq_semicircle:2:Deltax} cannot be fixed by this simple heuristic argument. While conceptually quite important (see e.g. \cite{ref_semicircle: Serfaty}), this missing bit will prove rather inconsequential in the following. 

\clearpage

{\bf 7. Final expression}\\

Combining \eqref{eq_semicircle:jpdgaussboltzafterunityBis}, \eqref{eq_semicircle:2:energycontinuumscaled}, \eqref{eq_semicircle:Shannon} and \eqref{eq_semicircle:2:Deltax}, the partition function eventually reads
\begin{equation}
\mathcal{Z}_{N,\beta} \simeq C_{N,\beta}\int \mathcal{D}[n(\llambda)] e^{-\beta N^2\mathcal{F}_0[n(\llambda)]+\frac{\beta}{2}N\ln N+\left(\frac{\beta}{2}-1\right)N \mathcal{F}_1[n(\llambda)]-\frac{\beta}{2}N\ln c+o(N)}\ ,\label{eq_semicircle:jpdgaussboltzafterunity2laterFinalExpression}
\end{equation}
where 
\begin{align}
\mathcal{F}_0[n(\llambda)] &=\frac{1}{2}\int d\llambda\ \llambda^2 n(\llambda)-\frac{1}{2}\iint d\llambda d\llambda^\prime
n(\llambda)n(\llambda^\prime)\ln |\llambda-\llambda^\prime|\ ,\label{eq_semicircle:fof2}\\
\mathcal{F}_1[n(\llambda)] &=\int d\llambda\  n(\llambda)\ln\ n(\llambda)\label{eq_semicircle:fof1}\ .
\end{align}

Note that the term $(\beta/2)N\ln N$ is essentially independent of the potential, and can be absorbed into the overall normalization constant. The $\mathcal{O}(N)$ contribution is composed by i) the self-energy term, ii) the entropic term, and iii) a contribution coming from the unknown constant $c$ in \eqref{eq_semicircle:2:Deltax}.\\

{\bf 8. Flash-forward: cross-check with finite-$N$ result}\\

We now cheat a bit.\\

Let us use some information we will actually prove later, namely that the equilibrium density of the fluid is Wigner's semicircle law $n^\star(\llambda)\equiv \rho_{\mathrm{SC}}(\llambda)=\frac{1}{\pi}\sqrt{2-\llambda^2}$.\\

Inserting the semicircle law into \eqref{eq_semicircle:fof2} and \eqref{eq_semicircle:fof1} - and evaluating the corresponding integrals - we obtain
\begin{align}
\mathcal{F}_0[n^\star(\llambda)] &=\frac{3}{8}+\frac{\ln 2}{4}\ ,\\
\mathcal{F}_1[n^\star(\llambda)] &=\frac{1}{2} (1-\ln (2)-2 \ln (\pi ))\ .
\end{align}

Therefore, the partition function \eqref{eq_semicircle:jpdgaussboltzafterunity2laterFinalExpression} reads for large $N$
\begin{align}
\nonumber\mathcal{Z}_{N,\beta} &\simeq C_{N,\beta}\int \mathcal{D}[n(\llambda)] e^{-\beta N^2\mathcal{F}_0[n(\llambda)]+\frac{\beta}{2}N\ln N+\left(\frac{\beta}{2}-1\right)N \mathcal{F}_1[n(\llambda)]-\frac{\beta}{2}N\ln c+o(N)}\\
\nonumber &\approx C_{N,\beta} e^{-\beta N^2\mathcal{F}_0[n^\star(\llambda)]+\frac{\beta}{2}N\ln N+\left(\frac{\beta}{2}-1\right)N \mathcal{F}_1[n^\star(\llambda)]-\frac{\beta}{2}N\ln c+o(N)}\\
&\approx\boxed{\exp\left[\frac{\beta}{4}N^2\ln N+a_\beta N^2+ \frac{1}{2} \left (1 + \frac{\beta}{2} \right )N\ln N+b_\beta N+o(N)\right]}\ ,
\label{eq_semicircle:jpdgaussboltzafterunity2laterV2}
\end{align}
where we used the easy asymptotics
\begin{equation}
\ln C_{N,\beta}\sim \frac{\beta}{4}N^2\ln N+\frac{\beta}{4}(\ln\beta) N^2+\frac{1-\beta/2}{2}N\ln N+\frac{(1-\beta/2)\ln\beta}{2}N\ .\label{eq_valuetheeigenvalue:asymptoticsCN} 
\end{equation}

The constants $a_\beta$ and $b_\beta$ are given as follows:
\begin{align}
a_\beta &=\frac{\beta}{4}\ln\beta-\beta \mathcal{F}_0[n^\star(\llambda)]\ ,\\
b_\beta &=\left(\frac{\beta}{2}-1\right)\mathcal{F}_1[n^\star(\llambda)] +\frac{1-\beta/2}{2}\ln\beta-\frac{\beta}{2}\ln c\ .
\end{align}

Can we check that this result is plausible?\\

Note that for $\beta=2$, the partition function $\mathcal{Z}_{N,\beta=2}$ from \eqref{eq_valuetheeigenvalue:partitionfunction} has a particularly simple expression at \emph{finite} $N$,
\begin{equation}
\mathcal{Z}_{N,\beta=2}=(2\pi)^{N/2} G(N+2)\ ,\label{eq_semicircle:Barnes}
\end{equation}
where $G(x)$ is a Barnes G-function\footnote{The Barnes G-function is defined via the recursion $G(z+1)=\Gamma(z)G(z)$, with $G(1)=1$.}. Hence, if everything was done correctly, the large-$N$ asymptotics of 
\eqref{eq_semicircle:Barnes} should precisely match the large-$N$ behavior \eqref{eq_semicircle:jpdgaussboltzafterunity2laterV2}.\\

Let us check.\\

Using known asymptotics of the Barnes G-function, we deduce that
\begin{equation}
\ln \mathcal{Z}_{N,\beta=2}\sim \frac{1}{2}N^2\ln N-\frac{3}{4}N^2+N\ln N+N\left(\ln(2\pi)-1\right)+\mathcal{O}(1)\ ,
\end{equation}
which coincides (up to the term $N\ln N$ included) with the asymptotics of $\mathcal{Z}_{N,\beta}$ in \eqref{eq_semicircle:jpdgaussboltzafterunity2laterV2} once $\beta$ is set to $2$.\\

 This check should convince you that the ``mean-field" approach - based on a continuum description of the charged fluid of eigenvalues - is indeed capable of capturing the first three terms of the free energy, and only fails at the level of $\mathcal{O}(N)$ contributions - as the renormalized self-energy term cannot be precisely determined by a simple-minded scaling argument.\\

{\bf 9. What's next?}\\

Let us recap what we have done so far. The normalization constant $\mathcal{Z}_{N,\beta}$ of the Gaussian model has been re-interpreted as the canonical partition function of a 2D static fluid of charged particles confined on a line, in equilibrium at inverse temperature $\beta$. For a large number of particles, among all possible configurations, the fluid will choose the one that minimizes its free energy, i.e. the logarithm of this partition function.\\

The partition function has been written as a functional integral over the space of normalized counting functions $n(\llambda)$, see \eqref{eq_semicircle:jpdgaussboltzafterunity2laterFinalExpression}. For large $N$, it lends itself to a saddle-point evaluation, which will be carried out in the next Chapter.

\chapter{Saddle-point-of-view}\label{chap:SaddlePoint}

Let us continue the study of the Coulomb gas method for large random matrices.

\section{Saddle-point. What's the point?}

Earlier we showed that the partition function for the Gaussian model could be represented as
\begin{equation}
\mathcal{Z}_{N,\beta} \simeq C_{N,\beta}\int \mathcal{D}[n(\llambda)] e^{-\beta N^2\mathcal{F}_0[n(\llambda)]+\frac{\beta}{2}N\ln N+\left(\frac{\beta}{2}-1\right)N \mathcal{F}_1[n(\llambda)]-\frac{\beta}{2}N\ln c+o(N)}\ ,\label{eq_semicircle:jpdgaussboltzafterunity2later}
\end{equation}
where 
\begin{align} \label{eq:saddlepoint_functional}
\mathcal{F}_0[n(\llambda)] &=\frac{1}{2}\int d\llambda\ \llambda^2 n(\llambda)-\frac{1}{2}\iint d\llambda d\llambda^\prime
n(\llambda)n(\llambda^\prime)\ln |\llambda-\llambda^\prime|\ ,\\
\mathcal{F}_1[n(\llambda)] &=\int d\llambda\  n(\llambda)\ln\ n(\llambda)\ .
\end{align}

Quite interestingly, the leading term in the exponential is of order $\sim\mathcal{O}(N^2)$ and not of $\sim\mathcal{O}(N)$ as in standard short-range models. As a consequence of the all-to-all coupling between the charged particles, the free energy per particle is dominated by the ``energetic" component at the expenses of the ``entropic" part (sub-leading for large $N$).\\

Recall now that the functional integral runs over functions $n(\llambda)$ that are normalized, i.e. $\int_{\mathbb{R}}d\llambda\ n(\llambda)=1$. We can enforce this constraint introducing another delta function
\begin{equation}
\delta\left[\int_{\mathbb{R}}d\llambda\ n(\llambda)-1\right]=\int_{\mathbb{R}}\frac{dk}{2\pi}e^{\mathrm{i}k \left(\int_{\mathbb{R}}d\llambda\ n(\llambda)-1\right) }\ .
\end{equation}

Rescaling $\mathrm{i}k\to \beta N^2 \kappa$ and ignoring sub-leading terms, you end up with the truly appealing representation
\begin{equation}
\mathcal{Z}_{N,\beta}\approx C_{N,\beta}\int\mathcal{D}[n(\llambda)]\int_{\mathbb{R}} d \kappa\ e^{-\beta N^2 \mathcal{S}[n(\llambda),\kappa]+\mathcal{O}(N)}\ , \label{eq_semicircle:2:funcintegralZ2}
\end{equation}
where the \emph{action} is
\begin{equation}
\mathcal{S}[n(\llambda),\kappa]=\mathcal{F}_0[n(\llambda)] -\kappa\left(\int d\llambda\ n(\llambda)-1\right)\ .\label{eq_semicircle:actionSP}
\end{equation}

A saddle-point evaluation yields\footnote{The pre-factor $C_{N,\beta}$ has the large-$N$ behavior \eqref{eq_valuetheeigenvalue:asymptoticsCN}, whose logarithm is $\sim\mathcal{O}(N^2\ln N)$ and thus strictly speaking leading with respect to $N^2$. However, it is just an overall constant term, and the 'dynamical' part of the free energy is of $\sim\mathcal{O}(N^2)$.}
\be
\mathcal{Z}_{N,\beta}\approx\exp(-\beta N^2 \mathcal{S}[n^\star(\llambda),\kappa^\star])\ .\ee 
Here, $n^\star(\llambda)$ is the minimizer of the functional \eqref{eq:saddlepoint_functional} in the space of normalizable and non-negative functions $n(\llambda)$.\\

We set up the minimization problem by searching for the critical points\footnote{Note that the factor $1/2$ in front of the double integral
disappears because the functional differentiation picks up \emph{two} counting functions, as in
the integrand we have $n(\llambda)n(\llambda^\prime)$. An interesting account on functional differentiation can be found at \cite{ref_semicircle:functional}.}
\begin{equation}
\begin{cases}
0 &=\frac{\delta}{\delta n(\llambda)}\mathcal{S}[n(\llambda),\kappa]\Big|_{{n=n^\star\atop \kappa=\kappa^\star}}=\frac{\llambda^2}{2}-\int_{\mathbb{R}}
d\llambda^\prime n^\star (\llambda^\prime)\ln |\llambda-\llambda^\prime|-\kappa^\star\ ,\\
0 &=\frac{\partial}{\partial \kappa} \mathcal{S}[n(\llambda),\kappa]\Big|_{{n=n^\star\atop \kappa=\kappa^\star}}
\Rightarrow \int_{\mathbb{R}} d\llambda\ n^\star(\llambda)=1\ ,
\end{cases}
\label{eq_semicircle:intcarleman}
\end{equation}
for $\llambda$ in the support of $n^\star(\llambda)$.\\

Effectively, $\kappa^\star$ (hereafter renamed $\kappa$ for simplicity) is just a Lagrange multiplier enforcing the normalization $\int_{\mathbb{R}} d\llambda\ n^\star(\llambda)=1$.\\

What is then the intensive free energy 
\begin{equation}
f=-(1/\beta N^2)\ln \mathcal{Z}_{N,\beta}\label{eq_semicircle:intfreeenergy}
\end{equation}
of our Coulomb gas for $N\to\infty$? It is just given by $f=\mathcal{S}[n^\star(\llambda),\kappa]\equiv \mathcal{F}_0[n^\star(\llambda)]$ - the action evaluated at the saddle-point density. \\

To summarize, the main task is now to find the solution of the integral equation \eqref{eq_semicircle:intcarleman}

\begin{equation}
\boxed{\frac{\llambda^2}{2}-\int_{\mathbb{R}}
d\llambda^\prime n^\star (\llambda^\prime)\ln |\llambda-\llambda^\prime|-\kappa=0}\ ,
\label{eq_semicirclesequel:intcarleman2initial}
\end{equation}
satisfying $n^\star(\llambda)\geq 0$ everywhere, and $\int_{\mathbb{R}}n^\star(\llambda)d\llambda=1$.

\section{Disintegrate the integral equation}

...or (in more academic terms), solve it.\\

As a preliminary observation, note that the \emph{support} of $n^\star(\llambda)$ (i.e. the set of $\llambda$-values for which $n^\star(\llambda)>0$) cannot be the full real line. In the limit $\llambda\to\infty$, the integral term
\begin{equation}
\int_{\mathbb{R}}
d\llambda^\prime n^\star (\llambda^\prime)\ln |\llambda-\llambda^\prime| \sim \ln\llambda\int_{\mathbb{R}}
d\llambda^\prime n^\star (\llambda^\prime)=\ln\llambda\ ,
\end{equation}
- where we used normalization of the density - which is clearly incompatible with the behavior $\sim x^2/2$ of the known term in the equation\footnote{This is true in general for potentials growing super-logarithmically at infinity - not just for the quadratic potential corresponding to Gaussian ensembles.}.\\ 

Therefore, we need to look for a solution over an interval $(a,b)$ of the real line. Indeed, a rather amusing feature of this type of integral equations - of the \emph{Carleman} class - is that the support over which the solution is to be found is itself unknown, and part of the problem! \\

The solution $n^\star\equiv n^\star(\llambda;a,b)$ we find will then be a parametric function of $a,b$. We will then fix the 'optimal' $a,b$ by requiring that the resulting free energy $f$ in \eqref{eq_semicircle:intfreeenergy} is minimized - i.e. any other choice of the support $(a,b)$ for normalized and non-negative function $\tilde{n}(\llambda)\neq n^\star(\llambda)$, once inserted into \eqref{eq_semicircle:intfreeenergy}, would produce a larger value for the free energy.\\

Let us now first convert the integral equation into a ``simpler" one.

\section{Better weak than nothing}

The solution to the integral equation \eqref{eq_semicirclesequel:intcarleman2initial} can be obtained by first differentiating both sides with respect to $\llambda$. Since $\ln |\llambda-\llambda'|$ is not (strictly speaking) differentiable at $\llambda=\llambda'$, we consider the derivative in the \emph{weak} sense.\\

Let $u$ be a function in $\mathcal{L}^1([a,b])$. We say that $v\in \mathcal{L}^1([a,b])$ is a \emph{weak derivative} of $u$ if
\begin{equation}
\int_a^b u(\llambda)\varphi'(\llambda)d\llambda=-\int_a^b v(\llambda)\varphi(\llambda)d\llambda\label{eq_semicirclesequel:weakderivative}
\end{equation}
for all infinitely differentiable functions $\varphi$ with $\varphi(a)=\varphi(b)=0$. The notion of weak derivative extends the standard (strong) derivative to functions that are not differentiable, but integrable in $[a,b]$. Also, if $u$ is differentiable in the standard sense, than its weak and strong derivatives coincide - just using integration by parts.\\

Setting $u(\llambda)=\int d\llambda' n^\star(\llambda')\ln |\llambda-\llambda'|$, we can write
\begin{align}
\nonumber &\int \varphi'(\llambda)\left[\int d\llambda' n^\star(\llambda')\ln |\llambda-\llambda'|\right]d\llambda=\frac{1}{2}\lim_{\epsilon\to 0}
\int \varphi'(\llambda)\left[\int d\llambda' n^\star(\llambda')\ln [(\llambda-\llambda')^2+\epsilon^2]\right]d\llambda\\
&=-\frac{1}{2}\int \varphi(\llambda)\left[\int d\llambda' n^\star(\llambda')\frac{2(\llambda-\llambda')}{(\llambda-\llambda')^2+\epsilon^2}\right]d\llambda
=-\int \varphi(\llambda)d\llambda \left[\mathrm{Pr}\int d\llambda'\frac{n^\star(\llambda')}{\llambda-\llambda'}\right]\ ,
\end{align}
where $\mathrm{Pr}$ stands for Cauchy's principal value\footnote{This means precisely the limit $\lim_{\epsilon\to 0}\left[\int^{x-\epsilon}F(x^\prime)dx^\prime + \int_{x+\epsilon}F(x^\prime)dx^\prime\right]$, if $x$ is a singular point of $F(x)$.}.\\

Comparing with \eqref{eq_semicirclesequel:weakderivative}, we obtain that the weak derivative of $u(\llambda)$ is $\mathrm{Pr}\int d\llambda'\frac{n^\star(\llambda')}{\llambda-\llambda'}$, therefore the new (singular) integral equation to be solved now is
\begin{equation}
\boxed{\mathrm{Pr}\int d\llambda^\prime \frac{n^\star(\llambda^\prime)}{\llambda-\llambda^\prime}=\llambda}\ .\label{eq_semicirclesequel:2tricomi}
\end{equation}

To solve \eqref{eq_semicirclesequel:2tricomi}, we invoke a theorem by Tricomi \cite{ref_semicirclesequel:Tricomi3}, stating that 
\begin{equation}
\mathrm{Pr}\int_a^b dx^\prime\frac{f(x^\prime)}{x-x^\prime}=g(x)\Rightarrow f(x)=\frac{C-\mathrm{Pr}\int_a^b
\frac{dt}{\pi}\frac{\sqrt{(t-a)(b-t)}}{x-t}g(t)}{\pi\sqrt{(x-a)(b-x)}}\ ,\label{eq_semicirclesequel:solutiontricomi}
\end{equation}
provided that $[a,b]$ is a single (compact) support and $C$ is an arbitrary constant.

\begin{center}
\fbox{\begin{minipage}{33em}
{\it Question.} Who tells me that the optimal counting function $n^\star(\llambda)$ is supported on a \emph{single} interval $[a,b]$? \\
$\blacktriangleright$ There is some nice physical intuition behind this. The ``thermodynamical" interpretation of the eigenvalues implies that the gas of particles is confined by a quadratic well with a \emph{single} minimum (see Fig. \ref{fig:thefluidsemicircle_confining_well}). It is then physically reasonable to foresee that the particles will fill the single minimum of the potential. If a potential has many minima, then it is possible that $n^\star(\llambda)$ ``splits" into as many connected components as the number of minima of the potential. Any attempt to use \eqref{eq_semicirclesequel:solutiontricomi} in these multiple-support cases will produce unphysical solutions. 
\end{minipage}}
\end{center}

Evaluating the principal value integral with $g(t)=t$ and imposing
the normalization $\int_a^b d\llambda\ n^\star(\llambda)=1$, we get
\begin{equation}
n^\star(\llambda)=\frac{1}{\pi\sqrt{(\llambda-a)(b-\llambda)}}\left[1-\llambda^2+\frac{1}{2}(a+b)\llambda+\frac{1}{8}(b-a)^2\right]\ .\label{eq_semicirclesequel:2:nstarsolutionab}
\end{equation}\\

Note that the density in \eqref{eq_semicirclesequel:2:nstarsolutionab} is a solution of the integral equation \eqref{eq_semicirclesequel:2tricomi} between $a$ and $b$ \emph{for any choice of $a$ and $b$}. How to fix the ``optimal'' $a$ and $b$ will be the subject of the next sections.\\

[Of course, do not even consider trusting us on this. You are not allowed to proceed until you have derived \eqref{eq_semicirclesequel:2:nstarsolutionab} yourself. Sorry.]

\section{Smart tricks}

Now, stare at \eqref{eq_semicirclesequel:2:nstarsolutionab} intensely. As promised, the function $n^\star(\llambda)$ (defined for $\llambda\in(a,b)$) indeed depends on two free parameters $a$ and $b$.\\

We need now to compute the intensive free energy 
\be
f=\mathcal{F}_0[n^\star(\llambda)]\ .
\ee
It will of course depend as well on the two free parameters $a$ and $b$, which arose as a Phoenix from the ashes of the integral equation \eqref{eq_semicirclesequel:2tricomi}.\\

A couple of smart tricks will make our life easier. First, we would really like to get rid of the double integral in 
\begin{equation}
f\equiv\mathcal{F}_0[n^\star(\llambda)]=\frac{1}{2}\int d\llambda\ \llambda^2 n^\star(\llambda)-\frac{1}{2}\iint d\llambda
d\llambda^\prime n^\star(\llambda)n^\star(\llambda^\prime)\ln |\llambda-\llambda^\prime|\ .\label{eq_semicirclesequel:2:energycontinuumscaledsameNv2}
\end{equation}
To do that, we multiply the saddle point equation \eqref{eq_semicirclesequel:intcarleman2initial}
\begin{equation}
\frac{\llambda^2}{2}-\int
d\llambda^\prime n^\star (\llambda^\prime)\ln |\llambda-\llambda^\prime|-\kappa=0
\label{eq_semicirclesequel:intcarlemanv2}
\end{equation}
by $n^\star(\llambda)$ and integrate over $\llambda$. This way we obtain
\begin{equation}
\iint d\llambda d\llambda^\prime n^\star(\llambda)n^\star(\llambda^\prime)\ln |\llambda-\llambda^\prime|=\frac{1}{2}\int d\llambda\ n^\star(\llambda)\llambda^2-\kappa\ ,
\end{equation}\\
where we used $\int n^\star(\llambda)d\llambda=1$.\\

Next, we fix the Lagrange multiplier $\kappa$ by setting $x=a$ in \eqref{eq_semicirclesequel:intcarlemanv2}. We obtain $\kappa=a^2/2-\int_a^b d\llambda\ n^\star(\llambda)\ln (\llambda-a)$. Combining everything, we get
\begin{equation}\label{eq_semicirclesequel:2:Vnstarsimplified}
f\equiv\mathcal{F}_0[n^\star(\llambda)]=\frac{1}{4}\int_a^b d\llambda\ n^\star(\llambda)\llambda^2+\frac{a^2}{4}-\frac{1}{2}\int_a^b d\llambda\ n^\star(\llambda)\ln (\llambda-a)\ .
\end{equation}\\

No more $\kappa$, and no more double integrals. Nice, uh?\\

\section{The final touch}

Inserting \eqref{eq_semicirclesequel:2:nstarsolutionab} into \eqref{eq_semicirclesequel:2:Vnstarsimplified} and computing the integrals with the help of an abacus\footnote{It may be useful to first change variables $z=(\llambda-a)/(b-a)$. The resulting integrals can then be handled by most symbolic computation programs. }, we obtain
\begin{align}
\nonumber f\equiv f(a,b) &=\frac{1}{512} \left(-9 a^4+4 a^3 b+2 a^2 \left(5 b^2+48\right)+4 a b
   \left(b^2+16\right)\right.\\
   &\left.-256 \ln (b-a)-9 b^4+96 b^2+512 \ln (2)\right)\ .\label{eq_semicirclesequel:3:freeenergyab}
\end{align}\\

We now have our (quite ugly) intensive free energy: In the code [$\spadesuit$ \verb"integral_check.m"] we provide a simple numerical confirmation that the above result is equivalent to \eqref{eq_semicirclesequel:2:Vnstarsimplified}.\\

All we need to do is to minimize it with respect to $a$ and $b$ - the (soft) edge points of the support of $n^\star(\llambda)$.\\

If you do that, you will obtain the solution\footnote{The fact that the soft edges are symmetrically located around the origin is a consequence of the symmetry of the confining potential under the exchange $x\to -x$.} $a=-\sqrt{2}$ and $b=\sqrt{2}$, which imply for $n^\star(\llambda)$ from \eqref{eq_semicirclesequel:2:nstarsolutionab} the following form
\begin{equation}
\boxed{n^\star(\llambda)\equiv \rho_{\mathrm{SC}}(\llambda)=\frac{1}{\pi}\sqrt{2-\llambda^2}}\ ,\label{eq_semicirclesequel:3:semicircle}
\end{equation}
the famous \emph{Wigner's semicircle law}. Very appropriate name, given that it is \emph{not} the equation of a semicircle, but rather of a semi-ellipse. The code [$\spadesuit$ \verb"Tricomi_check.m"] offers a numerical verification that the semicircle indeed solves equation \eqref{eq_semicirclesequel:2tricomi} for $a=-b=-\sqrt{2}$. \\

How to show this analytically, though?\\

We need to prove that
\begin{equation}
\mathrm{Pr}\int_{-\sqrt{2}}^{\sqrt{2}}d\llambda'\frac{\sqrt{2-\llambda'^2}}{\pi(\llambda-\llambda')}=\llambda\ .
\end{equation}

The primitive of the integrand is - ignoring an additive constant
\begin{equation}
F(y)=\frac{\sqrt{2-x^2} \ln \left(\varphi(x,y)\right)-\sqrt{2-x^2} \ln (x-y)+x \arcsin\left(\frac{y}{\sqrt{2}}\right)-\sqrt{2-y^2}}{\pi }\ ,
\end{equation} 
where
\begin{equation}
\varphi(x,y)=\sqrt{2-x^2} \sqrt{2-y^2}-x y+2\ .
\end{equation}

Hence all you have to show is
\begin{equation}
\lim_{\epsilon\to 0^+} \left[F(x-\epsilon )-F\left(-\sqrt{2}\right)+F\left(\sqrt{2}\right)-F(x+\epsilon )\right]=x \ , \qquad -\sqrt{2}\leq x\leq \sqrt{2}\ .
\end{equation}

Have a go at it!\\

\section{Epilogue}

What is again the interpretation of the ``semicircular" $n^\star(\llambda)$? It is just the equilibrium profile of a gas of many charged particles on a line, which minimizes the free energy of the gas. In the ``eigenvalue" language, it represents the normalized histogram of the $N$ eigenvalues of a \emph{single} (very big!) instance of the Gaussian ensemble. The property that this object also faithfully represents the spectrum \emph{averaged over many samples} (i.e. $n^\star(\llambda)=\langle n(x)\rangle=\rho(x)$) is called \emph{self-averaging} and we will assume it to hold.\\

\begin{figure}[t]
\centering
\includegraphics[width=.75\columnwidth]{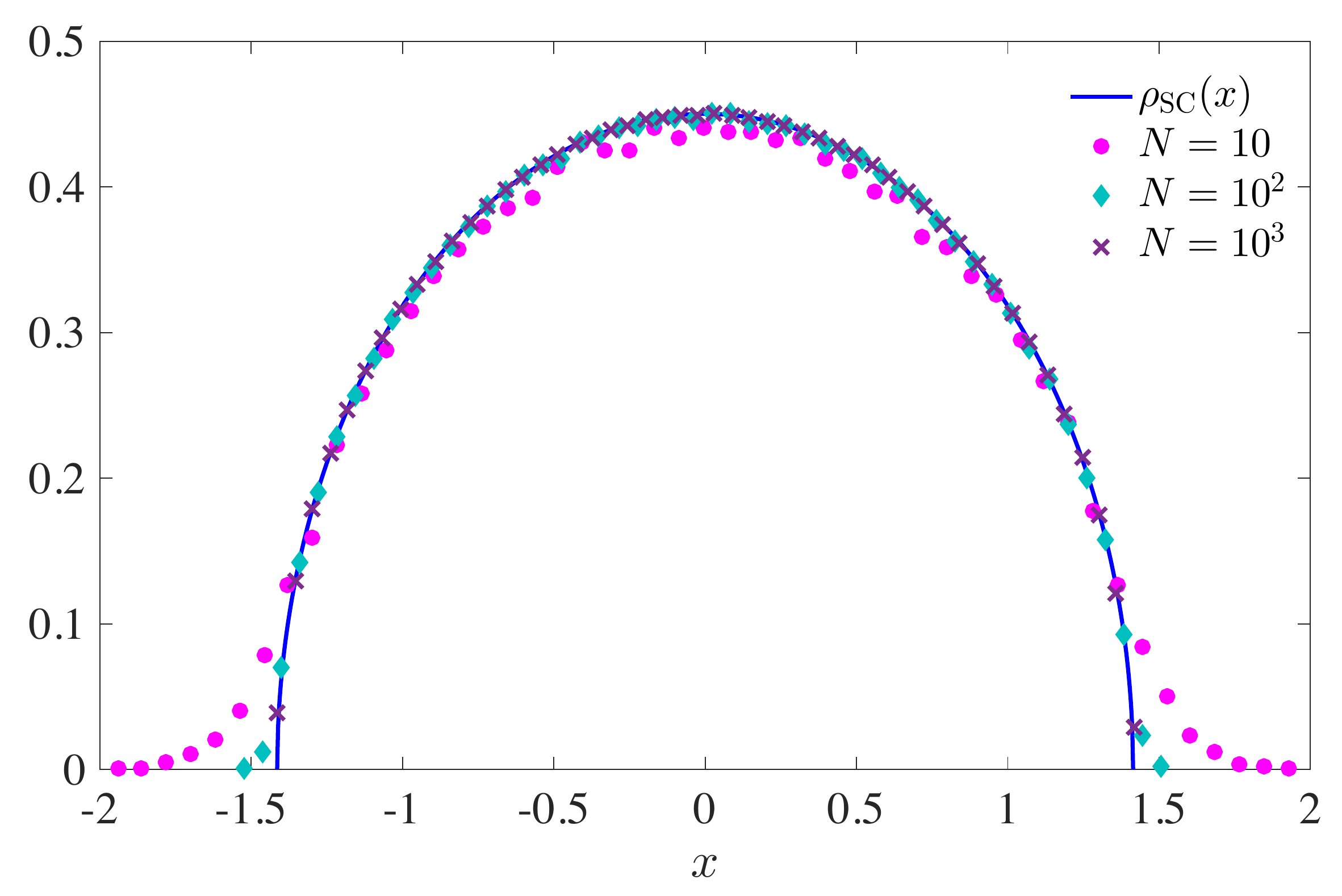}
\caption{Numerical check of the semicircle law for GOE. Increasing the value of $N$, after a suitable rescaling, the eigenvalue histograms collapse on top of the semicircle curve.}
\label{fig:saddlepointofview_numerical_check}
\end{figure}

The code [$\spadesuit$ \verb"Coulomb_gas.m"] provides a numerical verification of what we worked on in this Chapter and the previous one. It simulates the Coulomb gas through a simple Monte Carlo procedure, which produces the equilibrium density for long enough times. Also, a numerical check of the semicircle distribution can be performed directly, i.e. through the numerical diagonalization of random matrices, with the code [$\spadesuit$ \verb"Gaussian_finite_N_rescaled.m"] (see Fig. \ref{fig:saddlepointofview_numerical_check}). \\

Note that, at the very beginning of the derivation of \eqref{eq_semicirclesequel:3:semicircle}, we rescaled the eigenvalues by $\sqrt{\beta N}$ (Eq. \eqref{eq_semicircle:jpdgaussboltz}). Therefore, in the simulations we need to perform the same rescaling of our eigenvalues by $\sqrt{\beta N}$ before comparing the histogram to the theoretical semicircle. This is in agreement with the precise statement we made in the second Question in Chapter \ref{chap:Classifiedmaterial}, namely
\begin{equation}
\lim_{N\to\infty}\sqrt{\beta N}\rho(\sqrt{\beta N}\llambda)=\rho_{\mathrm{SC}}(\llambda)\ ,
\end{equation}
where the function $\rho_{\mathrm{SC}}(\llambda)=\frac{1}{\pi}\sqrt{2-\llambda^2}\equiv n^\star (\llambda)$ is $\beta$-independent.\\

As a final remark, what happens if the confining potential is not quadratic? In general, if our invariant ensemble is characterized by a joint probability density of the entries of the form
\begin{equation}
\rho[H]\propto\exp\left[-\mathrm{Tr}V(H)\right]\ ,
\end{equation}
then the joint law of the eigenvalues is of the form
\begin{equation}
\rho(\llambda_1,\ldots,\llambda_N)\propto \exp\left[-\sum_{i=1}^N V(\llambda_i)\right]\prod_{j<k}|\llambda_j-\llambda_k|^\beta\label{eq_saddlepoint:potential}
\end{equation}
and the analogue of the Tricomi equation for the spectral density is
\begin{equation}
\boxed{\mathrm{Pr}\int d\llambda^\prime \frac{n^\star(\llambda^\prime)}{\llambda-\llambda^\prime}=V'(\llambda)}\ .\label{eq_semicirclesequel:2tricomigeneral}
\end{equation}

Try to solve for $n^\star(\llambda)$ in the case $V(\llambda)=\llambda-\alpha\ln\llambda$ ($\llambda>0$). This will correspond to the \emph{Wishart-Laguerre ensemble} of random matrices, which will be extensively discussed in Chapter \ref{chap:Wishart}.

\begin{center}
\fbox{\begin{minipage}{33em}
{\it Question.} Do all existing random matrix ensembles have the semicircle as their average spectral density? \\ \\
$\blacktriangleright$ Certainly not! The spectral density is highly \emph{non-universal} - i.e. it strongly depends on the ensemble you consider. This said, it is true that many ensembles share it as their spectral density for large $N$. This is the case for instance of Wigner ensembles (non-invariant), when the distribution of entries decays sufficiently fast at infinity (see \cite{ref_semicirclesequel:Erdos}).
\end{minipage}}
\end{center}

\begin{center}
\fbox{\begin{minipage}{33em}
{\it Question.} What are the moments of the semicircle law? \\ \\
$\blacktriangleright$ They are given by the so called \emph{Catalan numbers}. More precisely, defining
\be
\langle\mathrm{Tr} X^k\rangle=\int d\llambda_1\cdots d\llambda_N \rho(\llambda_1,\ldots,\llambda_N)\sum_{i=1}^N\llambda_i^k=N\int d\llambda\ \llambda^k \rho(\llambda)\ ,
\ee
where $ \rho(\llambda_1,\ldots,\llambda_N)$ is the jpdf for the Gaussian ensemble \eqref{eq_valuetheeigenvalue:jpdgaussonehalf} and $\rho(\llambda)$ its one-point marginal for finite $N$, we have the relation
\be
\lim_{N\to\infty} \frac{\langle\mathrm{Tr} X^{2n}\rangle}{\beta^{n} N^{n+1}}=\frac{1}{\pi}\int_{-\sqrt{2}}^{\sqrt{2}}dy\ y^{2n} \sqrt{2-y^2}=\frac{C_n}{2^n}\ ,
\ee
where $C_n=\frac{1}{n+1} {2n \choose n}$ is the $n$th Catalan number. Catalan numbers occur in a variety of combinatorial problems, for example $C_n$ is the number of ways to correctly match $n$ pairs of brackets.
\end{minipage}}
\end{center}

\begin{center}
\fbox{\begin{minipage}{33em}
{\it Question.} I see that the Coulomb gas treatment is insensitive to the precise value of $\beta$. But is it possible to construct an explicit random matrix ensemble $\rho[H]$, whose eigenvalues are distributed according to a Coulomb gas with $\beta\neq 1,2,4$? \\ \\
$\blacktriangleright$ Yes! This has been achieved by Dumitriu and Edelman \cite{ref_semicirclesequel:Dumitriu}, who produced ensembles of tridiagonal matrices - hence non-invariant - with independent but not identically distributed nonzero entries, whose jpdf of eigenvalues can be nevertheless computed analytically. This jpdf turns out to be equal to the Gaussian or Wishart-Laguerre ones, albeit with a continuous Dyson index $\beta>0$ (it enters as a \emph{parameter} of the distribution of the nonzero entries). These ensembles are very useful also on the numerical side: they provide a much faster way to sample GXE-distributed eigenvalues (with X=O,U,S), without having to diagonalize full Gaussian matrices!
\end{minipage}}
\end{center}

\begin{center}
\fbox{\begin{minipage}{33em}
{\it Question.} If I drop the symmetry requirements on the entries of the ensemble $(H_{ij}\neq H_{ji})$, what is the resulting analogue of the semicircle law for \emph{complex} eigenvalues? \\ \\
$\blacktriangleright$ This is called the \emph{Girko-Ginibre} (or \emph{circular}) law. In essence, for any sequence of random $N\times N$ matrices whose entries are i.i.d. random variables, all with mean zero and variance equal to $1/N$, the limiting spectral density is the uniform distribution over the unit disc in the complex plane.
\end{minipage}}
\end{center}

\section{To know more...}

\begin{enumerate}
\item The Gaussian ensemble for $\beta=2$. The eigenvalues can be interpreted as the positions of fermions in a harmonic trap. To understand this mapping, have a look at \cite{ref_semicircle:Marino} and references therein. 
\item Recently, the Coulomb gas technique has been improved and modified to tackle a wealth of different problems. It all started with a beautiful calculation on the following problem: what is the probability that \emph{all} the eigenvalues of a Gaussian matrix are negative? Check this paper out \cite{ref_semicircle:MajumdarDean2}.
\item The Gaussian ensembles can also come in a variant called \emph{fixed-trace}: this means that one multiplies the jpdf \eqref{eq_semicircle:jpdgaussonehalf} by $\delta\left(\sum_{i=1}^N\llambda_i^2-t\right)$, which fixes the squared trace to the value $t$ (see \cite{ref_semicircle:Akemann1,ref_semicircle:Akemann2} for details).
\item The normalization constant $\mathcal{Z}_{N,\beta}$ for the Gaussian ensemble can be computed for finite $N$, with simple algebraic manipulations on the so called \emph{Selberg integral}
\be \int_0^1 d\bm\llambda |\Delta_N(\bm\llambda)|^\beta \prod_{i=1}^N
\llambda_i^{a-1}(1-\llambda_i)^{b-1}\ .\ee 
It was computed by the norwegian mathematician A. Selberg, who showed that, when
it exists, it is given by \be \prod_{j=1}^N \frac{\Gamma(1+\beta
j/2)\Gamma(a+(N-j)\beta/2)\Gamma(b+(N-j)\beta/2)}
{\Gamma(1+\beta/2)\Gamma(a+b+\beta(2N-j-1)/2)}\ .\ee
To know more about recent developments in the beautiful theory of Selberg integrals, have a look at \cite{ref_valuetheeigenvalue:Forrester_Warnaar}.
\end{enumerate}

\chapter{Time for a change}\label{chap:ChangeVar}

In this Chapter, we show how to compute the jpdf of eigenvalues for random matrix models - whenever possible.

\section{Intermezzo: a simpler change of variables}\label{sec:ChangeVar:Intermezzo}
Suppose we have to compute the following double integrals
\begin{equation}
I_1=\int_{\mathbb{R}^2} dx\ dy\ \rho_1(x,y) \qquad I_2=\int_{\mathbb{R}^2} dx\ dy\ \rho_2(x,y)\ ,
\end{equation}
with $\rho_1(x,y)=f(x^2+y^2)$ and $\rho_2(x,y)=x f(x^2+y^2)$. Here, $f(t)$ is a function of your choice that makes both integrals convergent.\\

A good strategy is to make the ``polar" change of variables $\{x,y\}=\{r\cos\theta,r\sin\theta\}$ to write
\begin{equation}
I_1=\int_0^\infty dr\int_0^{2\pi} d\theta \hat\rho_1(r,\theta),\qquad I_2=\int_0^\infty dr\int_0^{2\pi} d\theta \hat\rho_2(r,\theta)\ ,
\end{equation}
where $\hat\rho_1(r,\theta) = r f(r^2)$ and $\hat\rho_2(r,\theta)=r^2\cos\theta f(r^2)$. Obviously, we had to include here the extra Jacobian factor 
\begin{equation}
J(r,\theta)=\left(\begin{array}{cc}
\frac{\partial x}{\partial r} & \frac{\partial x}{\partial \theta}\\
\frac{\partial y}{\partial r} & \frac{\partial y}{\partial \theta}\\
\end{array}\right)=r\ .\label{eq_vandermonde:jacobianpolar}
\end{equation}
Therefore, we can formally write $\rho_1(x,y)dx dy=\hat\rho_1(r,\theta)dr d\theta$ (and similarly for $\rho_2$), meaning that the two expressions give the same result once integrated over ``corresponding" domains (e.g. $\mathbb{R}^2 \to (0,\infty)\times (0,2\pi)$).\\

This is all trivial and easy. But together with the following two remarks, it is all you need to know to fully understand what happens in the RMT case, with jpdf of entries and eigenvalues all over the place.

\begin{enumerate}
\item $\hat\rho_1(r,\theta)$ (the \emph{new} integrand) is nothing
 but $ \rho_1(r\cos\theta,r\sin\theta)\times |J(r,\theta)| $ (the \emph{old} integrand, written in terms of the new variables, times the Jacobian factor) - and similarly for $\hat\rho_2$.
\item The \emph{marginal} $\hat\rho_1(r)=\int_0^{2\pi}d\theta \hat\rho_1(r,\theta)$ is easier to compute than the corresponding $\hat\rho_2(r)$. This for two reasons: i) the original $\rho_1(x,y)$, once expressed in the new polar variables, no longer depends on one of them $(\theta)$, and ii) also the Jacobian does not depend on $\theta$. So the integration in $\theta$ becomes trivial and gives just a constant factor $2\pi$.
\end{enumerate}

\section{...that is the question}

Take the case of real symmetric matrices for simplicity - call them $H$ instead of $H_s$ from now on.\\
 
Look again at the jpdf of \emph{eigenvalues} \eqref{eq_valuetheeigenvalue:jpdgaussonehalf} for the GOE ensemble $(\beta=1)$
\begin{equation}
\rho(\llambda_1,\ldots,\llambda_N)=\frac{1}{\mathcal{Z}_{N,\beta=1}}
e^{-\frac{1}{2}\sum_{i=1}^N
\llambda_i^2}\prod_{j<k}|\llambda_j-\llambda_k|\ .\label{eq_vandermonde:jpdgaussonehalf}
\end{equation}
We gave it without proof.\\

How to obtain it from the jpdf of \emph{entries} in the upper triangle, $\rho[H]$
\begin{equation}
\rho[H]=\prod_{i=1}^N \frac{e^{-H_{ii}^2/2}}{\sqrt{2\pi}}\prod_{i<j}\frac{e^{-H_{ij}^2}}{\sqrt{\pi}}\ ?
\end{equation}

In this Chapter, we provide an answer to this outstanding question.

\section{Keep your volume under control}

A real symmetric matrix can be diagonalized by an orthogonal matrix $O$ as $H=OXO^T$, with $X=\mathrm{diag}(\llambda_1,\ldots,\llambda_N).$\\

Orthogonal $N\times N$ matrices are characterized by the property that $OO^T=\mathbbm{1}$, where $\mathbbm{1}$ is the identity matrix. As a subspace  of $\mathbb{R}^{N^2}$, these matrices form a 
sub-manifold $\mathbb{V}_N$ of dimension $N(N-1)/2$, called the \emph{Stiefel manifold}. $dO$ is precisely its ``volume element" - the analog of $d\theta$ in the warm-up example above.\\

We know that $\int_0^{2\pi}d\theta=2\pi$. It is perhaps intuitive to give this number $2\pi$ the meaning of ``volume" occupied while $d\theta$ spans the entire one-dimensional manifold (the circumference of the unit circle). What is, then, the ``volume" occupied by orthogonal matrices in $\mathbb{R}^{N^2}$?\\

A relatively simple calculation \cite{ref_vandermonde:muirhead} shows that
\begin{equation}
\mathrm{Vol}(\mathbb{V}_N)=\int_{\mathbb{V}_N}dO=\frac{2^N\pi^{N^2/2}}{\Gamma_N(N/2)}\ ,
\label{eq_vandermonde:VolVN}
\end{equation}
where 
\begin{equation}
\Gamma_m(a)=\pi^{m(m-1)/4}\prod_{i=1}^m\Gamma(a-(i-1)/2)\ .
\end{equation}
We will use this result in a minute.\\

If we call
\begin{equation}
\mathcal{D}O=\frac{dO}{\mathrm{Vol}(\mathbb{V}_N)}\ ,
\end{equation}
this defines the so-called \emph{Haar measure} on the orthogonal group. The Haar measure is invariant under orthogonal conjugation, and defines a probability space on orthogonal matrices. For further information, consult \cite{ref_vandermonde:muirhead,ref_vandermonde:edelmanthesis,ref_vandermonde:edelmanweb,ref_vandermonde:mathai}.

\section{For doubting Thomases...}

Let us compute the volume $\mathrm{Vol}(\mathbb{V}_2)$ for $2\times 2$ orthogonal matrices ``from first principles''\footnote{Alternatively, one may notice that the elements of $\mathbb{V}_2$ can be written either in the form $\begin{pmatrix}\cos\theta&\sin\theta\\-\sin\theta&\cos\theta\end{pmatrix}$ (rotations in the plane by an angle $\theta$) or in the form $\begin{pmatrix}\cos\theta&\sin\theta\\\sin\theta&-\cos\theta\end{pmatrix}$ (rotations followed by a reflection). That is, this group has two disconnected components. Clearly, each of these components has a volume $2\pi$, so the volume of $\mathbb{V}_2$ is $4\pi$.}.\\

Let 
\begin{equation}
O=\begin{pmatrix}
o_{11} & o_{12}\\
o_{21} & o_{22}
\end{pmatrix}\ .
\end{equation}

The $\{o_{ij}\}$ are real variables. The volume we are after is
\begin{equation}
\mathrm{Vol}(\mathbb{V}_2)=\int \prod_{i,j=1}^2 do_{ij} \delta\left(\sqrt{o_{11}^2+o_{21}^2}-1\right)\delta\left(\sqrt{o_{12}^2+o_{22}^2}-1\right)\delta(o_{11}o_{12}+o_{21}o_{22})\ ,
\end{equation}
where the delta functions enforce the constraints on the columns of $O$ being orthogonal with each other, and each having unit norm.\\

Changing to polar coordinates, we get
\begin{align}
\nonumber \mathrm{Vol}(\mathbb{V}_2) &=\int_0^{2\pi}d\theta\int_0^{2\pi}d\phi\int_0^\infty dr\ r\delta(r-1)\int_0^\infty dR\ R\delta(R-1)\delta\left(r R\cos(\theta-\phi)\right)\\
&=\int_0^{2\pi}d\theta\int_0^{2\pi}d\phi \delta\left(\cos(\theta-\phi)\right)=4\pi\ ,
\end{align}
in agreement with \eqref{eq_vandermonde:VolVN} for $N=2$ as it should.

\section{Jpdf of eigenvalues and eigenvectors}

As in section \ref{sec:ChangeVar:Intermezzo} - but this time with more variables - we are after the change of variables $H\to \{\bm\llambda,O\}$
\begin{equation}
\rho(H_{11},\ldots,H_{NN})\prod_{i\leq j} d H_{ij} =\underbrace{\rho(H_{11}(\bm\llambda,O),\ldots,H_{NN}(\bm\llambda,O))\Big|J(H\to \{\bm\llambda,O\})\Big|}_{\hat\rho(\llambda_1,\ldots,\llambda_N,O)}dO\prod_{i=1}^N d\llambda_i\ .\label{eq_vandermonde:changeofvariablesjpds}
\end{equation}
On the left hand side, the jpdf of the $N(N+1)/2$ entries of $H$ in the upper triangle, including the diagonal. On the right hand side, the jpdf $\hat{\rho}$ of \emph{both} eigenvalues $(N)$ \emph{and} independent eigenvector components ($N(N-1)/2$, the dimension of the Stiefel manifold spanned by the orthogonal group over the reals). The number of ``degrees of freedom" is OK, thanks to the mind-wrecking and highly nontrivial identity $N(N+1)/2=N+N(N-1)/2$.\\

Clearly, on the right hand side we had to include the Jacobian of the change of variables, which we are going to compute below. While in principle this Jacobian could depend on the \emph{full} set of variables $\{\bm\llambda,O\}$, it turns out that it \emph{only} depends on the eigenvalues $\{\bm\llambda\}$, exactly as it happens for the change to polar coordinates \eqref{eq_vandermonde:jacobianpolar}.\\

In our RMT case, this Jacobian is precisely the so-called \emph{Vandermonde determinant}\footnote{Why this is indeed a determinant in disguise will become clearer very shortly.}, 
\be
J(H\to \{\bm\llambda,O\})=\prod_{j>k}(\llambda_j-\llambda_k)\ .\label{eq_vandermonde:fundamentalJacobianrelation}
\ee

This can be generalized to the hermitian and quaternion self-dual cases. The only difference is that the Vandermonde is then raised to the power $\beta=2,4$ respectively. We will prove this in the next Chapter.\\

\section{Leave the eigenvalues alone}

Now, stare at the right hand side of \eqref{eq_vandermonde:changeofvariablesjpds} carefully.\\

The \emph{joint} probability density of eigenvalues \emph{and} eigenvectors $\hat\rho(\llambda_1,\ldots,\llambda_N,O)$ is the product of two terms: the jpdf of entries - written as a function of eigenvalues \emph{and} eigenvectors - \emph{times} the Jacobian - which is a function of the eigenvalues \emph{alone}.\\

Then the next question is: how can I get the jpdf of eigenvalues \emph{alone}? Well, you will need to integrate out the eigenvector components $\{O\}$ in \eqref{eq_vandermonde:changeofvariablesjpds}. More precisely
\begin{equation}
\hat\rho(\llambda_1,\ldots,\llambda_N)d\bm\llambda =d\bm\llambda \int_{\mathbb{V}_N} dO  \hat\rho(\llambda_1,\ldots,\llambda_N,O)\ ,\label{eq_vandermonde:marginalizejpdf}
\end{equation}
exactly as we did earlier on to find $\hat\rho_{1,2}(r)$ from $\hat\rho_{1,2}(r,\theta)$. And exactly as in that case, this integration over $\mathbb{V}_N$ may or may not be easy/possible to perform explicitly.\\

It is certainly possible when the \emph{original} jpdf of entries, once expressed in terms of eigenvalues and eigenvector components, is itself independent of eigenvectors - in complete analogy with our previous example with $r$ and $\theta$. In this case, we would get
\begin{align}
\nonumber\hat\rho(\llambda_1,\ldots,\llambda_N,O) &\equiv\rho(H_{11}(\bm\llambda,\xcancel{O}),\ldots,H_{NN}(\bm\llambda,\xcancel{O}))\Big|J(H\to \{\bm\llambda,\xcancel{O}\})\Big|\\
&=\text{function of }\bm\llambda\text{ alone}\ ,\label{eq_vandermonde:favorable}
\end{align}
and all is left to do in \eqref{eq_vandermonde:marginalizejpdf} is the ``volume" integral $\int_{\mathbb{V}_N} dO$, yielding the simple constant in \eqref{eq_vandermonde:VolVN} - much like $2\pi$ in the warm-up example with $r,\theta$ above.\\

The prototypes of this favorable case are the rotationally invariant ensembles, see the next section\footnote{Instead, for models with independent entries, the jpdf of entries cannot be - in general - written in terms of the eigenvalues alone. For such models, the jpdf of eigenvalues is therefore not generally known.}.

\section{For invariant models...}\label{sec_vandermonde:invariant}

We can now formulate a cute little theorem for invariant models \cite{ref_vandermonde:edelmanthesis}. The proof is given below.\\

Let the real symmetric $N\times N$ matrix $H$ have a jpdf of entries $\rho[H]=\phi\left(\mathrm{Tr}\ H,\ldots,\mathrm{Tr}\ H^N\right)$, which is evidently invariant under orthogonal similarity transformations\footnote{Recall Weyl's lemma \eqref{eq_classified:Weyl}.}. Then the jpdf of the
$N$ \emph{ordered} eigenvalues of $H$ $(\llambda_1\geq \llambda_2\geq\cdots\geq \llambda_N)$ is
\begin{equation}
\rho_{ord}(\llambda_1,\ldots,\llambda_N)=\frac{\pi^{N^2/2}}{\Gamma_N (N/2)}\phi\left(\sum_i \llambda_i,\ldots,\sum_i \llambda_i^N\right)\prod_{i<j}(\llambda_i - \llambda_j)\ .
\label{eq_vandermonde:Edelmantheorem}
\end{equation}

Note that there is no absolute value around the Vandermonde, as the eigenvalues are ordered.\\

Let us see how this theorem works in practice for the GOE case. We have
\begin{equation}
\rho[H]=\prod_{i=1}^N \frac{e^{-H_{ii}^2/2}}{\sqrt{2\pi}}\prod_{i<j}\frac{e^{-H_{ij}^2}}{\sqrt{\pi}} = \frac{1}{(2\pi)^{N/2} \pi^{\frac{N^2-N}{4}}}\exp\left(-\frac{1}{2}\mathrm{Tr}\ H^2\right)\ .
\end{equation}

Therefore, applying the theorem above
\begin{equation}
\rho_{ord}(\llambda_1,\ldots,\llambda_N)=\underbrace{\frac{\pi^{N^2/2}}{\Gamma_N (N/2)}\times \frac{1}{(2\pi)^{N/2} \pi^{\frac{N^2-N}{4}}}}_{\frac{1}{\mathcal{Z}^{(ord)}_{N,\beta=1}}} e^{-\frac{1}{2}\sum_{i=1}^N\llambda_i^2}\prod_{i<j}(\llambda_i - \llambda_j)\ ,\label{eq_vandermonde:Edelman}
\end{equation}
which needs to be compared with Eq. \eqref{eq_valuetheeigenvalue:jpdgaussonehalf} for $\beta=1$ - given without proof at the time
\begin{equation}
\rho(\llambda_1,\ldots,\llambda_N)=\frac{1}{\mathcal{Z}_{N,\beta=1}}
e^{-\frac{1}{2}\sum_{i=1}^N
\llambda_i^2}\prod_{j<k}|\llambda_j-\llambda_k|\ ,\label{eq_vandermonde:jpdfeigGOEunordered}
\end{equation}
with
\begin{equation}
\mathcal{Z}_{N,\beta=1}=(2\pi)^{N/2}\prod_{j=1}^N\frac{\Gamma(1+j/2)}{\Gamma(3/2)}\ .
\end{equation}

Do the two equations \eqref{eq_vandermonde:Edelman} and \eqref{eq_vandermonde:jpdfeigGOEunordered} indeed agree, as they should? Almost.\\

Notice that \eqref{eq_vandermonde:Edelman} holds for \emph{ordered} eigenvalues, while \eqref{eq_vandermonde:jpdfeigGOEunordered} holds for \emph{unordered} eigenvalues (hence the need to include the absolute value). The two normalization constants differ indeed by a factor $N!$
\begin{equation}
\mathcal{Z}_{N,\beta=1}=N! \mathcal{Z}^{(ord)}_{N,\beta=1}\ .
\end{equation}

\section{The proof}

Where does the normalizing factor 
\begin{equation}
\frac{\pi^{N^2/2}}{\Gamma_N (N/2)}\label{eq_vandermonde:normalizingfactor}
\end{equation} 
in \eqref{eq_vandermonde:Edelmantheorem} come from? It is instructive to look at this derivation more closely.\\

Recall from \eqref{eq_vandermonde:marginalizejpdf} and \eqref{eq_vandermonde:favorable} that (for the favorable case where one can integrate out the eigenvectors)
\begin{equation}
\text{jpdf eigenv.} = \text{jpdf entries (as function of eigenv. alone)}\times |\text{Vandermonde}|\times\!\!\!\! \int_{\mathbb{V}_N}dO\ .
\end{equation}
This means that \emph{morally} the normalizing factor \eqref{eq_vandermonde:normalizingfactor} should corresponds to the volume integral $\int_{\mathbb{V}_N} dO$ as in \eqref{eq_vandermonde:VolVN} (for $\beta=1$, or $\int_{\tilde{\mathbb{V}}_N}dU$ over unitary matrix elements for $\beta=2$ etc.).\\

There is a subtlety though: the change of variables between entries and eigenvalues ($H\to OXO^T$) must be one-to-one. But eigenvectors are defined up to a phase, e.g. if $\mathbf{v}$ is a real eigenvector, so is $-\mathbf{v}$. To guarantee the uniqueness of the eigen-decomposition, it is sufficient to fix the sign of the first row of the matrix $O$, or the phases of the first row of the matrix $U$. This reduces the volume integral $\int_{\mathbb{V}_N} dO$ by a factor $2^N$ in the orthogonal case, and the volume integral $\int_{\tilde{\mathbb{V}}_N} dU$ by $(2\pi)^N$ in the unitary case. And the proof is complete.

\chapter{Meet Vandermonde}\label{chap:Vander}

The ``repulsive" term between eigenvalues of invariant models $\prod_{i<j}(\llambda_j-\llambda_i)$ can be written as a determinant, called \emph{Vandermonde} in honor of the French mathematician Alexandre-Th\'eophile Vandermonde (who never wrote it \cite{ref_vandermonde:Ycart4}).

\section{The Vandermonde determinant}

We have the following identity
\begin{equation}\label{vdm1}
\Delta_N(\bm\llambda):=\prod_{i<j}^N(\llambda_j-\llambda_i)=\det(\llambda_j^{i-1})=\det{\left(\begin{array}{ccc}1&\ldots & 1\\ \llambda_1&\ldots & \llambda_N\\
.&.&.\\ .&.&.\\.&.&.\\\llambda^{N-1}_1&\ldots & \llambda^{N-1}_N
\end{array}\right)}\ .
\end{equation}
The Vandermonde is clearly a completely anti-symmetric polynomial in $N$ variables: take for example $N=3$. We have $\Delta_3(\bm\llambda)=(\llambda_2-\llambda_1)(\llambda_3-\llambda_1)(\llambda_3-\llambda_2)$. Now, exchange any two $\llambda_j$s: for example, $\llambda_3\leftrightarrow\llambda_2$. We get $-\Delta_3(\bm\llambda)$ (we pick up a  minus sign any time we make \emph{any} exchange of two $\llambda_j$s). \\

The Vandermonde has a quite funny property: we can understand it already on a $2\times 2$ matrix. Take
 \begin{equation}\label{vdm2}
\det{\left(\begin{array}{cc}1& 1\\ \llambda_1& \llambda_2\\
\end{array}\right)}=\llambda_2-\llambda_1\qquad 
\det{\left(\begin{array}{cc}1& 1\\ 3\llambda_1+17& 3\llambda_2+17\\
\end{array}\right)}=3(\llambda_2-\llambda_1)\ .
\end{equation}\\

Stare at these two determinants carefully. We have just replaced the second row of the first matrix (containing first powers of $\llambda_1$ and $\llambda_2$) with a first degree polynomial. The result is just $3$ times the Vandermonde on the left. The $17$ has disappeared altogether! This means that you have a lot of freedom in devising a matrix whose determinant gives the Vandermonde.\\ 

More formally, the entries $\llambda_i^k$ in the $(k+1)$th row can be replaced,
up to a constant factor $a_0a_1 \cdots a_{N-1}$, by a polynomial of degree $k$ of the form:
$\pi_k(\llambda_i)=a_k\llambda_i^k+\cdots$, where we omit terms of lower order in
$\llambda_i$. The important point is that these lower order terms can be absolutely
anything. The result is that:
\begin{equation}\label{eq_vandermonde:VanAsPi}
\Delta_N(\bm\llambda)=\frac{1}{a_0a_1\cdots a_{N-1}}\det{
\left(\begin{array}{ccc}\pi_0(\llambda_1)&\ldots& \pi_0(\llambda_N)
\\ \pi_1(\llambda_1)&\ldots &\pi_1(\llambda_N)\\
.&.&.\\ .&.&.\\.&.&.\\ \pi_{N-1}(\llambda_1)&\ldots& \pi_{N-1}(\llambda_N)
\end{array}\right)}\ .
\end{equation}\\

\emph{Orthogonal polynomials} are an important class of polynomials $\pi_k(\llambda)$ that can be especially useful to play this trick. We will discuss in Chapter \ref{chap:FiniteN} how this simple property can actually turn seemingly impossible calculations into feasible ones.\\

For instance, let us show how the Hermite and Laguerre orthogonal polynomials can be used to express the Vandermonde. For $N=3$ it is easy to see that \be \det{
\left(\begin{array}{ccc}H_0(\llambda_1)&H_0(\llambda_2)& H_0(\llambda_3)
\\ H_1(\llambda_1)&H_1(\llambda_2)& H_1(\llambda_3)\\
H_2(\llambda_1)&H_2(\llambda_2)& H_2(\llambda_3)
\end{array}\right)}=\det{
\left(\begin{array}{ccc}1&1& 1
\\ \llambda_1&\llambda_2& \llambda_3\\
\llambda_1^2-1&\llambda_2^2-1& \llambda_3^2-1
\end{array}\right)}=\Delta_3(\bm\llambda),\ee and that $-\Delta_3(\bm\llambda)/2=$ \begin{align} \det{
\left(\begin{array}{ccc}L_0(\llambda_1)&L_0(\llambda_2)& L_0(\llambda_3)
\\ L_1(\llambda_1)&L_1(\llambda_2)& L_1(\llambda_3)\\
L_2(\llambda_1)&L_2(\llambda_2)& L_2(\llambda_3)
\end{array}\right)}=\det{
\left(\begin{array}{ccc}1&1& 1
\\ -\llambda_1+1&-\llambda_2+1& -\llambda_3+1\\
\frac{\llambda_1^2}{2}-2\llambda_1+1&\frac{\llambda_2^2}{2}-2\llambda_2+1& \frac{\llambda_3^2}{2}-2\llambda_3+1
\end{array}\right)}\ .\end{align}

\section{Do it yourself}

We now derive the nontrivial relation \eqref{eq_vandermonde:fundamentalJacobianrelation} $J(H\to \{\bm\llambda,O\})=\Delta_N(\bm\llambda)$ for real symmetric matrices $H$. We stress that this proof does not require any assumption on the rotational invariance of the ensemble.\\

These can be
diagonalized through an orthogonal transformation ${H}={O}X{O}^T$, where
$X={\mathrm{diag}}(\llambda_1,\ldots,\llambda_N)$. To find the Jacobian, we formally differentiate\footnote{The matrix element $H_{ij}$ can be written as $H_{ij}=\sum_{\ell,m}O_{i\ell}X_{\ell m}O_{jm}=\sum_{\ell}O_{i\ell}\llambda_\ell O_{j\ell}$. The infinitesimal matrix $\delta H$ has entries $(\delta H)_{ij}=dH_{ij}$ given by the differential of $H_{ij}$. Eq. \eqref{infiniH} is a shorthand of this explicit differentiation w.r.t. $O_{i\ell}$, $\llambda_\ell$ and $O_{j\ell}$.} $H$,
\begin{equation}
\delta{H}=(\delta{O})X{O}^T+ {O}(\delta X){O}^T+{O}X(\delta{O}^T)\ ,\label{infiniH}
\end{equation} and use $\delta{O}^{T}=-{O}^T (\delta{O}){O}^T $, which follows from ${O}{O}^T=\mathbbm{1}$. We get
\begin{equation}
\delta{H}=(\delta{O})X{O}^T+{O}(\delta X){O}^T-{O}X{O}^T (\delta{O}) {O}^T\ .
\end{equation}
Pulling out a factor ${O}$ to the left and ${O}^T$ to the right we obtain $\delta{H}={O}
(\delta\hat{H}){O}^T$, where \be \delta\hat{H}=(\delta{\Omega})X-X(\delta{\Omega})+\delta X\ .\ee
Here, $\delta{\Omega}={O}^T \delta {O}$ is an antisymmetric matrix\footnote{Obviously, you need to prove it before proceeding.}. Since $\delta{H}$ and
$\delta\hat{H}$ are related via an orthogonal transformation, we only have to find the
Jacobian of $\delta\hat{H}\to\{\delta X,\delta{\Omega}\}$.\\

Noting that $\delta X$ is
diagonal, we can write
\begin{equation}
d\hat{H}_{ij}=d{\Omega}_{ij}(\llambda_j-\llambda_i)+d\llambda_i \delta_{ij}\ .
\end{equation}
This is equivalent to the following differential relations: \be
\frac{d\hat{H}_{ij}}{d\llambda_k} =\delta_{ij}\delta_{ik},\quad
\frac{d\hat{H}_{ij}}{d\Omega_{k\ell}} =\delta_{ik}\delta_{j\ell}(\llambda_j-\llambda_i)\ .
\ee
Don't you see the Vandermonde trying hard to crop up here \Smiley{}?\\

Let us now construct the Jacobian matrix $J$ for a concrete $3\times 3$ case. The
generalization to the $N\times N$ case will then appear obvious. The matrix $J$ has
dimension $\frac{N(N+1)}{2}$, so it is a $6\times 6$ matrix for $N=3$. We parametrize the
antisymmetric matrix $\delta\Omega$ as follows:
\begin{equation}
\delta\Omega=\left(
\begin{array}{ccc}
0 &\Omega_{12} & \Omega_{13}\\
 -\Omega_{12} & 0 & \Omega_{23}\\
  -\Omega_{13} & -\Omega_{23} & 0
\end{array}
\right)\ .
\end{equation}
Then the Jacobian matrix becomes:
\begin{equation}
\renewcommand\arraystretch{2}
\begin{pmatrix}
\frac{d\hat{H}_{11}}{d\Omega_{12}} & \frac{d\hat{H}_{11}}{d\Omega_{13}} & \frac{d\hat{H}_{11}}{d\Omega_{23}} & \frac{d\hat{H}_{11}}{d\llambda_{1}} & \frac{d\hat{H}_{11}}{d\llambda_{2}} & \frac{d\hat{H}_{11}}{d\llambda_{3}}\\
\frac{d\hat{H}_{12}}{d\Omega_{12}} & \frac{d\hat{H}_{12}}{d\Omega_{13}} & \frac{d\hat{H}_{12}}{d\Omega_{23}} & \frac{d\hat{H}_{12}}{d\llambda_{1}} & \frac{d\hat{H}_{12}}{d\llambda_{2}} & \frac{d\hat{H}_{12}}{d\llambda_{3}}\\
\frac{d\hat{H}_{13}}{d\Omega_{12}} & \frac{d\hat{H}_{13}}{d\Omega_{13}} & \frac{d\hat{H}_{13}}{d\Omega_{23}} & \frac{d\hat{H}_{13}}{d\llambda_{1}} & \frac{d\hat{H}_{13}}{d\llambda_{2}} & \frac{d\hat{H}_{13}}{d\llambda_{3}}\\
\frac{d\hat{H}_{22}}{d\Omega_{12}} & \frac{d\hat{H}_{22}}{d\Omega_{13}} & \frac{d\hat{H}_{22}}{d\Omega_{23}} & \frac{d\hat{H}_{22}}{d\llambda_{1}} & \frac{d\hat{H}_{22}}{d\llambda_{2}} & \frac{d\hat{H}_{22}}{d\llambda_{3}}\\
\frac{d\hat{H}_{23}}{d\Omega_{12}} & \frac{d\hat{H}_{23}}{d\Omega_{13}} & \frac{d\hat{H}_{23}}{d\Omega_{23}} & \frac{d\hat{H}_{23}}{d\llambda_{1}} & \frac{d\hat{H}_{23}}{d\llambda_{2}} & \frac{d\hat{H}_{23}}{d\llambda_{3}}\\
\frac{d\hat{H}_{33}}{d\Omega_{12}} & \frac{d\hat{H}_{33}}{d\Omega_{13}} &
\frac{d\hat{H}_{33}}{d\Omega_{23}} & \frac{d\hat{H}_{33}}{d\llambda_{1}} &
\frac{d\hat{H}_{33}}{d\llambda_{2}} & \frac{d\hat{H}_{33}}{d\llambda_{3}}
\end{pmatrix}
=
\renewcommand\arraystretch{1}
\begin{pmatrix}
0 & 0 & 0 & 1 & 0 & 0\\
\llambda_2-\llambda_1 & 0 & 0 & 0 & 0 & 0\\
0 & \llambda_3-\llambda_1 & 0 & 0 & 0 & 0\\
0 & 0 & 0 & 0 & 1 & 0\\
0 & 0 & \llambda_3-\llambda_2 & 0 & 0 & 0\\
0 & 0 & 0 & 0 & 0 & 1
\end{pmatrix}\ .
\end{equation}
Swapping rows and columns, it is possible to bring this to the diagonal form, so that the
determinant becomes trivial to compute. In the general $N$ case, one has:
\begin{equation}
|\det J| = \prod_{j<k}|\llambda_j-\llambda_k|\ ,\label{4:jacobianidentity}
\end{equation}
as expected. The proof in the complex hermitian and quaternion self-dual cases is
analogous and is left as an exercise.\\

For a nice numerical test of the Jacobian
identity \eqref{4:jacobianidentity}, we refer to \cite{ref_vandermonde:edelmanrao4}, Section 3.2, while for a ``back-of-the-envelope"
derivation based on counting degrees of freedom, see \cite{ref_vandermonde:Zee4}.\\

We will make extensive use of the Vandermonde determinant and its properties in Chapter \ref{chap:FiniteN}.

\chapter{Resolve(nt) the semicircle}\label{chap:resolvent}

In this Chapter, we introduce the so called \emph{resolvent}, a complex function from which the spectral density\footnote{Unless otherwise stated, we will no longer make a distinction between $n^\star(\llambda)$, $\langle n(\llambda)\rangle$ and $\rho(\llambda)$.} can be calculated. The advantage of the resolvent approach is that one has to solve an \emph{algebraic} equation (like $a x^2+bx+c=0$) instead of a (singular) \emph{integral} equation (like $\mathrm{Pr}\int d\llambda^\prime\frac{n^\star(\llambda^\prime)}{\llambda-\llambda^\prime}=\llambda$, see \eqref{eq_semicirclesequel:2tricomi}). The disadvantage is that you need to know a bit of complex analysis.

\section{A bit of theory}\label{sec_resolvent:bitoftheory}

We introduce the complex function $G_N(z)$, with $z\in\mathbb{C}\setminus\{\llambda_i\}$ 
\begin{equation}
G_N(z)=\frac{1}{N}{\rm Tr}\frac{1}{z-H}=\frac{1}{N}\sum_{i=1}^N
\frac{1}{z-\llambda_i}\ , \label{eq_resolvent:Stieltjesdef}
\end{equation}\\
where the notation $1/(z-H)$ means the matrix inverse of $z\mathbbm{1} - H$, and $\mathbbm{1}$ is the $N \times N$ identity matrix. \\

If $H$ is a random matrix, then $G_N(z)$ is a random complex function that has \emph{poles} at the locations $\llambda_i$ of each eigenvalue.\\

The second ingredient we need is the \emph{Sokhotski-Plemelj formula}
\begin{equation}
\label{eq_resolvent:Sokhotsky} \lim_{\epsilon \rightarrow 0^+} \frac{1}{y \pm \mathrm{i}\epsilon} =
\mathrm{Pr} \left ( \frac{1}{y} \right ) \mp\mathrm{i} \pi \delta(y),
\end{equation}
which should be interpreted as the integral relation (for a real-valued test function $\varphi(x)$ such that the integrals make sense)
\begin{equation}
\lim_{\epsilon\to 0^+}\left[\left(\int_{-\infty}^{-\epsilon}+\int_{\epsilon}^\infty\right)\frac{\varphi(y)}{y}dy\right]\mp\mathrm{i}\pi\varphi(0)=\lim_{\epsilon\to 0^+}\int_{-\infty}^\infty\frac{\varphi(y)}{y\pm\mathrm{i}\epsilon}dy\ .\label{eq_resolvent:Sokhotsky2} 
\end{equation}

For a one-liner proof, see below (around \eqref{eq_resolvent:Gav}).

\begin{center}
\fbox{\begin{minipage}{33em}
{\it Question.} What is the point of introducing this identity? \\ \\
$\blacktriangleright$ First, stare at \eqref{eq_resolvent:Sokhotsky} carefully. You see that, on the right hand side, the imaginary part is just a delta function. So, this identity is (yet another) way of representing a delta function, as the imaginary part of a rational function (the left hand side). Knowing that the spectral density is defined in terms of a delta function $\rho(\llambda)=\langle (1/N)\sum_{i=1}^N\delta(\llambda-\llambda_i)\rangle$, you should be spotting an interesting connection here. More on this later.
\end{minipage}}
\end{center}

\section{Averaging}\label{sec_resolvent:averaging}

Imagine now to take the limit $N\to\infty$ of $\langle G_N(z)\rangle$, where we average over the distribution of the matrix $H$. This average is called \emph{resolvent}, or $Green's function$, or \emph{Stieltjes transform}. It is natural to assume (and can be mathematically justified) that:
\begin{itemize}
\item the sum in \eqref{eq_resolvent:Stieltjesdef} 
\begin{equation}
G_N(z)=\frac{1}{N}{\rm Tr}\frac{1}{z-H}=\frac{1}{N}\sum_{i=1}^N
\frac{1}{z-\llambda_i}\ .\label{eq_resolvent:StieltjesdefBis}
\end{equation}\\
gets converted into an integral,
\item the poles at $\llambda_i$ merge into a continuous ``cut" on the real line,
\item we have to ``weigh" the integrand with the average density of eigenvalues $\rho(x)$ at point $\llambda$.
\end{itemize}

The cut on the real line is therefore nothing but the support of the spectral density, and the average resolvent is defined for all complex values $z$ \emph{outside} this cut (for example, outside the interval $[-\sqrt{2},\sqrt{2}]$ on the real line for the Gaussian ensemble).\\

In formulae
\be
\langle G_N(z)\rangle\to G_\infty^{(av)}(z)=\int dx^\prime\frac{\rho(x^\prime)}{z-x^\prime}\ ,\qquad\mbox{for }N\to\infty\ .\label{eq_resolvent:convergenceavaragedStieltjes}
\ee\\

If you are inclined to believe that \eqref{eq_resolvent:convergenceavaragedStieltjes} is very plausible (to say the least), we can now proceed smoothly.\\

Compute now $G_\infty^{(av)}(z)$ (the averaged resolvent in the large $N$ limit) at $z=\llambda-\mathrm{i}\epsilon$. Carrying out this herculean task, we get 
\be
G_\infty^{(av)}(\llambda-\mathrm{i}\epsilon)=\int dx^\prime\frac{\rho(x^\prime)}{\llambda-\mathrm{i}\epsilon-x^\prime}=\int d\llambda^\prime \frac{\rho(\llambda^\prime)(\llambda-\llambda^\prime)}{(\llambda-\llambda^\prime)^2+\epsilon^2}+\mathrm{i}\int d\llambda^\prime \rho(\llambda^\prime)\frac{\epsilon}{(\llambda-\llambda^\prime)^2+\epsilon^2}\ ,\label{eq_resolvent:Gav}
\ee\\
where we have multiplied up and down by $\llambda-\llambda^\prime+\mathrm{i}\epsilon$ and separated the real and imaginary parts.\\

Sending now $\epsilon\to 0^+$, we are basically proving the Sokhotski-Plemelj formula: the real part becomes a principal value integral (the so called \emph{Hilbert transform}), $\mathrm{Pr}\int d\llambda^\prime \frac{\rho(\llambda^\prime)}{\llambda-\llambda^\prime}$, while the imaginary part (\emph{with the sign reversed with respect to the argument of $G_\infty^{(av)}$, $\pm\to\mp$}) becomes $\pi \rho(x)$, using the following representation for the delta function 
\be
\delta(x)=\frac{1}{\pi}\lim_{\epsilon\to 0^+}\frac{\epsilon}{x^2+\epsilon^2}\ .\ee

In summary
\begin{equation}
\label{eq_resolvent:rhofromG} \rho(x)= \frac{1}{\pi} \lim_{\epsilon \rightarrow 0^+}
\mathrm{Im} \ G_\infty^{(av)}(\llambda - \mathrm{i} \epsilon)\ .
\end{equation}\\
So, if you know (or can calculate) the resolvent in the complex plane, you can from it deduce the spectral density.\\

All this in theory. Practice in the next section.

\begin{center}
\fbox{\begin{minipage}{33em}
Are there important properties of the resolvent $G_\infty^{(av)}(z)$ in \eqref{eq_resolvent:convergenceavaragedStieltjes} that are worth remembering? \\ \\
$\blacktriangleright$ First of all, if you send $|z|\to\infty$ in \eqref{eq_resolvent:convergenceavaragedStieltjes}, you get
\be
G_\infty^{(av)}(z)=\int dx^\prime\frac{\rho(x^\prime)}{z-x^\prime}\approx \frac{1}{z}\int d x^\prime\rho(x^\prime)=\frac{1}{z}+....\ ,
\ee
where we have used normalization of $\rho(x)$. This asymptotic $\sim 1/z$ behavior can be important in applications.\\

Next, expanding the denominator in \eqref{eq_resolvent:convergenceavaragedStieltjes} to all orders, we observe that the resolvent is the generating function of \emph{moments} $\mu_k=\langle\mathrm{Tr}(H^k)\rangle=\int d x\rho(x)x^k$ 
\be
G_\infty^{(av)}(z)=\int dx^\prime\frac{\rho(x^\prime)}{z-x^\prime}=\frac{1}{z}\int dx^\prime\frac{\rho(x^\prime)}{1-x^\prime/z}=\frac{1}{z}\sum_{k=0}^\infty\int dx^\prime \rho(x^\prime)\left(\frac{x^\prime}{z}\right)^k=\sum_{k=0}^\infty\frac{\mu_k}{z^{k+1}}\ ,
\ee
with $\mu_0=1$.
\end{minipage}}
\end{center}

\section{Do it yourself}
We propose here a truly elementary derivation of the algebraic equation satisfied by the resolvent for the Gaussian ensemble.\\

Consider the partition function of the standard Gaussian
ensemble, after a further rescaling $\llambda_i\to\llambda_i\sqrt{N}$ and ignoring prefactors
\begin{equation}
\mathcal{Z}_{N,\beta}\propto\int_{\mathbb{R}} \prod_{j=1}^N d\llambda_j\ 
e^{-\frac{\beta N}{2}\sum_{i=1}^N
\llambda_i^2}\prod_{j<k}|\llambda_j-\llambda_k|^\beta=\int_{\mathbb{R}} \prod_{j=1}^N d\llambda_j\ 
e^{-\beta N\mathcal{V}[\bm\llambda]}\ ,\label{eq_resolvent:partitionfunctionGaussian}
\end{equation}
with
\be 
\mathcal{V}[\bm\llambda]=\frac{1}{2}\sum_i \llambda_i^2-\frac{1}{2N} \sum_{i\neq j}\ln
|\llambda_i-\llambda_j|\ .
\ee \\

Compared to our earlier Coulomb gas treatment, we have pulled out a factor $N$ (not $N^2$), so that the $\llambda_i$ are now of $\mathcal{O}(1)$ for large $N$. Instead of introducing a continuous counting function $n(\llambda)$ (as we did in Chapter \ref{chap:semicircle}), we can directly perform the saddle point evaluation of the $N$-fold integral \eqref{eq_resolvent:partitionfunctionGaussian}, obtaining for each variable $\llambda_i$ the equation

\begin{equation}
\frac{\partial \mathcal{V}[\bm\llambda]}{\partial\llambda_i}=0\Rightarrow \llambda_i =\frac{1 }{N}\sum_{j\neq i}\frac{1}{\llambda_i-\llambda_j}\ . \label{eq_resolvent:SP}
\end{equation}

Multiplying \eqref{eq_resolvent:SP} by $\frac{1}{N(z-\llambda_i)}$ and summing over $i$, we get:
\begin{equation}
\frac{1}{N}\sum_i \frac{\llambda_i}{z-\llambda_i}=\frac{1}{N}\sum_{i}\sum_{j\neq
i}\frac{1}{\llambda_i-\llambda_j}\frac{1}{N (z-\llambda_i)}\ .
\end{equation}
Adding and subtracting $z$ in the numerator, the left-hand-side $L$ becomes
\begin{equation}
L=\frac{1}{N}\sum_i \frac{\llambda_i-z+z}{z-\llambda_i}=-1+z\frac{1}{N}\sum_i
\frac{1}{z-\llambda_i}=-1+z  G_N(z)\ .
\end{equation}
As for the right-hand-side, let us define
$
R = \frac{1}{N^2}\sum_{i=1}^N \sum_{j\neq
i}\frac{1}{z-\llambda_i}\frac{1}{\llambda_i-\llambda_j}\ .
$
Writing 
\be
\frac{1}{(z-x_i)(x_i-x_j)}=\frac{1}{z-x_j}\left(\frac{1}{z-x_i}+\frac{1}{x_i-x_j}\right)\ ,
\ee
one obtains the following self-consistency equation for $R$
\be
R=G_N^2(z)+\frac{1}{N}G_N^\prime(z)-R\qquad\Rightarrow\qquad R = \frac{1}{2}G_N^2(z)+\frac{1}{2N}G_N^\prime (z)\ .
\ee\\

Equating $L$ to $R$, we obtain as promised that the saddle-point condition \eqref{eq_resolvent:SP} gets converted into an equation for the resolvent
\begin{equation}
-1+z  G_N(z)=\frac{1}{2}G_N^2(z)+\frac{1}{2N} G_N^\prime (z)\ .\label{eq_resolvent:resolventdifferential}
\end{equation}\\

This is good, but is still a \emph{differential} equation for $G_N(z)$, while we promised an even simpler \emph{algebraic} equation. It is actually easy to get rid of the differential term in \eqref{eq_resolvent:resolventdifferential} by noticing that, with $\llambda_i$ of $\mathcal{O}(1)$, the resolvent as defined in \eqref{eq_resolvent:Stieltjesdef} is itself of $\mathcal{O}(1)$ and therefore the term $\frac{1}{2N} G_N^\prime (z)$ is subleading for large $N$.\\

Taking the average, the surviving algebraic (at long last!) equation for $N\to\infty$ reads
\begin{equation}
G_\infty^{(av)2}(z)-2 z G_\infty^{(av)}(z)+2=0\ .\label{eq_resolvent:quadraticGav}
\end{equation}

It is instructive to solve \eqref{eq_resolvent:quadraticGav} directly as a quadratic equation (recall that quadratic equations for complex variables admit the same solving formula as their real counterparts), yielding
\begin{equation}
G_\infty^{(av)}(z)=z\pm\sqrt{z^2-2}\ .\label{eq_resolvent:Gsol}
\end{equation}

Setting now, $z=\llambda-\mathrm{i} \epsilon$, we obtain $G_\infty^{(av)}(\llambda-\mathrm{i}\epsilon)=\llambda-\mathrm{i}\epsilon\pm\sqrt{(\llambda^2-\epsilon^2-2)+\mathrm{i}(-2\llambda\epsilon)}$. The square root (with positive real part) of a complex number $a+\mathrm{i}b$ can be written as \cite{ref_resolvent:Rabinowitz} $\sqrt{a+\mathrm{i}b}=p+\mathrm{i}q$, with 
\begin{equation}
p =\frac{1}{\sqrt{2}}\sqrt{\sqrt{a^2+b^2}+a}\qquad
q = \frac{\mathrm{sign}(b)}{\sqrt{2}}\sqrt{\sqrt{a^2+b^2}-a}\ ,
\end{equation}\label{eq_resolvent:squareroot}
where $\mathrm{sign}(x)=1$ if $x>0$ and $=-1$ if $x<0$.\\

Hence we obtain (recalling \eqref{eq_resolvent:rhofromG})
\begin{align}
\nonumber\frac{1}{\pi}\mathrm{Im} G_\infty^{(av)}(\llambda-\mathrm{i}\epsilon) &=-\frac{\epsilon}{\pi}\pm\frac{\mathrm{sign}(-2 \llambda \epsilon)}{\pi\sqrt{2}}
\sqrt{\sqrt{(\llambda^2-\epsilon^2-2)^2+4\llambda^2\epsilon^2}-\llambda^2+\epsilon^2+2}\\
&\stackrel{\epsilon\to 0^+}{\longrightarrow}\pm\frac{\mathrm{sign}(-\llambda)}{\pi\sqrt{2}}\sqrt{|\llambda^2-2|-\llambda^2+2}\ .
\end{align}

From this expression, you see that i) for $|x|>\sqrt{2}$ you obtain that the density is $0$, and ii) for $|x|<\sqrt{2}$, you need to select the $(-)$ or $(+)$ sign in front, according to whether $x>0$ or $x<0$ respectively. After choosing the right sign, you get $\rho(\llambda)=(1/\pi)\sqrt{2-x^2}$ as expected.\\

To double-check this result, we can insert the semicircle $\rho_{\mathrm{SC}}(\llambda)=\frac{1}{\pi}\sqrt{2-\llambda^2}$ back into the definition \eqref{eq_resolvent:convergenceavaragedStieltjes} and perform the numerical integration for $z=x-\mathrm{i}\epsilon$, with $\epsilon$ a small positive number and separately for the two cases, $0<x<\sqrt{2}$ and $-\sqrt{2}<x<0$. This is done with the code [$\spadesuit$ \verb"check_resolvent.m"], where the results are compared with the two choices of sign in \eqref{eq_resolvent:Gsol}. You see that the $(+)$ choice in \eqref{eq_resolvent:Gsol} only works with $-\sqrt{2}<x<0$, and the $(-)$ choice in \eqref{eq_resolvent:Gsol} only works with $0<x<\sqrt{2}$.\\

\section{Localize the resolvent} \label{sec_resolvent:resolvent_matrix}

Let us now take a step back and ``unpack'' the definition of the resolvent in equation \eqref{eq_resolvent:Stieltjesdef}.\\

We wrote that definition as the trace of the matrix $(z-H)^{-1}$, with $z\in\mathbb{C}\setminus\{\llambda_i\}$, $\llambda_i$s being the eigenvalues of $H$. If we write the trace explicitly, i.e. as a sum over the diagonal elements of the resolvent, we have $G_N(z) = (1/N) \sum_{i=1}^N G_{N,ii}(z)$, where

\be \label{eq_resolvent:diag_elements}
G_{N,ii}(z) = [(z-H)^{-1}]_{ii} = \sum_{j=1}^N \frac {(c_i^j)^2} {z-x_j} \ ,
\ee
and $c_i^j$ is the $i$th component of the normalized eigenvector associated with the $j$th eigenvalue of $H$. \\

So, you may now ask, what should I make of these matrix elements? Well, it turns out that they contain precious information about the \emph{localization} properties of the matrix ensemble they are associated with.\\

But, first of all, what is localization?\\

Simply put, the term localization refers to how ``spread out'' over their components the eigenvectors of a matrix are. Let us define the \emph{inverse participation ratio} (IPR) of a normalized eigenvector as

\be \label{eq_resolvent:IPR}
\mathcal{I}_{N,j} = \sum_{i=1}^N (c_i^j)^4 \ .
\ee 

Now, when the eigenvector's components are all roughly of the same magnitude, then we must have $c_i^j \simeq 1/\sqrt{N}$, $\forall \ i$, due to normalization. Hence, we will have $\mathcal{I}_{N,j} \simeq 1/N$, and the IPR will vanish in the large $N$ limit.\\

If, on the other hand, the eigenvector is significantly different from zero only on a number $s$ of sites, then for those sites we will have $c_i^j \simeq 1/\sqrt{s}$, and the IPR will remain roughly equal to $1/s$ in the large $N$ limit. So, all in all, the IPR is a handy tool that tells us whether certain eigenvectors of a matrix are extended (i.e. have an extensive number of non zero components) or instead localized on a finite number of sites. \\

Although this may sound like a mathematical curiosity, the localization properties of matrix ensembles are related to a number of relevant features of the physical systems they describe. In particular, it is often crucial to detect the so called \emph{mobility edge}, i.e. the critical eigenvalue that separates the part of the spectrum associated with extended states from the one associated with localized states. For example, it has famously been shown that the mobility edge determines the Anderson transition in electronic systems \cite{ref_resolvent:Anderson}. \\

All in all, it should be now clear that having analytical access to the distributional properties of the IPRs corresponding to different segments of a given ensemble's eigenvalue spectrum is a valuable thing. Luckily, this is where our diagonal elements \eqref{eq_resolvent:diag_elements} come to the rescue. Indeed, it has been shown in \cite{ref_resolvent:Metz} that the average value $P(x)$ of IPRs associated with states whose corresponding eigenvalues lie between $x$ and $x+dx$ can be written in the large $N$ limit as

\be \label{eq_resolvent:avg_IPR}
P(x) =\lim_{\epsilon \rightarrow 0^+} \lim_{N \rightarrow \infty} \frac{\epsilon}{\pi N \rho(x)}  \sum_{i=1}^N \left \langle  | G_{N,ii}(x - \mathrm{i} \epsilon) |^2 \right \rangle =
\lim_{\epsilon \rightarrow 0^+} \frac{\epsilon}{\pi \rho(x)}  | G_\infty^{(av)}(x - \mathrm{i} \epsilon) |^2\ .
\ee

\section{To know more...}

\begin{enumerate}

\item The saddle-point evaluation \eqref{eq_resolvent:SP} based on the partition function \eqref{eq_resolvent:partitionfunctionGaussian} is clearly valid when the neglected terms in the exponent are indeed subleading $(\mathcal{O}(N))$. There are models - rotationally invariant by construction - where the Dyson index $\beta$ is allowed to scale with $N$ \cite{ref_resolvent:Allez1,ref_resolvent:Allez2}. These models provide explicit realizations of invariant $\beta$-ensembles, for which the resolvent equation is necessarily more involved. Ref. \cite{ref_resolvent:Allez2} is also suggested for an elementary derivation of this ``improved" resolvent equation in the presence of a hard wall in the spectrum.

\item Matrix models such as the Gaussian can be constructed introducing a fictitious time evolution (stochastic) of the entries. In this case, it is possible to show that the resolvent satisfies a partial differential equation of the \emph{Burgers} type (see the beautiful paper \cite{ref_resolvent:blaizot}). 

\item The equation for the resolvent can be given a pretty interpretation in terms of \emph{planar diagrams}. Diagrammatic methods are at the heart of many beautiful results in RMT (see \cite{ref_resolvent:Jurk} and \cite{ref_resolvent:Bouttier}).

\end{enumerate}

\chapter{One pager on eigenvectors}\label{chap:Eigenvectors}

Take the GUE ensemble of $N\times N$ hermitian matrices. Any given matrix in the ensemble will have unit-norm eigenvectors having in general complex components. What is the statistics of such components?\\

Since eigenvalues and eigenvectors of invariant matrix models are decoupled, the only constraint on the $N$ components of an eigenvector is that its norm must be one, therefore their jpdf reads
\begin{equation}
P_{GUE}({\bm c})=C_N\delta\left(1-\sum_{n=1}^N |c_n|^2\right)\ ,\label{sec_Eigenvectors2:jpdfPGUE}
\end{equation}
where $C_N$ is a normalization constant.\\

It is convenient to compute the marginal distribution of a single component, say $|c_1|^2$, given by
\begin{equation}
P_{GUE}(y)=\int d^2 c_1\cdots d^2 c_N \delta(y-|c_1|^2)P_{GUE}({\bm c})\ .\label{sec_Eigenvectors2:PGUE}
\end{equation}

Similarly, we can compute the jpdf of eigenvector components (this time all real numbers) of a GOE matrix.\\ 

The calculation in \eqref{sec_Eigenvectors2:PGUE} is carried out by first defining an auxiliary object
\begin{equation}
P_{GUE}(y;t)=\int d^2 c_1\cdots d^2 c_N \delta(y-|c_1|^2)C_N\delta\left(t-\sum_{n=1}^N |c_n|^2\right)\ ,\label{sec_Eigenvectors2:PGUEaux}
\end{equation}
such that $P_{GUE}(y)=P_{GUE}(y;1)$. Then, taking the Laplace transform with respect to $t$ to kill the delta function in \eqref{sec_Eigenvectors2:PGUEaux}
\begin{equation}
\int_0^\infty dt\ e^{-s t}P_{GUE}(y;t)=C_N\int d^2 c_1\delta(y-|c_1|^2)e^{-s |c_1|^2}\left(\int d^2 c\ e^{-s|c|^2}\right)^{N-1}\ ,
\end{equation}
and finally converting the 2d integrals in polar coordinates
\begin{equation}
\int_0^\infty dt\ e^{-s t}P_{GUE}(y;t) =\hat{C}_N\int_0^\infty dr\ r\delta(y-r^2)e^{-s r^2}\left(\int_0^\infty d\rho\ \rho\ e^{-s\rho^2}\right)^{N-1}\propto \frac{e^{-s y}}{s^{N-1}}\ ,
\end{equation}
where we have absorbed the angular constants in the overall normalization.\\

Inverting the Laplace transform, we obtain
\begin{equation}
P_{GUE}(y;t)\propto (t-y)^{N-2}\theta(t-y)\ ,
\end{equation}
where $\theta(z)$ is the Heaviside step function. Setting $t=1$ and normalizing, we obtain
\begin{equation}
P_{GUE}(y)=(N-1)(1-y)^{N-2}\qquad\text{ for }0\leq y\leq 1\ .
\end{equation}
Similarly, for the GOE one obtains
\begin{equation}
P_{GOE}(y)=\frac{1}{\sqrt{\pi}}\frac{\Gamma(N/2)}{\Gamma((N-1)/2)}\frac{(1-y)^{(N-3)/2}}{\sqrt{y}}\qquad\text{ for }0\leq y\leq 1\ .
\end{equation}
Computing the average $\langle y\rangle$ in both cases
\begin{equation}
\langle y\rangle_{GOE}=\int_0^1 dy\ y P_{GOE}(y)=\frac{\Gamma(N/2)}{2\Gamma(1+N/2)}\sim \frac{1}{N}\qquad\langle y\rangle_{GUE}=\int_0^1 dy\ y P_{GUE}(y)=\frac{1}{N}\ ,
\end{equation}
leads us to consider the scaled variable $\eta=yN$ and take the limit $N\to\infty$. This produces the scaled densities
\begin{align}
P_{GOE}(\eta) &=\lim_{N\to\infty} \frac{1}{N}P_{GOE}\left(\frac{\eta}{N}\right)=\frac{1}{\sqrt{2\pi\eta}}e^{-\eta/2}\ ,\\
P_{GUE}(\eta) &=\lim_{N\to\infty} \frac{1}{N}P_{GUE}\left(\frac{\eta}{N}\right)=e^{-\eta}\ .
\end{align}
The first of these densities is called the \emph{Porter-Thomas} distribution \cite{ref_Eigenvectors2:Porter,ref_Eigenvectors2:Brody}. Note also that the Gaussian nature of the matrix ensembles has not been used anywhere in the derivation (the same densities would be obtained for any orthogonal or unitary ensemble). \\

The study of eigenvectors of random matrices has been recently revived due to their importance in quantum systems (see, e.g.,\cite{ref_Eigenvectors2:chalker,ref_Eigenvectors2:mehlig,ref_Eigenvectors2:Fyodorov_arxiv,ref_Eigenvectors2:Truong})

\chapter{Finite $N$}\label{chap:FiniteN}

Look back at Chapter \ref{chap:GettingStarted}, where we constructed Gaussian matrices and histogrammed their eigenvalues. For $N\to\infty$, we showed in various ways that the average spectral density converges to the semicircle law. But what happens for \emph{finite} $N$? Can we compute analytically the shape of the histogram for, say, a $13\times 13$ Gaussian matrix? The answer is Yes - and not only for Gaussian matrices, but for any rotationally invariant ensemble! This is done here. We start from the case $\beta=2$, as it is much easier.

\section{$\beta=2$ is easier}

Already in Chapter \ref{chap:Vander}, we mentioned that the Vandermonde determinant has some funny properties: in particular, each row in the Vandermonde matrix can be replaced by a polynomial of suitable degree, with many \emph{a priori} unspecified coefficients. The freedom in choosing these polynomials is enormous. A judicious choice is the key of the celebrated \emph{orthogonal polynomial technique}.\\

Take the jpdf of the $N$ real eigenvalues of a rotationally invariant ensemble with $\beta=2$

\begin{equation} \rho(\llambda_1,\ldots,\llambda_N)=\frac{1}{\mathcal{Z}_N} \prod_{i=1}^Ne^{-V(\llambda_i)}
|\Delta_N(\bm\llambda)|^2\ ,\label{eq_finiteN:jpdf1}
\end{equation} 
which is written in the `potential' form (see eq. \eqref{eq_saddlepoint:potential}). For example, for the Gaussian ensemble $V(x)=x^2/2$.\\

What is the goal then? To compute the average spectral density \emph{for finite $N$}, i.e. the $N-1$-fold integral
\be
\boxed{\rho(\llambda_1)=\int d\llambda_2 \cdots d\llambda_N \rho(\llambda_1,\llambda_2,\ldots,\llambda_N)=\frac{1}{\mathcal{Z}_N}\int d\llambda_2 \cdots d\llambda_N  \prod_{i=1}^Ne^{-V(\llambda_i)}|\Delta_N(\bm\llambda)|^2}\ ,
\label{eq_finiteN:goal}
\ee
where the partition function is $\mathcal{Z}_N=\int  d\llambda_1 \cdots d\llambda_N \prod_{i=1}^Ne^{-V(\llambda_i)}|\Delta_N(\bm\llambda)|^2$.\\

Note that in \eqref{eq_finiteN:goal} we are integrating over all variables \emph{but} one. These integrals are nasty, though! The integrand does not factorize at all, so we need to find some smart trick to carry out the integration. It took a while even to the pioneers of these calculations (for instance, Gaudin and Mehta) to figure out how to proceed. The steps are as follows:\\

{\bf Step 1:}\\

Rewrite the Vandermonde $\Delta_N(\bm\llambda)$ as a determinant of the matrix $A$, whose entries are polynomials $\pi_k(\llambda)$ (to be determined), as in \eqref{eq_vandermonde:VanAsPi}
\begin{equation}\label{eq_finiteN:VanAsPi}
\Delta_N(\bm\llambda)=\frac{1}{a_0a_1\cdots a_{N-1}}\det{
\left(\begin{array}{ccc}\pi_0(\llambda_1)&\ldots& \pi_0(\llambda_N)
\\ \pi_1(\llambda_1)&\ldots &\pi_1(\llambda_N)\\
.&.&.\\ .&.&.\\.&.&.\\ \pi_{N-1}(\llambda_1)&\ldots& \pi_{N-1}(\llambda_N)
\end{array}\right)}\ .
\end{equation}

{\bf Step 2:}\\

Use the general relation\footnote{Hereafter, inside a determinant the indices of the entries will run from $1$ to $N$.}
\be
(\det A)^2=\det(A^T A)=\det\left(\sum_{j=1}^N A_{ji}A_{jk}\right)\ ,
\ee
applied to the matrix $A$ from step $1$, to write
\be
\Delta_N^2(\bm\llambda)=\frac{1}{(\prod_{j=0}^{N-1}a_j)^2}\det\left(\sum_{j=1}^N \pi_{j-1}(\llambda_i)\pi_{j-1}(\llambda_k)\right)\ .\label{eq_finiteN:vandermondesquared}
\ee\\

{\bf Step 3:}\\

Pull the weight $\exp(-\sum_i V(\llambda_i))$ inside the determinant\footnote{Use $\left(\prod_\ell\alpha_\ell\right)\det(f(i,j))=\det(\sqrt{\alpha_i\alpha_j}f(i,j))$.} and shift the index $j\to j-1$, to write eventually
\be
\rho(\llambda_1,\ldots,\llambda_N)=\frac{1}{\mathcal{Z}_N (\prod_{j=0}^{N-1}a_j)^2}\det\left(\sum_{j=0}^{N-1}\phi_{j}(\llambda_i)\phi_{j}(\llambda_k)\right)=\frac{\det\left(K_N(\llambda_i,\llambda_k)\right)}{\mathcal{Z}_N (\prod_{j=0}^{N-1}a_j)^2}\ ,\label{eq_finiteN:rhoasdet}
\ee
where $\phi_i(\llambda)=e^{-V(\llambda)/2}\pi_i(\llambda)$ and
\be
\boxed{K_N(\llambda,\llambda^\prime)=e^{-\frac{1}{2}(V(\llambda)+V(\llambda^\prime))}\sum_{j=0}^{N-1}\pi_j(\llambda)\pi_j(\llambda^\prime)}\ ,\label{eq_finiteN:kernel}
\ee
which is a central object in RMT: the \emph{kernel}.\\

\clearpage 

{\bf Step 4:}\\

Choose judiciously the (so far undetermined) polynomials $\pi_j(\llambda)$. A great choice is to pick them \emph{orthonormal} with respect to the weight\footnote{Note that there is a factor $(1/2)$ multiplying $V(\llambda)$ in the kernel \eqref{eq_finiteN:kernel}, while there is none in the weight function of the orthonormal polynomials in \eqref{eq_finiteN:orthonormality}.} $\exp(-V(\llambda))$
\be
\int e^{-V(\llambda)}\pi_i(\llambda)\pi_j(\llambda)d\llambda=\delta_{ij}\ .\label{eq_finiteN:orthonormality}
\ee

For instance, for the Gaussian (unitary) ensemble ($V(\llambda)=\llambda^2/2$) the corresponding orthonormal polynomials are
\begin{equation}
\pi_j(\llambda)=\frac{H_j(\llambda/\sqrt{2})}{\sqrt{\sqrt{2\pi}\ 2^j j!}}\ ,\label{eq_finiteN:pis}
\end{equation}
if $H_j(\llambda)$ are Hermite polynomials satisfying $\int_{-\infty}^\infty d\llambda\ H_j(\llambda)H_k(\llambda)\exp(-\llambda^2)=\sqrt{\pi}2^j j!\delta_{jk}$.

\begin{center}
\fbox{\begin{minipage}{33em}
{\it Question.} What is the advantage of choosing polynomials with this ``orthonormality'' property? \\ \\
$\blacktriangleright$ Well, the reason is that the kernel $K_N(\llambda,\llambda^\prime)$ in \eqref{eq_finiteN:kernel}, \emph{if} the polynomials are chosen this way, satisfies a quite amazing ``reproducing" property
\be
\int dy K_N(\llambda,y)K_N(y,\llambda^\prime)=K_N(\llambda,\llambda^\prime)\ .\label{eq_que_finiteN:reproducing}
\ee
The proof is very simple: just insert \eqref{eq_finiteN:kernel} into \eqref{eq_que_finiteN:reproducing} and use the orthonormality relation \eqref{eq_finiteN:orthonormality}. This property has a quite unexpected consequence, which eventually allows to carry out the multiple integrations in \eqref{eq_finiteN:goal} in a very elegant, iterative way. Another ingredient is necessary, though, and is presented in the next Section.
\end{minipage}}
\end{center}

\section{Integrating inwards}

Summarizing, we have to carry out the multiple integration in \eqref{eq_finiteN:goal} over a jpdf, which can be written as the determinant of a kernel (see \eqref{eq_finiteN:rhoasdet}), something like
\be
\int d\llambda_2\cdots d\llambda_N\det(K_N(\llambda_j,\llambda_k)) =?\label{eq_finiteN:questionmark}
\ee

In normal situations, this would seem a rather hopeless task. But the reproducing property of the kernel offers an unexpected way around.\\

First, an illuminating $2\times 2$ example, and then the full-fledged (though dry) theory. Imagine the following $2\times 2$ matrix $J_2(\bm{x})$, depending on the vector $\bm{x}=\{x_1,x_2\}$ through a function $f(x,y)$ as follows
\be
J_2(\bm{x})=\left(\begin{array}{cc} f(x_1,x_1)& f(x_1,x_2)\\
 f(x_2,x_1) & f(x_2,x_2)
\end{array}\right)\ .
\ee\\

Suppose now that the function $f$ satisfies the `'reproducing" property \eqref{eq_que_finiteN:reproducing}, namely $\int f(x,y)f(y,z)\,d\mu(y)=f(x,z)$ for a certain measure $\mu(y)$. What happens to the following integral
\be
\int d \mu(x_2) \det(J_2(\bm{x})) ?
\ee\\

Well, we have
\begin{align}
\nonumber \int d \mu(x_2) \det(J_2(\bm{x})) &=\int d \mu(x_2) \left[f(x_1,x_1)f(x_2,x_2)-f(x_1,x_2)f(x_2,x_1)\right]=\\
&=q f(x_1,x_1)-f(x_1,x_1)=(q-1)f(x_1,x_1)\ ,\label{eq_finiteN:2x2}
\end{align}
where $q=\int d \mu(x_2) f(x_2,x_2)$. We used the reproducing property to evaluate the second integral.\\

Maybe this short calculation is not particularly revealing, but it can be actually extended to the $N\times N$ case as follows: let $J_N(\bm{x})$ be an $N\times N$ matrix
whose entries depend on a real vector $\bm{x}=(x_1,x_2,\ldots,x_N)$ and have the form
$J_{ij}=f(x_i,x_j)$, where $f$ is some function satisfying the ``reproducing kernel"
property
$
\int f(x,y)f(y,z)\,d\mu(y)=f(x,z)\ ,
$
for some measure $d\mu(y)$. Then the following holds:
\begin{equation}\label{eq_finiteN:reproducingfdet}
\int \det[J_N(\bm{x})]\,d\mu(x_N)=[q-(N-1)]\det(J_{N-1}(\tilde{\bm{x}}))\ ,
\end{equation}
where $q=\int f(x,x)\,d\mu(x)$, and the matrix $J_{N-1}$ has the same functional form as
$J_N$ with $\bm{x}$ replaced by $\tilde{\bm{x}}=(x_1,x_2,\ldots,x_{N-1})$. A friendly proof can be found in \cite{ref_valuetheeigenvalue:fyodreview2}. \\

This is a quite spectacular result, which is commonly referred to as \emph{Dyson-Gaudin integration lemma}\footnote{The most accurate reference seems however to be \cite{ref_finiteN:Mahoux}.}. First of all, note that the $2\times 2$ result \eqref{eq_finiteN:2x2} is in agreement with the general statement. Second, comparing \eqref{eq_finiteN:questionmark} and \eqref{eq_finiteN:reproducingfdet}, we see that this lemma actually allows to integrate $\det(K_N(\llambda_j,\llambda_k))$ (essentially, the jpdf) over the last variable $x_N$, producing as a result a determinant of a \emph{smaller} kernel matrix
\begin{equation}\label{eq_finiteN:firststepinduction}
\int \mbox{det}\left(K_N(x_i,x_j)\right)_{1\le i,j\le N}
\,dx_N=\mbox{det}\left(K_N(x_i,x_j)\right)_{1\le i,j\le N-1}\ ,
\end{equation}
where we have used $q=\int d x K_N(x,x)=N$ (immediate from the definition of the kernel \eqref{eq_finiteN:kernel}). Basically, the reproducing property \emph{carries over} from the kernel to the \emph{determinant} of the kernel!\\

Therefore, we can iterate the process $N-k$ times, killing one integral at a time and reducing the dimension of the determinant by one, with a remarkable domino effect
\begin{equation}\label{eq_finiteN:kstepinduction}
\int\ldots \int\mbox{det} \left(K_N(x_i,x_j)\right)_{1\le i,\,j\le N} \,dx_{k+1}\cdots
dx_{N}= (N-k)!\ \mbox{det}\left(K_N(x_i,x_j)\right)_{1\le i,\,j\le k}\ .
\end{equation}

In particular, setting $k=0$ we can normalize the jpdf \eqref{eq_finiteN:jpdf1} as
\begin{equation} 
1 =\int d\bm\llambda\rho(\bm\llambda)=\frac{1}{\mathcal{Z}_N (\prod_{j=0}^{N-1}a_j)^2} \int d\bm\llambda\det\left(K_N(\llambda_i,\llambda_k)\right)\Rightarrow \mathcal{Z}_N \left(\prod_{j=0}^{N-1}a_j\right)^2=N!\ ,
\end{equation} 

so that the \emph{two-point} marginal $\rho(x_1,x_2)$
\be
\rho(x_1,x_2)=\int \prod_{j=3}^N d x_j\rho(x_1,\ldots,x_N)=\frac{1}{N(N-1)}\det
\left(\begin{array}{cc} K_N(x_1,x_1)& K_N(x_1,x_2)\\
 K_N(x_2,x_1) & K_N(x_2,x_2)\label{eq_finiteN:twopointmarginal}
\end{array}\right)\ ,
\ee
(where one uses $(N-2)!/N!=1/[N(N-1)]$), while the \emph{one-point} marginal (the average spectral density \emph{}) is simply
\be
\rho(x_1)=\int d x_2\cdots d x_N \rho(x_1,\ldots,x_N)=\frac{1}{N}K_N(x_1,x_1)\ .
\ee

And the problem is solved not just for the one-point marginal, but for any $k$-point correlation function - once the kernel is built out of suitable polynomials, orthonormal with respect to the weight $V(\llambda)$. The fact that all such functions can be expressed in terms of determinants is usually referred to as \emph{determinantal structure} of the unitarily invariant ensembles.\\

For modern extensions of the ``integrate-out" lemma and applications, have a look at \cite{ref_finiteN:Kanzieper,ref_finiteN:Akemann}. 

\section{Do it yourself}

Let us apply the general formalism to the GUE case, for which the orthonormal polynomials are $\pi_j(\llambda)=H_j(\llambda/\sqrt{2})/\sqrt{\sqrt{2\pi}2^j j!}$, where $H_j(\llambda)$ are Hermite polynomials. Then, we obtain immediately the spectral density at finite $N$ as
\begin{equation}
\boxed{\rho(\llambda)=\frac{1}{N\sqrt{2\pi}}e^{-\llambda^2/2}\sum_{j=0}^{N-1}\frac{H_j^2(\llambda/\sqrt{2})}{2^j j!}}\ .\label{eq_finiteN:theordensityGUE}
\end{equation}
In fig. \ref{fig_finiteN} of Chapter \ref{chap:FiniteNbeta14} we show a comparison between a numerically generated
histogram of GUE eigenvalues, and the corresponding theoretical result in
\eqref{eq_finiteN:theordensityGUE}.

\begin{center}
\fbox{\begin{minipage}{33em}
{\it Question.} If I send $N\to\infty$ in \eqref{eq_finiteN:theordensityGUE}, shouldn't I recover the semicircle? I do not see how. \\ \\
$\blacktriangleright$ Yes, you should, and you will! The precise statement is
\be
\lim_{N\to\infty}\sqrt{2N}\rho(z\sqrt{2N})=\frac{1}{\pi}\sqrt{2-z^2}\ ,\qquad\mbox{for }-\sqrt{2}<z<\sqrt{2}\ ,
\ee
which requires a bit of work on the asymptotics of Hermite polynomials. We will give a flavor of the steps you need just below.
\end{minipage}}
\end{center}

\section{Recovering the semicircle}

First, one injects the so called \emph{Christoffel-Darboux formula} \cite{ref_finiteN:VanAssche} into the game, a quite spectacular relation that hugely simplifies sums of orthogonal polynomials. Specialized to the Hermite polynomials, it reads
\be
\sum_{k=0}^n\frac{H_k(x)H_k(y)}{k! 2^k}=\frac{1}{n! 2^{n+1}}\frac{H_n(y)H_{n+1}(x)-H_n(x)H_{n+1}(y)}{x-y}\ .
\ee\\

With an eye towards \eqref{eq_finiteN:theordensityGUE}, with a few manipulations and taking the limit $x\to y$, we obtain the relation
\be
\sum_{k=0}^{N-1}\pi_k^2(x)=\sqrt{N}\left[\pi_{N-1}(x)\pi_N^\prime(x)-\pi_N(x)\pi_{N-1}^\prime(x)\right]\ ,
\ee\\
where the orthonormal polynomials with respect to the Gaussian weight were defined in \eqref{eq_finiteN:pis}.\\

After huge simplifications, the GUE spectral density for finite $N$ - suitably rescaled - can be rewritten in the form
\be
\boxed{\sqrt{2N}\rho(z\sqrt{2N})=\frac{2 e^{-N z^2}}{\sqrt{\pi N}2^N\Gamma(N)}\left[N H_{N-1}^2(z\sqrt{N})-(N-1)H_N(z\sqrt{N})H_{N-2}(z\sqrt{N})\right]}\ .\label{eq_kernel:tobeasymptotics}
\ee 

We should now analyze \eqref{eq_kernel:tobeasymptotics} in the limit $N\to\infty$ for $z\sim\mathcal{O}(1)$. To do so, we need to use the following asymptotic formula for Hermite polynomials \emph{in the bulk}\footnote{What does \emph{in the bulk} mean? The point is that Hermite polynomials (and other classical orthogonal polynomials) have \emph{two} different asymptotics, according to the way their argument and parameter scale with $N$. This in turns corresponds to different \emph{regimes}, namely different locations $\llambda$ where the spectrum is looked at, and different zooming resolutions.}

\begin{align}
H_{N+m}(X\sqrt{2N})=\left(\frac{2}{\pi}\right)^{1/4}\frac{2^{m/2+N/2}N^{m/2-1/4}(N!)^{1/2}e^{N X^2}}{(1-X^2)^{1/4}}g_{m,N}(X)\left[1+\mathcal{O}\left(\frac{1}{N}\right)\right]\ ,\label{eq_kernel:asymptoticHermite}
\end{align}
valid for $-1<X<1$, $m\sim\mathcal{O}(1)$ and $g_{m,N}(x)$ given by the following expression
\be
g_{m,N}(x)=\cos\left(N x\sqrt{1-x^2}+(N+1/2)\arcsin(x)-N\pi/2-m\arccos(x)\right)\ .\label{eq_kernel:gmN}
\ee

We can now apply this asymptotic expansion to \eqref{eq_kernel:tobeasymptotics}, with $m=0,-1,-2$ as needed, after the identification $X=z/\sqrt{2}$. \\

The two terms $H_{N-1}^2$ and $H_N\times H_{N-2}$ produce the same $N$-dependent prefactor $\frac{(2/\pi)^{1/2} 2^{N-1}N^{-3/2}(N!)e^{N z^2}}{(1-z^2/2)^{1/2}}$ (check it!), and after simplifications we get to
\be
\sqrt{2N}\rho(z\sqrt{2N})\sim\frac{2}{N\pi\sqrt{2-z^2}}\left[N \cos^2(\alpha+\phi)-(N-1)\cos(\alpha)\cos(\alpha+2\phi)\right]\ ,
\ee
where the cosine terms come from $g_{m,N}(x)$ in \eqref{eq_kernel:gmN}. Here 
\begin{align}
\alpha &=N (z/\sqrt{2})\sqrt{1-z^2/2}+(N+1/2)\arcsin(z/\sqrt{2})-N\pi/2\ ,\\
\phi &=-\arccos(z/\sqrt{2})\ .
\end{align}

Keeping only the leading $\propto N$ terms in the square bracket, and using the identity \\ $\cos ^2(\alpha +\phi )-\cos (\alpha ) \cos (\alpha +2 \phi )=\sin^2(\phi)$, we finally get
\be
\sqrt{2N}\rho(z\sqrt{2N})\sim \frac{2}{\pi\sqrt{2-z^2}}\sin^2(-\arccos(z/\sqrt{2}))=\frac{1}{\pi}\sqrt{2-z^2}\ ,\label{eq_kernel:bulkscalingfinal}
\ee
as expected.

\chapter{Meet Andr\'eief}\label{chap:Andrejef}

In this Chapter, we present a couple of very useful integral identities involving the Vandermonde determinant, and one cute application.

\section{Some integrals involving determinants}\label{sec_Andrejef:cintsdets}

We start with the Andr\'eief identity - also called sometimes the Gram or Heine identity
\cite{ref_Andrejef:Andrejef}. It states that a certain multiple integral involving the product of two
determinants can be written as the determinant of a matrix whose entries are
\emph{single} integrals.\\

Confused? Let's have a closer look.\\ 

We are given two sets of $N$ functions, $\{f_k(x)\}$ and
$\{g_k(x)\}$. We also have an integration measure $\mu(x)$. We then have
\begin{equation}
\boxed{\int \prod_{j=1}^N d\mu(x_j) \det(f_j(x_k))\det(g_j(x_k)) = N!\ \det\left(\int d\mu(x)
f_j(x)g_k(x)\right)}\ .\label{eq_Andrejef:theorem}
\end{equation} 
This can be proved by just expanding
the left hand side as a double sum over permutations, performing the integrals and then
folding the result back into a single sum. Try to prove it yourself - for example, right now.\\

If you think about it for a second, this identity seems too good to be true. On the left hand side, you have, say, a $20$-fold integral of a truly nasty object, and on the right hand side a $20\times 20$ determinant, which can be easily handled by any scientific software - when not explicitly computable in closed form!\\

This identity is especially useful for unitary invariant ensembles $(\beta=2)$, because there you can write the square of the Vandermonde determinant as $\prod_{j<k}(\llambda_j-\llambda_k)^2=\det(\llambda_j^{k-1})\det(\llambda_j^{k-1})$. For example, the partition function of the GUE can be written as a determinant
\begin{align}
\nonumber \mathcal{Z}_{N,\beta} &=\int_{(-\infty,\infty)^N}\prod_{j=1}^N d\llambda_j e^{-\frac{1}{2}\llambda_j^2}\prod_{j<k}(\llambda_j-\llambda_k)^2=
N! \det\left(\int_{-\infty}^\infty d\llambda\ e^{-\frac{1}{2}\llambda^2}\llambda^{j+k-2}\right)\\
&= N! \det\left(2^{\frac{1}{2} (j+k-3)} \left((-1)^{j+k}+1\right) \Gamma
   \left(\frac{1}{2} (j+k-1)\right)\right)\ ,
\end{align}
which can also be evaluated in closed form as a Selberg-like integral \cite{ref_valuetheeigenvalue:Forrester_Warnaar}.\\

A nice feature of the final determinant - and this happens for all $\beta=2$ calculations - is that it is of the form $\det(M_{i+j})$, i.e. it is a \emph{Hankel determinant} (the matrix $M$ is constant along the skew-diagonals). This happens because the two determinants in the integrand on the left hand side are equal, and this produces a factor $(x^{j-1})(x^{k-1})$ on the right hand side.\\

There are two other identities that are similar in spirit to the Andr\'eief identity (the
de Brujin identities \cite{ref_Andrejef:deBrujin}). They read as follows: 
\be\int_{x_1\leq
x_2\leq\ldots\leq x_N}d \mu(\bm x) \det [\phi_i (x_j)] =\mathrm{Pf}
\left[\iint\mathrm{sign}(x-y)\phi_i (x)\phi_j(y) d\mu(x) d\mu(y)\right]\ ,\label{eq_Andrejef:deBrujin}
\ee 
where $i$ and
$j$ run from $1$ to $N$, and 
\be\int d\mu(\bm x)\det \left[\phi_i(x_j)\quad
\psi_i(x_j) \right] =(2N)! \mathrm{Pf}\left[\int d\mu (x)
(\phi_i(x)\psi_j(x)-\phi_j(x)\psi_i(x))\right]\ , \label{eq_Andrejef:deBrujin2}
\ee where $i$ and $j$ run from $1$ to
$2N$.\\

In both the equations above, $\mathrm{Pf}$ denotes a \emph{Pfaffian}. Just like the determinant can be written as a sum over permutations, a Pfaffian is written as a sum over pairings.\\

Given a set $S$ with an even number of elements, $\{1,...,2n\}$, a pairing of $S$ is a collection of $n$ pairs of elements from $S$. For instance, the set $\{1,2,3,4\}$ has three possible pairings: $\{\{1,2\},\{3,4\}\}$, $\{\{1,3\},\{2,4\}\}$, and $\{\{1,4\},\{2,3\}\}$.\\

We can realize pairings as permutations acting of the trivial pairing $\{\{1,2\},\{3,4\}\}$. The previous pairings then correspond to the identity permutation, the transposition $(23)$ and the cycle $(243)$.\\

In terms of these permutations, we have
\be {\rm Pf}(A)=\sum_{P} s(P) \prod_{j=1}^{n}A_{P(2j-1),P(2j)}\ , \label{eq_Andrejef:pfaffian} \ee
where $s(P)$ is the signature of the permutation and $A$ is a even-dimensional skew-symmetric matrix.\\

For example, take $n=2$ and $A$ the following $4\times 4$ matrix
\begin{equation}
A=\begin{pmatrix}
0 & A_{12} & A_{13} & A_{14}\\
-A_{12} & 0 & A_{23} & A_{24}\\
-A_{13} & -A_{23} & 0 & A_{34}\\
-A_{14} & -A_{24} & -A_{34} & 0\ 
\end{pmatrix} \ .
\end{equation}
We have $ {\rm Pf}(A)=A_{12}A_{34}-A_{13}A_{24}+A_{14}A_{23}$ (compare with the pairings listed above for the set $\{1,2,3,4\}$). Note also that ${\rm Pf}(A)=\sqrt{\det(A)}$.\\

We will encounter Pfaffians again in Chapter \ref{chap:FiniteNbeta14}.

\section{Do it yourself}
Let us see a simple example where the Andr\'eief formula turns a nasty problem into a doable one.\\

Question: what is the probability that a $9\times 9$ GUE matrix has $N_+=7$ positive eigenvalues? From first principles, we have in general
\begin{equation}
P_N(N_+=n)=\int d\llambda_1\cdots d\llambda_N \rho(\llambda_1,\ldots,\llambda_N)\delta\left(n - \sum_{i=1}^N \theta(\llambda_i)\right)\ ,\label{eq_Andrejef:probNplus}
\end{equation}
where $\theta(x)$ is the Heaviside step function, $=1$ if $x>0$ and $0$ otherwise.\\

Note that the delta function in \eqref{eq_Andrejef:probNplus} is more correctly a Kronecker delta $\delta_{n,\sum_{i=1}^N \theta(\llambda_i)}$. We can introduce the generating function 
\begin{equation}
\varphi_N(z)=\sum_{n=0}^N P_N(N_+=n)z^n=\frac{1}{\mathcal{Z}_{N,\beta=2}}\int_{-\infty}^\infty \prod_{j=1}^N d\llambda_j
e^{-\frac{1}{2}\sum_{i=1}^N\llambda_i^2+(\ln z)\sum_{i=1}^N\theta(\llambda_i)}\prod_{j<k}(\llambda_j-\llambda_k)^2\ .\label{eq_andrejef:phiNz}
\end{equation}

This multiple integral seems hopelessly complicated. But spotting that $\prod_{j<k}(\llambda_j-\llambda_k)^2=\det(\llambda_i^{j-1})\det(\llambda_i^{j-1})$, we can use the Andr\'eief formula to write
\begin{equation}
\varphi_N(z)=\frac{\det\left(\int_{-\infty}^\infty d\llambda\ e^{-\frac{1}{2}\llambda^2+(\ln z)\theta(\llambda)}\llambda^{i+j-2}\right)}{\det\left(\int_{-\infty}^\infty d\llambda\ e^{-\frac{1}{2}\llambda^2}\llambda^{i+j-2}\right)}=\frac{\det\left( \left((-1)^{i+j}+z\right) c_{i+j}\right)}{\det\left(\left((-1)^{i+j}+1\right) c_{i+j}\right)}\ ,
\end{equation}
where $c_k=2^{\frac{k-3}{2}} \Gamma \left(\frac{k-1}{2}\right)$. We have used Andr\'eief also to express $\mathcal{Z}_{N,\beta=2}$ as a determinant, and erased a common $N!$ factor.\\

Evaluating the integrals, we got a ratio of Hankel determinants, which can easily be evaluated \emph{exactly} with a symbolic software.\\

Note that $\varphi_N(1)=1$, as it should by normalization of $P_N(N_+=n)$ (see \eqref{eq_andrejef:phiNz}). The probabilities $P_N(N_+=n)$ can then be reconstructed by differentiation
\begin{equation}
P_N(N_+=n)=\frac{1}{n!}\partial_z^{(n)}\varphi_N(z)\Big|_{z\to 0}\ .\label{eq_andrejef:probnplus}
\end{equation}

Carrying out this program, we may find that for a $9\times 9$ GUE matrix, 
\begin{align}
\nonumber P_{N=9}(N_+=7) &=\frac{161229045760-20942589825 \pi ^2-9172989000 \pi ^3+3386880000 \pi ^4}{48168960000
   \pi ^4}\\
   &\simeq 5.67686\times 10^{-6}\ .
\end{align}
Note that the evaluation is \emph{exact}, and can be extended to many other values of $n,N$. Of course, it would be desirable to have an exact and explicit formula for these probabilities at arbitrary $n,N$ (see  \cite{ref_Andrejef:TodaForrester}).\\

For a numerical check of \eqref{eq_andrejef:probnplus}, see [$\spadesuit$ \verb"Andreief_check.m"].

\section{To know more...}

\begin{enumerate}
\item There is an interesting connection between Hankel determinants, the so-called \emph{Toda equation} on a semi-infinite lattice, and \emph{Painlev\'e functions}. Define
\begin{equation}
\tau_n=\det(a_{i+j-2})_{i,j=1,\ldots,n}\ ,
\end{equation}
with ``initial conditions" $\tau_{-1}=0$, $\tau_{0}=1$ and $\tau_1=a_0$. Imagine that the entries $a_k$ of this Hankel matrix are functions of $x$ (and so is $\tau_n$, for any fixed $n$). If the $a_k$ satisfy the following relation, $a_k=a_{k-1}^\prime$ (where $^\prime$ denotes differentiation with respect to $x$), then the following hierarchy of equations holds
\begin{equation}
\tau_n^{\prime\prime}\tau_n-(\tau_n^\prime)^2=\tau_{n+1}\tau_{n-1}\ .
\end{equation} 
In [$\spadesuit$ \verb"Toda.m"] we test this property for $n=3$.\\

Quite amazingly, the same Toda lattice equation is obeyed by so-called $\tau$-functions, which arise in the Hamiltonian formulation of the six Painlev\'e equations (PI - PVI), other fundamental objects in the theory of nonlinear integrable systems \cite{ref_Andrejef:ForresterWitte}.\\

These deep connections between Andr\'eief evaluations, Hankel determinants, Toda lattice and Painlev\'e functions are at the root of quite spectacular results (see e.g. \cite{ref_Andrejef:Kanzieper1} and \cite{ref_Andrejef:Kanzieper2}).

\end{enumerate}

\chapter{Finite $N$ is not finished}\label{chap:FiniteNbeta14}

In this short Chapter, we compute in the quickest way the spectral density for the GOE ($\beta=1$) and GSE ($\beta=4$). The symmetry classes beyond the Unitary have a reputation for being ``unfriendly". We do not aim at giving the most general treatment of correlation functions for such cases. The goal of this Chapter is just to provide a smooth and gentle appetizer, allowing you to tackle the nastier bits with your back covered.

\section{$\beta=1$}

Let us assume $N$ is even for simplicity. Indices $i,j$ run from $1$ to $N$, while $k,\ell$ run from $0$ to $N-1$.\\ 

Suppose the jpdf of eigenvalues is given by \be \rho(\llambda_1,\ldots,\llambda_N)=\frac{1}{\mathcal{Z}}|\Delta_N(\bm\llambda)|\prod_{i=1}^N w(\llambda_i)\ .\ee

For $w(x)=\exp(-x^2/2)$, we recover the jpdf for the GOE.\\

Let's compute the normalization factor, a.k.a. the partition function, \be \label{eq:finiteNbeta14_Z} \mathcal{Z}=\int
d\bm\llambda|\Delta_N(\bm\llambda)|\prod_{i=1}^N w(\llambda_i)=|\hat{a}_N|\int d\bm\llambda|\det(R_{j-1}(\llambda_i)w(\llambda_i))|\ ,\ee where
$R_k(x)=a_k x^{k}+\cdots$ is a family of polynomials through which the Vandermonde materializes, and 
\be \hat{a}_N= \left (\prod_{k=0}^{N-1}a_{k} \right)^{-1}\ .\ee To get rid of the absolute value, we restrict integration to the domain where the variables are ordered: \be \mathcal{Z}
=N! |\hat{a}_N|\int_{-\infty< \llambda_1<\llambda_2<\cdots<\infty} d\bm \llambda\det(R_{j-1}(\llambda_i)w(\llambda_i))\ . \label{eq_FiniteNbeta14:int} \ee

We may now use the de Brujin identity \eqref{eq_Andrejef:deBrujin} to get 
\be \mathcal{Z}=N! |\hat{a}_N| {\rm Pf}(A_{i,j})\ , \label{eq_FiniteNbeta14:pfaffian} \ee 
where 
\be
A_{i,j}=\int_{-\infty}^\infty dx\int_{-\infty}^{\infty}dy R_{i-1}(y)R_{j-1}(x)w(x)w(y){\rm
sign}(x-y)\ ,\label{eq_FiniteNbeta14:defA}\ee 
and Pf denotes the Pfaffian of the skew-symmetric matrix $A_{ij}$.\\

Let us now stop for a second to check on a $2 \times 2$ example that, indeed, the expressions in \eqref{eq_FiniteNbeta14:int} and \eqref{eq_FiniteNbeta14:pfaffian} coincide. Starting from the integral in \eqref{eq_FiniteNbeta14:int}, specialized to a $2 \times 2$ case, we have
\begin{align} \label{eq_FiniteNbeta14:int_check}
\nonumber &\int_{-\infty}^{+\infty} dy \int_{-\infty}^y dx R_0(x) R_1(y) w(x) w(y) - \int_{-\infty}^{+\infty} dy \int_{-\infty}^y dx R_0(y) R_1(x) w(x) w(y) \\ 
&=\int_{-\infty}^{+\infty} dx \int_{-\infty}^x dy R_0(y) R_1(x) w(x) w(y) - \int_{-\infty}^{+\infty} dy \int_{-\infty}^y dx R_0(y) R_1(x) w(x) w(y) \ ,
\end{align}
where we have simply renamed the variables $x\to y$ and $y\to x$ in the first integrals.\\

If we now expand the Pfaffian in equation \eqref{eq_FiniteNbeta14:pfaffian} we obtain
\begin{align}
 \nonumber {\rm Pf}(A_{i,j}) &= \int_{-\infty}^{+\infty} dy \int_{y}^{+\infty} dx R_0(y) R_1(x) w(x) w(y) - \int_{-\infty}^{+\infty} dy \int_{-\infty}^y dx R_0(y) R_1(x) w(x) w(y) \\
&=\int_{-\infty}^{+\infty} dx \int_{-\infty}^x dy R_0(y) R_1(x) w(x) w(y) - \int_{-\infty}^{+\infty} dy \int_{-\infty}^y dx R_0(y) R_1(x) w(x) w(y) \ ,
\end{align}
where in the first integral we have simply rewritten the integration domain $-\infty < y < x < \infty$. The above expression coincides with \eqref{eq_FiniteNbeta14:int_check}. \\

Now, stare at \eqref{eq_FiniteNbeta14:defA} for a few seconds. To simplify the notation slightly, we may define the following \emph{skew-symmetric inner product} 
\be \langle
f,g\rangle_1=\frac{1}{2}\int_{-\infty}^\infty dx\int_{-\infty}^{\infty}
 f(y)g(x)w(x)w(y){\rm
sign}(x-y)dy\ ,
\ee 
so that we can write $A_{i,j}=2\langle
R_{i-1},R_{j-1}\rangle_1$. Note that in general $\langle
f,g\rangle_1=-\langle
g,f\rangle_1$ - we wouldn't call it skew-symmetric otherwise, would we?\\

Now, in complete analogy with what we did for $\beta=2$ - identifying \emph{specific} polynomials, orthonormal with respect to the given weight - we may choose the polynomials $R$ that ``behave nicely" with respect to this inner product. The nice properties we require are: evens and
 odds are orthogonal among themselves, \be\label{eq_FiniteNbeta14:ortho1} \langle R_{2k},R_{2\ell}\rangle_1=\langle
 R_{2k+1},R_{2\ell+1}\rangle_1=0\ ,\ee and evens are orthogonal to odds unless they are adjacent,
\be\label{eq_FiniteNbeta14:ortho2} \langle R_{2k},R_{2\ell+1}\rangle_1=-\langle
R_{2\ell+1},R_{2k}\rangle_1=\delta_{k\ell}\ .\ee 

With this particular choice, the $R$'s are called \emph{skew-orthogonal polynomials}. The matrix $A$ in \eqref{eq_FiniteNbeta14:defA} acquires a simple form, \be \label{eq:finiteNbeta14_matrix} A=\begin{pmatrix}
0&2&&&\\-2&0&&&\\&&0&2&\\&&-2&0&\\&&&& \ddots
\end{pmatrix}\ ,\ee and the
expression for $\mathcal{Z}$ drastically simplifies: the determinant of $A$ becomes simply $2^N$, hence its Pfaffian becomes $2^{N/2}$, and all the information about the \emph{specific} weight function $w(x)$ is contained in $\hat{a}_N$. As a consequence, from \eqref{eq_FiniteNbeta14:pfaffian} we get for the partition function in \eqref{eq:finiteNbeta14_Z} $\mathcal{Z} = N! |\hat{a}_N| 2^{N/2}$.\\

Let us now generalize this calculation slightly. Consider the quantity 
\begin{align}
\nonumber \mathcal{Z}[f] &= |\hat{a}_N|\int d\bm\llambda|\det(R_{j-1}(\llambda_i)w(\llambda_i)f(\llambda_i))| \\
&= \mathcal{Z}[f = 1] \int d\bm \llambda \rho(\llambda_1,\ldots,\llambda_N)\prod_{i=1}^Nf(\llambda_i) \ ,
\end{align} where we introduced an arbitrary
function $f(x)$ in the game, such that the integral is convergent. Note that $\mathcal{Z}[f = 1]$ coincides with $\mathcal{Z}$.\\
 
From this new partition function, we can recover the density of eigenvalues for finite $N$
\be \rho(\llambda)=\int d\llambda_2d\llambda_3\cdots d\llambda_N \rho(\llambda,\llambda_2,\ldots,\llambda_N)\ee by means of a functional
derivative. This is the operator $\frac{\delta}{\delta f}$, which satisfies all the properties of a derivative, plus the condition
\be\frac{\delta}{\delta f(x)}f(y)=\delta(y-x)\ .\ee Then, \be \left.\frac{\delta}{\delta
f(x)}\mathcal{Z}[f]\right|_{f=1}= \mathcal{Z}[f=1]\int d\bm \llambda \rho(\llambda_1,\ldots,\llambda_N)\sum_{i=1}^N\delta(\llambda_i-\llambda)=N \mathcal{Z}[f=1] \rho(\llambda) \ . \label{eq:finiteNbeta14_funcder} \ee

Following a calculation perfectly analogous to the previous one, we arrive at 
\be \mathcal{Z}[f]=
N! |\hat{a}_N| {\rm Pf}(A_{ij}[f])\ ,\label{eq_FiniteNbeta14:Zf}
\ee
where \be A_{ij}[f]=\int_{-\infty}^\infty
dx\int_{-\infty}^{\infty}dyR_{i-1}(y)R_{j-1}(x)w(x)w(y)f(x)f(y){\rm sign}(x-y)\ . \label{eq:finiteNbeta14_Aelement} \ee
Computing the functional derivative, and recalling the definition of Pfaffian in equation \eqref{eq_Andrejef:pfaffian}, we have 
\be \left.\frac{\delta}{\delta
f(x)}\mathcal{Z}[f]\right|_{f=1}=N! |\hat{a}_N| \sum_{P} s(P) \left[\frac{\delta}{\delta
f(x)}\prod_{k=1}^{N/2}A_{P(2k-1),P(2k)}[f]\right]_{f=1}. \label{eq_finiteNbeta14:zeta} \ee

When we apply the product rule for the derivative, and set $f=1$, for each term in the sum over permutations we get
\begin{align} \label{eq_finiteNbeta14:expansion}
\nonumber & \left[\frac{\delta}{\delta f(x)}A_{P(1),P(2)}[f]\right]_{f=1} A_{P(3),P(4)}[1] \cdots A_{P(N-1),P(N)}[1] + \ldots \\ \nonumber
& + A_{P(1),P(2)}[1] \left[\frac{\delta}{\delta f(x)}A_{P(3),P(4)}[f]\right]_{f=1} \cdots A_{P(N-1),P(N)}[1] + \ldots \\
& + A_{P(1),P(2)}[1] \cdots \left[\frac{\delta}{\delta f(x)}A_{P(N-1),P(N)}[f]\right]_{f=1} \ .
\end{align}
The orthogonality
relations (\ref{eq_FiniteNbeta14:ortho1}), (\ref{eq_FiniteNbeta14:ortho2}) imply that the products in the above expressions are different from zero only when $P$ is the identity permutation, i.e. when $P(2j-1)$ and $P(2j)$ are adjacent numbers for each matrix element in the product.\\

Hence, the sum in \eqref{eq_finiteNbeta14:zeta} reduces to the expression in \eqref{eq_finiteNbeta14:expansion} where $P(j) = j$, $\forall \ j$. Each element $A_{P(2j-1),P(2j)}$ yields a factor $2$ from the matrix $A$ in \eqref{eq:finiteNbeta14_matrix}, so that each product in \eqref{eq_finiteNbeta14:expansion} reduces to an expression of the type $2^{N/2-1} \left[\frac{\delta}{\delta f(x)}A_{2k-1,2k}[f]\right]_{f=1}$. Comparing \eqref{eq:finiteNbeta14_funcder} and \eqref{eq_finiteNbeta14:zeta}, we eventually find that

\be \rho(x) = \frac{N! |\hat{a}_N| 2^{N/2-1}}{N \mathcal{Z}[f=1]} \sum_{k=1}^{N/2} \left[\frac{\delta}{\delta f(x)}A_{2k-1,2k}[f]\right]_{f=1} \ . \ee

Making the result of the functional differentiation of \eqref{eq:finiteNbeta14_Aelement} explicit, and rearranging indices, we finally obtain

\be 
\label{eq_finiteNnotfinished:theordensityGOE} 
\boxed{\rho(x)= \frac{1}{2N}\sum_{k=0}^{N/2-1}
w(x)[R_{2k}(x)\Phi_{2k+1}(x)-R_{2k+1}(x)\Phi_{2k}(x)]}\ ,
\ee where \be \Phi_k(x)=\int_{-\infty}^\infty
dyR_k(y)w(y){\rm sign}(x-y)\ \ee

For the Gaussian case, it can be shown \cite{ref_FiniteNbeta14:Ghosh,ref_FiniteNbeta14:Mehta} that we can choose

 \begin{align} 
 R_{2k}(x) &=\frac{\sqrt{2}}{\pi^{\frac{1}{4}}2^k(2k)!!}H_{2k}(x)\ ,\\
R_{2k+1}(x) &=\frac{\sqrt{2}}{\pi^{\frac{1}{4}}2^{k+2}(2k-1)!!}\left[-H_{2k+1}(x)+4kH_{2k-1}(x)\right]\ ,
\end{align}
where the $H_k(x)$ are Hermite polynomials.  This gives, for example, 
\be  R_0(x)=\frac{\sqrt{2}}{\pi^{\frac{1}{4}}},\quad R_1(x)=-\frac{\sqrt{2}x}{2\pi^{\frac{1}{4}}},\quad R_2(x)=\frac{\sqrt{2}(2x^2-1)}{2\pi^{\frac{1}{4}}},\quad R_3(x)=-\frac{\sqrt{2}x(2x^2-5)}{2\pi^{\frac{1}{4}}}\ .\ee Since the leading coefficient of $H_k(x)$ is $2^k$, we have \be \hat{a}_N=\frac{(-1)^{\frac{N}{2}}2^\frac{N(N-2)}{4}}{\pi^{\frac{N}{4}}\prod_{k=0}^{N/2-1}(2k)!}\ ,\ee even though this quantity has completely dropped out from the final expression for the density \eqref{eq_finiteNnotfinished:theordensityGOE}.

For a numerical check of \eqref{eq_finiteNnotfinished:theordensityGOE}, see Fig. \ref{fig_finiteN} below, which was obtained with the \\ code
[$\spadesuit$ \verb"Gaussian_finite_density_check.m"].

\section{$\beta=4$}

Now 
\be 
\rho(\llambda_1,\ldots,\llambda_N)=\frac{1}{\mathcal{Z}}|\Delta_N(\bm\llambda)|^4\prod_{i=1}^N w(\llambda_i)\ .
\ee 
Start by writing $|\Delta_N(\bm\llambda)|^4$ as a determinant of size $2N$, in which two columns depend on each variable. This is
\be
 |\Delta_N(\bm\llambda)|^4=\det [x_i^k\quad kx_i^{k-1}]\ ,
 \ee where $1\leq i\leq N$ and $0\leq k\leq 2N-1$. For instance, for $N=2$ we have 
 \be 
 (x_2-x_1)^4=\det\begin{pmatrix}1&0&1&0\\x_1&1&x_2&1\\
x_1^2&2x_1&x_2^2&2x_2\\x_1^3&3x_1^2&x_2^3&3x_2^2\end{pmatrix},
\ee 
which can be verified if you have 10 minutes to spare.\\

We can change $x_i^k$ by any family of polynomials $Q_k(x_i)=b_kx_i^k+\cdots$ that produce the Vandermonde, and $kx_i^{k-1}$ by its derivative $Q_k'(x_i)$. So\be 
|\Delta_N(\bm\llambda)|^4= |\hat{b}_N| \det [Q_{j-1}(x_i)\quad Q_{j-1}'(x_i)]\ ,
\ee 
where $1\leq j\leq 2N$ and 
\be \hat{b}_N=\left (\prod_{k=0}^{N-1}b_k^2 \right)^{-1} \ .
\ee

In this case, Eq. \eqref{eq_FiniteNbeta14:Zf} gets modified as
 \be \mathcal{Z}[a]=(2N)!|\hat{b}_N| {\rm Pf}(B_{i,j}[a])\ , 
 \ee 
 where \be
B_{i,j}[a]=\int_{-\infty}^\infty dx[Q_{i-1}(x)Q'_{j-1}(x)-Q'_{i-1}(x)Q_{j-1}(x)]w(x)a(x)\ .\ee
We have used here the second De Brujin identity  \eqref{eq_Andrejef:deBrujin2}.\\

We may consider the above integral as another skew-symmetric inner product \be \langle
f,g\rangle_4=\frac{1}{2}\int_{-\infty}^\infty dx
 [f(x)g'(x)-f'(x)g(x)]w(x)\ ,\ee and we may choose the polynomials $Q$
 to be skew-orthogonal with relation to this: evens and
 odds are orthogonal among themselves, \be\label{ortho3} \langle Q_{2k},Q_{2\ell}\rangle_4=\langle
 Q_{2k+1},Q_{2\ell+1}\rangle_4=0\ ,\ee and evens are orthogonal to odds unless they are adjacent,
\be\label{ortho4} \langle Q_{2k},Q_{2\ell+1}\rangle_4=-\langle
Q_{2\ell+1},Q_{2k}\rangle_4=\delta_{k\ell}\ .\ee

Computing the functional derivative as before, we have 
\be \label{eq_finiteNnotfinished:theordensityGSE}
\boxed{\rho(x)=\frac{1}{2N}\sum_{k=0}^{N-1}
w(x)[Q_{2k}(x)Q'_{2k+1}(x)-Q_{2k+1}(x)Q'_{2k}(x)]}\ .\ee
In the Gaussian case, we can choose \be Q_{2k}=\frac{\sqrt{2}}{\pi^{\frac{1}{4}}2^{k}(2k)!!}\left[4kQ_{2k-2}(x)+H_{2k}(x\sqrt{2})\right]\ ,\quad Q_{2k+1}(x)=\frac{\sqrt{2}}{\pi^{\frac{1}{4}}2^{k+1}(2k+1)!!}H_{2k+1}(x\sqrt{2})\ .\ee This gives, for example, \be Q_0=\frac{\sqrt{2}}{\pi^{\frac{1}{4}}}\ ,\quad Q_1(x)=\frac{2x}{\pi^{\frac{1}{4}}}\ ,\quad Q_2(x)=\frac{\sqrt{2}(4x^2+1)}{2\pi^{\frac{1}{4}}}\ ,\quad Q_3(x)=\frac{2x(4x^2-3)}{3\pi^{\frac{1}{4}}}\ .\ee Also, we have 
\be \hat{b}_N=\frac{2^{\lfloor \frac{N-1}{2} \rfloor}2^{N(N-1)}}{\pi^{\frac{N}{2}}\prod_{k=0}^{N-1}(k!!)^2}\ .\ee 

\begin{figure}[t]
\centering
\includegraphics[width=1.\columnwidth]{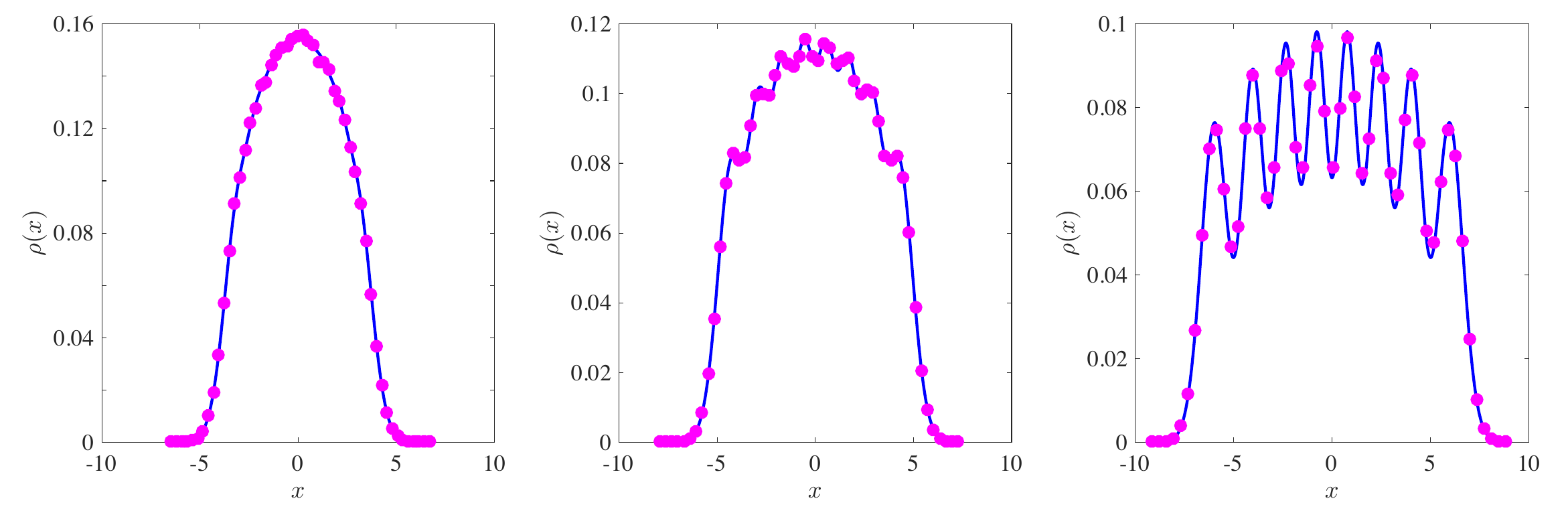}
\caption{Comparison between numerically generated eigenvalue histograms of $50000$ matrices of size $N=8$ belonging to the Gaussian ensembles (GOE on the left, GUE in the middle, and GSE on the right) and the corresponding theoretical densities.}
\label{fig_finiteN}
\end{figure}

Again, for a numerical check of \eqref{eq_finiteNnotfinished:theordensityGSE}, see Fig. \ref{fig_finiteN}, which was obtained with the \\ code [$\spadesuit$ \verb"Gaussian_finite_density_check.m"].

\chapter{Classical Ensembles: Wishart-Laguerre}\label{chap:Wishart}

In this Chapter, we present one of the ``classical" examples of rotationally invariant models: the Wishart-Laguerre (WL) ensemble.

\section{Wishart-Laguerre ensemble}\label{secWishart}

Historically, one of the earliest appearances of a random matrix ensemble\footnote{In Mathematics, however, the 1897 work by Hurwitz on the volume form of a general unitary matrix is of historical significance \cite{ref_Wishart:Hurwitz,ref_Wishart:Diaconis}.} occurred in 1928, when the Scottish mathematician John Wishart published a paper on multivariate data analysis in the journal \emph{Biometrika} \cite{ref_Wishart:WishartX}. \\

Wishart matrices are square $N\times N$ matrices $W$ with correlated entries. They are constructed as\footnote{Sometimes you find a normalized version of it, with a $1/M$ factor in front.}
$W=HH^{\dagger}$, where $H$ is a $N\times M$ matrix $(M\geq N)$ filled with
i.i.d. Gaussian entries\footnote{The notation $W(N,M)$ is also used.}. These entries may be real, complex or quaternion (we shall use again the Dyson index $\beta=1,2,4$ for the three cases, respectively), and $^\dagger$ stands for the transpose or hermitian conjugate of the matrix $H$. For example, for a $2\times 3$ \emph{complex} matrix $H$
\begin{equation}
W=
\left(
\begin{array}{ccc}
x_{11}+\mathrm{i}y_{11} & x_{12}+\mathrm{i}y_{12} & x_{13}+\mathrm{i}y_{13}\\
x_{21}+\mathrm{i}y_{21} & x_{22}+\mathrm{i}y_{22} & x_{23}+\mathrm{i}y_{23}\\
\end{array}
\right)
\left(
\begin{array}{cc}
x_{11}-\mathrm{i}y_{11} & x_{21}-\mathrm{i}y_{21}\\
x_{12}-\mathrm{i}y_{12} & x_{22}-\mathrm{i}y_{22} \\
x_{13}-\mathrm{i}y_{13} & x_{23}-\mathrm{i}y_{23}\\
\end{array}
\right)\ .\label{eq_Wishart:Wishart2x3}
\end{equation}\\

Work out the matrix product, and convince yourself that $W$ is hermitian, therefore has real eigenvalues.\\

The Wishart ensemble is also referred to as ``Laguerre", since its spectral properties involve Laguerre
polynomials, and also ``chiral" in the context of
applications to Quantum Chromodynamics (QCD) \cite{ref_Wishart:Verbaarschot}. They are often called LOE, LUE and LSE,
for $\beta=1,2,4$, respectively.\\

While the Gaussian eigenvalues can in principle be anywhere on the real axis, Wishart matrices have $N$ \emph{non-negative} eigenvalues,
$\{\llambda_1,\llambda_2,\ldots,\llambda_N\}$. Indeed, Wishart matrices $W$ are \emph{positive semidefinite}. This means that (e.g. for $\beta=2$) $\mathbf{u}^\star W\mathbf{u}\geq 0$ for \emph{all} nonzero column vectors $\mathbf{u}$ of $N$ complex numbers. The proof is not hard, have a go at it! 

\begin{center}
\fbox{\begin{minipage}{33em}
{\it Question.} What is the jpdf of the \emph{entries} of WL ensemble $W$? \\ \\
$\blacktriangleright$ With some effort (see below), it can be computed as
\begin{equation}
\rho[W]\propto e^{-\frac{1}{2} \mathrm{Tr} W} (\det
W)^{\frac{\beta}{2}(M-N+1)-1}\ ,\label{eq_que_Wishart:jpdfentriesW}
\end{equation}
from which you immediately see that i) the entries are correlated\footnote{Unless for specific combinations of $\beta,M,N$ for which the determinant disappears.} (the determinant easily kills any hope of factorizing this jpdf), and ii) the model is rotationally invariant\footnote{The jpdf \eqref{eq_que_Wishart:jpdfentriesW} is not in contrast with Weyl's lemma (see Eq. \eqref{eq_classified:Weyl}). The determinant of a $N\times N$ matrix $W$ can be indeed written as a function of the traces of the first $N$ powers of $W$ (see \cite{ref_Wishart:Kleefeld}).}.
\end{minipage}}
\end{center}

From \eqref{eq_que_Wishart:jpdfentriesW}, the jpdf of \emph{eigenvalues} can be written down immediately (just express everything in terms of the eigenvalues and append a Vandermonde at the end)
\begin{equation}
\rho(\llambda_1,\ldots,\llambda_N)= \frac{1}{\mathcal{Z}_{N,\beta}^{(L)}}\, e^{-\frac{1}{2}\sum_{i=1}^N
\llambda_i}\, \prod_{i=1}^N \llambda_i^{\alpha\beta/2}\,\prod_{j<k}
|\llambda_j-\llambda_k|^{\beta}, \label{eq_Wishart:jpdfeigenvalues}
\end{equation}
where $\alpha= (1+M-N)-2/\beta$ and the normalization constant $\mathcal{Z}_{N,\beta}^{(L)}$ can be computed again
using modifications of the Selberg integral\footnote{Note that while
for Wishart matrices $M-N$ is a non-negative {\em integer} and $\beta=1$, $2$ or $4$, the jpdf in \eqref{eq_Wishart:jpdfeigenvalues} is well defined for any $\beta>0$ and any
$\alpha>-2/\beta$ (this last condition is necessary to ensure that the jpdf is
normalizable). When these parameters take continuous values, this jpdf defines the
so-called $\beta$-Laguerre ensemble.} \cite{ref_valuetheeigenvalue:Forrester_Warnaar}. As for the Gaussian ensembles,
one may sometimes find in the literature an extra factor $\beta$ in the exponential.\\

The confining potential for the Wishart-Laguerre ensemble is thus $V(x)=\frac{1}{2}x-\frac{\alpha}{2}\ln x$, and this clearly motivates the use of (associated) Laguerre polynomials $L^{(\alpha)}_n(x)$, which are orthogonal with respect to this precise weight (after a simple rescaling),
\begin{equation}
\int_0^\infty dx\ x^\alpha e^{-x}L_n^{(\alpha)}(x)L_m^{(\alpha)}(x)=\frac{\Gamma(n+\alpha+1)}{n!}\delta_{m,n}\ .
\end{equation}

\begin{center}
\fbox{\begin{minipage}{33em}
\label{que_Wishart:AntiWishart}
{\it Question.} What happens if I take $M<N$? \\ \\
$\blacktriangleright$ This situation defines the so-called \emph{Anti-Wishart ensemble} $\tilde{W}$. In this case, one can
show that $N-M$ eigenvalues are exactly $0$. The jpdf is similar to \eqref{eq_Wishart:jpdfeigenvalues}, but some of the matrix elements of
$\tilde{W}$ are non-random and deterministically related to the first $M$ rows of $\tilde{W}$ \cite{ref_Wishart:Janik}.
\end{minipage}}
\end{center}

The code [$\spadesuit$ \verb"Wishart_check.m"] produces instances of Wishart matrices for different $\beta$s, as well as normalized histograms of their eigenvalues. You can start having a look at it now, but please come back to it after reading the next chapter.

\begin{center}
\fbox{\begin{minipage}{33em}
\label{que_Wishart:SpectralDensityWL}
{\it Question.} What is the limiting spectral density of the WL ensembles for $N\to\infty$? \\ \\
$\blacktriangleright$ It is called the \emph{Mar\v{c}enko-Pastur} density \cite{ref_Wishart:Marcenko}, which is superimposed to the histograms produced with the code above in Fig. \ref{fig:MP_check} of the next chapter. We are going to derive it using the resolvent method very shortly.
\end{minipage}}
\end{center}

\section{Jpdf of entries: matrix deltas...}

The calculation of the jpdf of entries \eqref{eq_que_Wishart:jpdfentriesW} proceeds through a few simple steps. Set for simplicity $\beta=2$ (hermitian matrices). We can formally write
\begin{equation}
\rho[W]=\int dH\rho[H]\delta\left(W-H H^\dagger\right)\ .\label{eq_Wishart:jpdfWformal}
\end{equation}
As usual, the measure $dH$ means that we are integrating over the $2NM$ degrees of freedom (dof)\footnote{With ``degrees of freedom" we mean the independent real parameters that are necessary to define a matrix. For example, a hermitian matrix has $N^2$ degrees of freedom - the $N$ real entries on the diagonal, and the real and imaginary parts of the entries in the upper triangle.} of $H$: each entry of the $N\times M$ matrix $H$ is a complex number, so it is parametrized by two real numbers\footnote{Note, in particular, that for $N=M$ the square matrix $H$ is \emph{not} hermitian, and has $2N^2$ ``degrees of freedom" (dof) instead of $N^2$.}. Therefore, $dH=\prod_{i=1}^N\prod_{j=1}^M d\mathrm{Re}[H_{ij}]d\mathrm{Im}[H_{ij}]$.\\

The matrix delta $\delta(W-H H^\dagger)$ enforces the constraint that a certain matrix $W$ must be equal to another matrix $HH^\dagger$. We do have an integral representation for the \emph{scalar} delta function, which does the same job for real numbers. It should then be easy to work out the corresponding integral representation for the delta function of, say, a $N\times N$ hermitian matrix $K$ - after all, it will just be the product of scalar deltas, one for each of the real dof
\begin{align}
\nonumber &\delta(K) =\prod_{i=1}^N\delta(K_{ii})\prod_{i=1}^N\prod_{j>i}\delta(K_{ij}^{(R)})\delta(K_{ij}^{(I)})=\\ &\int \frac{d T_{11}}{2\pi}\cdots \frac{d T_{NN}}{2\pi}
\exp\left\{\mathrm{i}\sum_{i=1}^N T_{ii} K_{ii}\right\}
\int \prod_{i=1}^N\prod_{j> i}\frac{d T^{(R)}_{ij}}{2\pi}\frac{d T^{(I)}_{ij}}{2\pi}
e^{\mathrm{i}\sum_{i=1}^N\sum_{j> i} [T^{(R)}_{ij} K^{(R)}_{ij}+T^{(I)}_{ij} K^{(I)}_{ij}]}\ ,\label{eq_Techniques:uglydelta}
\end{align}
where we have introduced a set of $N(N+1)/2$ parameters $\{T\}$, one for each delta.\\

Arranging the parameters $\{T\}$ into a hermitian matrix, try to show that the ugly expression in \eqref{eq_Techniques:uglydelta} can be recast in the more elegant form
\begin{equation}
\delta(K)=\frac{1}{2^N \pi^{N^2}}\int dT\ e^{\mathrm{i}\mathrm{Tr}[T K]}\ .\label{eq_Techniques:elegant}
\end{equation}

We can now perform the multiple integral in \eqref{eq_Wishart:jpdfWformal}, with Gaussian distributed dof of $H$
\begin{align}
\nonumber\rho[H] &\equiv\rho(H_{11}^{(R)},H_{11}^{(I)},\ldots,H_{NM}^{(R)},H_{NM}^{(I)})=\prod_{i=1}^N\prod_{j=1}^M \left[\frac{1}{2\pi}\exp\left(-\frac{1}{2}H_{ij}^{(R)2}-\frac{1}{2}H_{ij}^{(I)2}\right)\right]\\
&=\left(\frac{1}{2\pi}\right)^{NM}e^{-\frac{1}{2}\mathrm{Tr}(H H^\dagger)}\ ,\label{eq_Techniques:rhoWexplicit}
\end{align}
where $^{(R)}$ and $^{(I)}$ denote the real and imaginary part of each of the $NM$ entries of $H$.\\

Combining \eqref{eq_Wishart:jpdfWformal}, \eqref{eq_Techniques:elegant} and \eqref{eq_Techniques:rhoWexplicit} we have
\begin{equation}
\rho[W]=\frac{1}{2^N \pi^{N^2}}\left(\frac{1}{2\pi}\right)^{NM}\int dT\int dH\ e^{-\frac{1}{2}\mathrm{Tr}(H H^\dagger)+\mathrm{i}\mathrm{Tr}T(W-H H^\dagger)}\ .\label{eq_Wishart:finalTH}
\end{equation}
Dividing all the dof of the hermitian matrix $T$ by $1/2$ (i.e. changing variables $T\to T/2$), we obtain
\begin{equation}
\label{eq_Wishart:finalTH}\rho[W]=\frac{1}{2^N \pi^{N^2}}\left(\frac{1}{2\pi}\right)^{NM}\left(\frac{1}{2}\right)^{N^2}\int dT\int dH\ e^{-\frac{1}{2}\mathrm{Tr}(H H^\dagger)+\frac{\mathrm{i}}{2}\mathrm{Tr}T(W-H H^\dagger)}\ .
\end{equation}

\section{...and matrix integrals}

Next, we use the following identity for $N\times N$ hermitian matrices $T$
\begin{equation}
[\det(\mu \mathbbm{1}-T)]^{-M}=\frac{1}{(4\pi\mathrm{i})^{M N}}\int\prod_{k=1}^M d^2\mathbf{s}_k\exp\left\{\frac{\mathrm{i}}{2}\mu\sum_{k=1}^M\mathbf{s}_k^\dagger\mathbf{s}_k-\frac{\mathrm{i}}{2}\sum_{k=1}^M \mathbf{s}_k^\dagger T \mathbf{s}_k\right\}\ ,\label{eq_Techniques:detgaussianfyod}
\end{equation}
where $\mathbbm{1}$ is the $N\times N$ identity matrix, $\mathbf{s}_k$ are $k=1,\ldots,M$ complex (column) vectors, so that \\ $d^2\mathbf{s}_k=\prod_{i=1}^N d s_{k,i}d s^\star_{k,i}$ and $\mu$ is such that $\mathrm{Im}[\mu]>0$.

\begin{center}
\fbox{\begin{minipage}{33em}
{\it Question.} Any hint on how to prove it? \\ \\
$\blacktriangleright$ Just write $T=U^\dagger\Lambda U$, with $U$ the unitary matrix diagonalizing $T$, and $\Lambda$ the diagonal matrix of eigenvalues $\lambda_i$. Then make the change of variables $U\mathbf{s}_k\to \tilde{\mathbf{s}}_k$, which is unitary and thus has Jacobian equal to $1$. The resulting integral factorizes as 
\begin{equation}
\int\prod_{k=1}^M d^2\tilde{\mathbf{s}}_k \to \left[\prod_{\ell=1}^N 2\int dx\ dy\ e^{\frac{\mathrm{i}}{2}(x-\mathrm{i}y)(\mu-\lambda_\ell)(x+\mathrm{i}y)}\right]^M\ ,
\end{equation}
where $x,y$ are real and imaginary part of the $\ell$th entry of $\mathbf{s}_k$, and the factor of $2$ is the Jacobian of the change of variables $\{\tilde{s}_{k,i},\tilde{s}^\star_{k,i}\} \to \{x,y\}$. The integral in $\{x,y\}$ yields $2\pi\mathrm{i}/(\mu-\lambda_\ell)$, from which the claim is immediate.
\end{minipage}}
\end{center}

We can now perform the $H$ integral in \eqref{eq_Wishart:finalTH}. How? Just imagine that the $k$th vector $\mathbf{s}_k$ is constructed as $\mathbf{s}_k=(H_{1k},\ldots,H_{Nk})^T$, i.e. it is basically the $k$th
column of the rectangular matrix $H$.\\

Hence, note the identity $-(1/2)\mathrm{Tr}(H H^\dagger)=(\mathrm{i}/2)\mu\sum_{k=1}^M\mathbf{s}_k^\dagger\mathbf{s}_k$, with $\mu=\mathrm{i}$. Finally, we have to calculate the Jacobian of the change of variables $\{H^{(R)}_{ik},H^{(I)}_{ik}\}\to \{s_{k,i},s^\star_{k,i}\}$. For each entry of $H$, we have
\begin{equation}
\begin{cases}
s_{k,i} &=H^{(R)}_{ik}+\mathrm{i}H^{(I)}_{ik}\\
s^\star_{k,i} &=H^{(R)}_{ik}-\mathrm{i}H^{(I)}_{ik}\ .
\end{cases}
\end{equation}
The Jacobian from $s\to H$ is 
\begin{equation}
\left(\begin{array}{cc} \frac{\partial s_{k,i}}{\partial H^{(R)}_{ik}} & \frac{\partial s_{k,i}}{\partial H^{(I)}_{ik}} \\
&\\
\frac{\partial s^\star_{k,i}}{\partial H^{(R)}_{ik}}  & \frac{\partial s^\star_{k,i}}{\partial H^{(I)}_{ik}} 
\end{array}\right)=
\left(\begin{array}{cc} 1 & \mathrm{i}\\
1 & -\mathrm{i}
\end{array}\right)=-2\mathrm{i}\ .
\end{equation}
Thus, the Jacobian from $H\to s$ (the one we need) is (in absolute value) equal to $1/2$ for each entry. In total, $(1/2)^{NM}$.\\

Therefore, using \eqref{eq_Techniques:detgaussianfyod}
\begin{equation}
\rho[W]=\frac{1}{2^N \pi^{N^2}}\left(\frac{1}{2\pi}\right)^{NM}\left(\frac{1}{2}\right)^{N^2+NM}(4\pi\mathrm{i})^{MN}\int dT\ e^{\frac{\mathrm{i}}{2}\mathrm{Tr}(TW)}\left[\det\left(\mathrm{i} \mathbbm{1}-T\right)\right]^{-M}\ .
\end{equation}

We now need another matrix integral, with the pompous name ``Ingham-Siegel integral of second type" \cite{ref_Wishart:Fyodorov}, whose general formula reads (see Appendix A in \cite{ref_Wishart:Kanzieper})
\begin{equation}
\mathcal{J}_{N,M}(Q,\mu)=\int dT\ e^{\mathrm{i}\mathrm{Tr}(TQ)}[\det(T-\mu \mathbbm{1})]^{-M}=C_{M,N}(\det Q)^{M-N}e^{\mathrm{i}\mu\mathrm{Tr}Q}\ ,\label{eq_Techniques:Ingham}
\end{equation}
with $C_{M,N}=2^N \pi^{N(N+1)/2}\mathrm{i}^{NM}/\prod_{j=M-N+1}^M \Gamma(j)$, and the matrix $Q$ is hermitian and positive definite, while $T$ is just hermitian. Both are $N\times N$. We also require $\mathrm{Im}(\mu)>0$ to ensure convergence, and $M\geq N$.\\

To use this integral \eqref{eq_Techniques:Ingham}, we need to multiply back again all the degrees of freedom of the matrix $T$ by $2$, and pull out a factor $(-2)$ from the determinant, resulting in
\begin{equation}
\rho[W]=2^{-N(1+M)}\pi^{-N^2}\mathrm{i}^{-MN}\int dT\ e^{\mathrm{i}\mathrm{Tr}(TW)}\left[\det\left(T-\frac{\mathrm{i}}{2}\mathbbm{1}\right)\right]^{-M}\ ,
\end{equation}

which can be evaluated using \eqref{eq_Techniques:Ingham} as
\begin{align}
\nonumber\rho[W] &=\frac{2^{-N(1+M)}\pi^{-N^2}\mathrm{i}^{-MN}\times 2^N \pi^{N(N+1)/2}\mathrm{i}^{NM}}{\prod_{j=M-N+1}^M \Gamma(j)} (\det W)^{M-N} e^{-(1/2)\mathrm{Tr}W}=\\
&=\frac{1}{2^{NM}\pi^{\frac{N(N-1)}{2}}\prod_{j=M-N+1}^M \Gamma(j)} (\det W)^{M-N} e^{-(1/2)\mathrm{Tr}W}\ ,\label{eq_wishart:finaljpdfentrieswithconstants}
\end{align}
i.e. the jpdf of the entries of Wishart matrices for $\beta=2$, with the correct normalization\footnote{A reliable source for such normalizations is \cite{ref_vandermonde:edelmanthesis}.} (note that all the imaginary factors have correctly disappeared). Well done!

\section{To know more...}

\begin{enumerate}

\item The spectral densities of the Wishart-Laguerre ensemble for \emph{finite $N$} and $\beta=1,2,4$ have been given explicitly in \cite{ref_Wishart:Livan}, together with numerical checks.

\item The large-$N$ behavior of the spectral density and two-point function for the Wishart-Laguerre ensemble is determined by the asymptotics of Laguerre polynomials (in complete analogy with the Gaussian case). These are explicitly given in \cite{ref_Wishart:Garoni}.

\item Non-hermitian analogues of the Wishart-Laguerre ensemble can also be defined (see \cite{ref_Wishart:Akemann} for a nice review).

\item Readers interested in the diagrammatic approach to fluctuations in the Wishart ensemble should have a look at \cite{ref_resolvent:Jurk}.

\item For a nice review on usefulness of Wishart-Laguerre ensemble in physics, see \cite{ref_Wishart:MajumdarWishart}. For specific applications to QCD, see \cite{ref_Wishart:Verbaarschot,ref_Wishart:Splittorf}.

\end{enumerate}

\chapter{Meet Mar\v{c}enko and Pastur}\label{chap:Marcenko}

In this Chapter, we investigate the average spectral density for the Wishart-Laguerre ensemble.

\section{The Mar\v{c}enko-Pastur density}

The average density of eigenvalues has the following scaling form for $N,M\to\infty$ (such that $c=N/M\leq 1$ is kept fixed)
\begin{equation}
\rho(\llambda) \to \frac{1}{\beta N}\, \rho_{\rm MP}\left(\frac{\llambda}{\beta N}\right)
,\label{eq_Wishart:scalingdens1}
\end{equation}
where the Mar\v{c}enko-Pastur scaling function (the analogue of the semicircle $\rho_{\mathrm{SC}}(\llambda)$ in Eq. \eqref{eq_classified:semicirclefirstdefinition} for the Gaussian ensemble) is independent of $\beta$ and given by~\cite{ref_Wishart:WishartX}
\begin{equation}
\rho_{\rm MP}(y) = \frac{1}{2\pi y}\, \sqrt{(y-\zeta_-)(\zeta_+-y)}\ , \label{eq_Wishart:MP1}
\end{equation} 
for $x\in[\zeta_-,\zeta_+]$. The edge-points $\zeta_\pm$ are given by $\zeta_-=(1-c^{-1/2})^2$ and $\zeta_+=(1+c^{-1/2})^2$.\\

This scaling function $\rho_{\rm MP}(y)$ has a compact support on the positive semi-axis for $c<1$ (with two soft edges), but becomes singular at the origin if $c\to 1$ (and the origin becomes a hard edge). This means that Wishart matrices constructed from square matrices $H$ exhibit an accumulation of eigenvalues very close to zero.\\

It is worth stressing that the typical scale of an eigenvalue is $\sim\mathcal{O}(N)$ in the WL case, as opposed to the scale $\sim\mathcal{O}(\sqrt{N})$ for the Gaussian ensemble.\\

\section{Do it yourself: the resolvent method}

Let us now derive the Mar\v{c}enko-Pastur density using the resolvent (or Stieltjes transform) method. The partition function (normalization constant) for the Wishart-Laguerre ensemble reads (after a rescaling $\llambda_i\to\beta N\llambda_i$)
\begin{equation}
\mathcal{Z}_{N,\beta}^{(L)} \propto\int_0^\infty \prod_{j=1}^N d\llambda_j\ 
e^{-\frac{\beta N}{2}\sum_{i=1}^N
\llambda_i}\prod_{i=1}^N\llambda_i^{\alpha\beta/2}\prod_{j<k}|\llambda_j-\llambda_k|^\beta=\int_0^\infty\prod_{j=1}^N d\llambda_j\ 
e^{-\beta N\mathcal{V}[\bm\llambda]}\ ,\label{eq_Wishart:partitionfunctionWishart}
\end{equation}
with
\be 
\mathcal{V}[\bm\llambda]=\frac{1}{2}\sum_i \llambda_i+\left[\frac{2/\beta-1}{2N}-\frac{M}{2N}+\frac{1}{2}\right]\sum_i\ln\llambda_i-\frac{1}{2N} \sum_{i\neq j}\ln
|\llambda_i-\llambda_j|\ .
\ee \\

As in Chapter \ref{chap:resolvent}, the $\llambda_i$ are now of $\mathcal{O}(1)$ for large $N$. We can again perform the saddle point evaluation of the $N$-fold integral \eqref{eq_Wishart:partitionfunctionWishart}, but this time there is an additional subtlety which, if overlooked, leads straight to a nonsensical answer.\\

The subtlety is that the minimization of the exponent should be carried out within the set of \emph{positive} $\bm{\llambda}$. In other words, on top of the saddle-point equation, there is an inequality constraint to satisfy as well, $\llambda_i>0\qquad \forall i$.\\

One way to handle this constraint is to introduce a \emph{penalty function} $-\mu\sum_i \ln (\llambda_i)$ in the ``action" $\mathcal{V}[\bm\llambda]$, with a Lagrange multiplier $\mu$. Since $-\ln(t)\to\infty$ for $t\to 0$, it acts as if each particle felt an extra ``infinite wall"-type of repulsion while approaching the origin, and thus helps confining the eigenvalues on the positive semi axis. The extra wall is then ``gently" removed $(\mu\to 0)$ at the end of the calculation.\\

The saddle-point equations now read for any $i$ ( and for $N\gg 1$ and $N/M=c\leq 1$)

\begin{equation}
\frac{1}{2}+\left(\frac{1}{2}-\frac{1}{2c}-\mu\right)\frac{1}{\llambda_i}   =  \frac{1 }{N}\sum_{j\neq i}\frac{1}{\llambda_i-\llambda_j}\ . \label{eq_Wishart:SP}
\end{equation}

Multiplying \eqref{eq_Wishart:SP} by $\frac{1}{N(z-\llambda_i)}$ and summing over $i$, we get in analogy with Eq. \eqref{eq_resolvent:resolventdifferential}
\begin{equation}
\frac{1}{2}G_N(z)+\left(\frac{1}{2}-\frac{1}{2c}-\mu\right)\frac{1}{N}\sum_i \frac{1}{\llambda_i (z-\llambda_i)}=\frac{1}{2}G_N^2(z)+\frac{1}{2N}G_N^\prime(z)\ .
\end{equation}

The second term can be expressed in terms of $G_N(z)$ using
\begin{equation}
\frac{1}{N}\sum_i \frac{1}{\llambda_i (z-\llambda_i)}=\frac{1}{z N}\left(\sum_i \frac{1}{\llambda_i}+\frac{1}{z-\llambda_i}\right)=\frac{K+G_N(z)}{z}\ ,
\end{equation}
and taking the average $G_\infty^{(av)}(z)=\langle G_N(z)\rangle$ in the limit $N\to\infty$, we obtain
\begin{equation}
\frac{1}{2}G_\infty^{(av)}(z)+\left(\frac{1}{2}-\frac{1}{2c}-\mu\right)\frac{K+G_\infty^{(av)}(z)}{z}=\frac{1}{2}G_\infty^{(av)2}(z)\ .\label{eq_Wishart:quadraticG}
\end{equation}

Here $K$ is a constant that we assume finite (by derivation, we have $K=\int d\llambda \rho(\llambda)/\llambda$).\\

Note that, had we \emph{not} included the penalty function parametrized by $\mu$ from the beginning, we would have landed for $c=1$ on the equation $\frac{1}{2}G_\infty^{(av)}(z)=\frac{1}{2}G_\infty^{(av)2}(z)$, from which no sensible spectral density could be extracted! This is because the Wishart eigenvalues \emph{cannot} equilibrate on the \emph{entire} real line under a potential $V(\llambda)=\llambda$ (which is not confining for $\llambda\to -\infty$).\\

It is convenient to set $\gamma=(1-c)/c>0$. Solving now the quadratic equation \eqref{eq_Wishart:quadraticG} for $\mu\to 0$, we get 
\begin{equation}
G_\infty^{(av)}(z) = \frac{1}{2} \left(\pm\frac{\sqrt{\gamma ^2-4 \gamma  K z+z^2-2 \gamma 
   z}}{z}-\frac{\gamma }{z}+1\right)\ .\label{eq_Wishart:Gplusminus}
\end{equation}
Setting now $z=\llambda-\mathrm{i}\epsilon$, multiplying up and down by $\llambda+\mathrm{i}\epsilon$ and using the real and imaginary part of the square root $p$ and $q$ as in \eqref{eq_resolvent:squareroot}, we obtain
\begin{equation}
\frac{1}{\pi}\mathrm{Im} G_\infty^{(av)}(\llambda-\mathrm{i}\epsilon) =\frac{-\epsilon \llambda\pm q\llambda}{2\pi (\llambda^2+\epsilon^2)}\stackrel{\epsilon\to 0^+}{\longrightarrow}\frac{\sqrt{(\llambda-\llambda_-(\gamma,K))(\llambda_+(\gamma,K)-\llambda)}}{2\pi\llambda}\ ,\label{eq_Wishart:rholambdafromGG}
\end{equation}
where it is understood that the $(\pm)$ sign in \eqref{eq_Wishart:Gplusminus} is to be chosen differently in different $\llambda$-intervals, in analogy with the Gaussian case. Of course, the right hand side of \eqref{eq_Wishart:rholambdafromGG} is only valid for $\llambda$ such that the square root exists. The constants $\llambda_\pm(\gamma,K)=\gamma  \left(-2 \sqrt{K^2+K}+2 K+1\right)$.\\

We now have to fix the constant $K$ by requiring normalization of $\rho(\llambda)$. Using the integral (for $b>a$)
\begin{equation}
\int_a^b dx\frac{\sqrt{(x-a)(b-x)}}{2\pi x}=\frac{1}{4} \left(-2 \sqrt{a b}+a+b\right)\ ,
\end{equation}
all we have to do is to assign $a\leftarrow \llambda_-(\gamma,K)$ and $b\leftarrow \llambda_+(\gamma,K)$, and to solve $\frac{1}{4} \left(-2 \sqrt{a b}+a+b\right)=1$ for $K$. This gives $K=1/\gamma$.\\

And for this value of $K$, the edge points become $\llambda_\pm(\gamma,1/\gamma)\to(1\pm 1/\sqrt{c})^2$, which means that we have recovered the MP law \eqref{eq_Wishart:MP1} using the resolvent method. Congratulations! \\

\begin{figure}[t]
\centering
\includegraphics[width=.75\columnwidth]{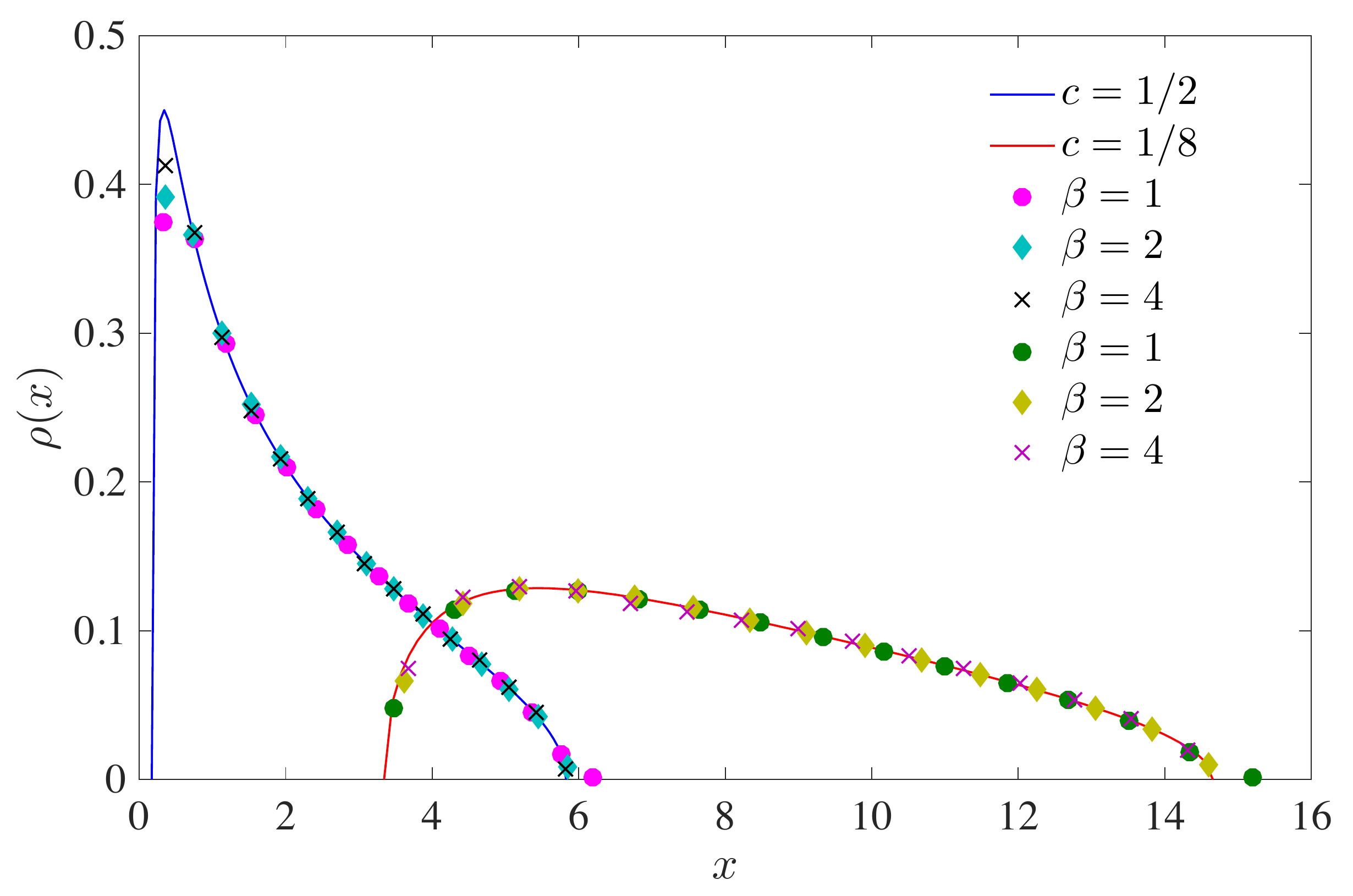}
\caption{Comparison between the Mar\v{c}enko-Pastur density for two different values of the rectangularity ratio $c$ and the corresponding histograms obtained from the numerical diagonalization of random Wishart matrices (for all possible values of $\beta$). All histograms are obtained from $5000$ Wishart matrices of size $N = 100$.}
\label{fig:MP_check}
\end{figure}

You can now fully enjoy Fig. \ref{fig:MP_check}, where we show a comparison between the Mar\v{c}enko-Pastur density and the histograms obtained by numerical diagonalization of WL random matrices for different $\beta$s.

\begin{center}
\fbox{\begin{minipage}{33em}
{\it Question.} Wait a second...In the derivation, we said that we had to assume $K$ finite and equal to $K=\int d\llambda \rho(\llambda)/\llambda$ (because the constant $K$ arises as the average $\langle\frac{1}{N}\sum_ i \frac{1}{\llambda_i}\rangle$). Shouldn't we check that this is consistent with the final result? \\ \\
$\blacktriangleright$ Yes, we should! The integral $\int d\llambda \rho(\llambda)/\llambda$ amounts to computing the following
\begin{equation}
\int_a^b d\llambda\frac{\sqrt{(\llambda-a)(b-\llambda)}}{2\pi\llambda^2}=\frac{-2 \sqrt{a b}+a+b}{4 \sqrt{a b}}\ ,
\end{equation}
and setting $a\leftarrow (1- 1/\sqrt{c})^2$ and $b\leftarrow (1+ 1/\sqrt{c})^2$. This gives $c/(1-c)$, which is precisely equal to $K=1/\gamma$. Bingo!
\end{minipage}}
\end{center}

\clearpage

\begin{center}
\fbox{\begin{minipage}{33em}
{\it Question.} What if I wanted to use the Coulomb gas technique to derive the Mar\v{c}enko-Pastur law? \\ \\
$\blacktriangleright$ The partition function (normalization constant) for the WL ensemble, after the rescaling $\llambda_i\to\llambda_i\beta$, reads 
\begin{equation}
\mathcal{Z}^{(L)}_{N,\beta}=C_{N,\beta}^{(L)}\int_{(0,\infty)^N} \prod_{j=1}^N d\llambda_j\ 
e^{-\beta\mathcal{V}[\bm\llambda]}\ ,\label{eq_Marcenko:partitionWishart}
\end{equation}
where the \emph{energy} is given this time by
$
\mathcal{V}[\bm\llambda]=\frac{1}{2}\sum_i \llambda_i-\frac{\alpha}{2}\sum_i \ln \llambda_i-\frac{1}{2}\sum_{i\neq j} \ln
|\llambda_i-\llambda_j|
$.\\

In the WL case, the gas is in equilibrium under the competing effect of a linear$+$logarithmic confining potential, and the 2D electrostatic repulsion.\\

Following the same procedure as in Chapter \ref{chap:semicircle} (but with the rescaling $n(\llambda)\to (1/N)n(\llambda/N)$), we obtain for the energy functional $\mathcal{V}[n(\llambda)] =N^2\mathcal{\hat{V}}$, with
\begin{equation}
\mathcal{\hat V}[n(\llambda)]=\int d\llambda\ v(\llambda)\ n(\llambda)-\frac{1}{2}\iint d\llambda
d\llambda^\prime n(\llambda)n(\llambda^\prime)\ln |\llambda-\llambda^\prime|\ ,\label{eq_Wishart:energycontinuumscaledsameN}
\end{equation}\\
with $v(\llambda)=x/2-\frac{1}{2}(1/c-1)\ln x$.\\

The singular integral equation for the equilibrium density readily follows
\begin{equation}
\mathrm{Pr}\int d\llambda^\prime\frac{n^\star(\llambda^\prime)}{\llambda-\llambda^\prime}=\frac{1}{2}-\frac{1}{2}\left(\frac{1}{c}-1\right)\frac{1}{x}\ .
\end{equation}\\

Try to apply Tricomi's formula \eqref{eq_semicirclesequel:solutiontricomi} - assuming a single-support solution on $[a,b]$ - and then determine $a,b$ in such a way that the free energy is minimized. You will discover that $n^\star(\llambda)=\rho_{\mathrm{MP}}(\llambda)$ as it should.
\end{minipage}}
\end{center}

\section{Correlations in the real world and a quick example: financial correlations}
A huge number of scientific disciplines, ranging from Physics to Economics, often need to deal with statistical systems described by a large number of degrees of freedom. Thus, understanding and describing the collective behavior of a large numbers of random variables is one of the most fundamental issues in multivariate Statistics. More often than not, the problem can be addressed in terms of correlations.\\

Suppose we are interested in understanding the correlation structure of a system described in terms of $N$ random variables $\{x_1, \ldots, x_N\}$, drawn from a - potentially unknown, but not changing in time - jpdf $p(\bm{x})$. In order to do so, one of the most obvious operations to perform is to collect, if possible, as many ``experimental observations'' of such variables. Such observations can then be used to compute empirical time averages of quantities expressed in terms of the random variables. So, let us assume we have collected $M$ observations - say, equally spaced in time - for each variable. Quite straightforwardly, one can collect all these numbers in a $N \times M$ matrix $X$ whose entries are $x_i^t$ ($i = 1, \ldots, N$, $t = 1, \ldots, M$).\\ 

Assuming all variables $x_i^t$ are adjusted in order that their sample mean\footnote{The sample mean is $\bar{x}_i=(1/M)\sum_{t=1}^M x_i^t$, not to be confused with the true mean $\langle x_i\rangle_{p(\bm x)}$, which is a property of the jpdf $p(\bm x)$.} is zero and their sample variance is $1$, then the quantity
\be c_{ij} = \frac{1}{M} \sum_{t=1}^M x_i^t x_j^t \label{eq:pearson} \ee
yields the well known Pearson estimator for the correlation between variables $x_i$ and $x_j$. This is an estimator of the \emph{true} (or \emph{population}) \emph{correlation}\footnote{The true correlation $\tilde{c}_{ij}$ is a property of the jpdf $p(\bm{x})$ of the random variables $\{x_1,\ldots,x_N\}$.} $\tilde{c}_{ij}$, which would be measured exactly for $M \rightarrow \infty$, i.e. as more and more observations are added to the data. However, real life practice always entails working with finite-sized datasets (i.e. with finite $M$), which introduces some degree of measurement error.\\

The estimators for each pair of variables in the system can be collected into a single $N \times N$ matrix ${C}= {X} {X}^T / M$, known as the \emph{sample correlation matrix} of the data in ${X}$, whose entries are given by Eq. \eqref{eq:pearson}. These amount to $N(N-1)/2$ real numbers (diagonal entries are equal to one), which for a large system represent a  whopping amount of information to process. So, what should we make of all this? Well, a reasonable first step could be to compare the empirical correlation matrix of the system we are interested in with the prediction of a suitably defined null hypothesis. In the first instance, we could for example look for a null model describing uncorrelated Gaussian random data and see how our empirical data differ from it.\\ 

By any chance, do we know a random matrix ensemble from which we can draw this kind of random correlation matrices? Well, of course we do! It is precisely the Wishart-Laguerre ensemble. As we discussed, the density of eigenvalues is well known for this ensemble, and it is given by the Mar\v{c}enko-Pastur law \eqref{eq_Wishart:MP1}. This means that a zero-th order assessment of the statistical significance of the correlations in a large system can be obtained from the comparison of the empirical eigenvalue spectrum of its correlation matrix with the Mar\v{c}enko-Pastur law for a system with the same rectangularity ratio $N/M$.\\

A prime example of the procedure outlined above is the analysis of financial correlations. Suppose you want to invest your money in $N$ stocks by forming an investment portfolio. As the old saying goes, ``don't put your eggs in one basket'', which in financial terms translates into ``don't invest all your money in a portfolio of highly correlated stocks'' - not the most effective punchline, admittedly. Hence, distinguishing signal from noise within financial correlation matrices is of paramount importance to build a well diversified portfolio, where the possible losses due to the adverse movement of a group of stocks can be offset by other groups of stocks.\\   

When an empirical financial correlation matrix is diagonalized, one usually finds that several eigenvalues are much larger than the expected upper bound of the Mar\v{c}enko-Pastur law. The information contained in the associated eigenvectors typically shows that these are due to the co-movements of groups of highly correlated stocks belonging to well defined market sectors (e.g. pharmaceutical, financial, etc). This kind of random matrix approach to financial correlations was initiated in \cite{ref_RMT:Bouchaud,ref_RMT:Stanley} and since then a considerable number of papers has been devoted to it (see \cite{ref_RMT:Bouchaud_2} for a recent account).

\chapter{Replicas...}\label{chap:replicas}

In this Chapter, we add one more powerful tool to our arsenal. The \emph{Edwards-Jones} formula, in conjunction with the celebrated \emph{replica trick}.

\section{Meet Edwards and Jones}

The Edwards-Jones formula \cite{ref_replicas:edwardsjones} allows to write down a \emph{formal}
expression for the average spectral density $\rho(\llambda)$ of a completely generic ensemble of real symmetric random
matrices $H$, taking as a starting point just the jpdf of the \emph{entries} in the upper triangle, $\rho[H]$.\\

The formula reads
\begin{equation}
\boxed{\rho(\llambda)=\frac{-2}{\pi N} \lim_{\varepsilon\to 0^+}\mathrm{Im}
\frac{\partial}{\partial\llambda}\Big\langle \mathrm{Log}\ Z(\llambda)\Big\rangle}\ ,\label{eq_replicas:rhoEJ}
\end{equation}
where
\be
Z(\llambda)=\int_{\mathbb{R}^N}d\bm y \exp\left[-\frac{\mathrm{i}}{2}\bm{y}^T \left(\llambda_\epsilon \mathbbm{1}-H\right)\bm{y}\right]\ ,\label{eq_replicas:Zllambda1v1}
\ee
where $\llambda_\epsilon=\llambda-\mathrm{i}\epsilon$.\\

The average $\langle\cdot\rangle$ is taken with respect to $\rho[H]$, i.e. $\langle\cdot\rangle=\int dH_{11}\cdots dH_{NN}\rho[H](\cdot)$.\\

This formula is remarkable: it allows to compute the spectral density - the marginal of the jpdf of the eigenvalues - \emph{without} knowing the jpdf of eigenvalues! Only the information about the \emph{entries} is required as input.\\

While the formula \eqref{eq_replicas:rhoEJ} is in principle valid for any finite $N$, in practice the calculations can be carried out until the end only in the limit $N\to\infty$, where several simplifications take place.

\section{The proof}

The proof is not complicated - even though there are several subtleties. Recall from Chapter \ref{chap:valuetheeigenvalue} how the average spectral density is defined
$
\rho(\llambda)=\Big\langle\frac{1}{N}\sum_{i=1}^N
\delta(\llambda-\llambda_i)\Big\rangle\ .
$

Recall also the Sokhotski-Plemelj identity: as $\epsilon\to 0^{+}$,
\begin{equation}
\frac{1}{x\pm {\mathrm{i}}\varepsilon}\to
{\mathrm{Pr}}\left(\frac{1}{x}\right)\mp{\mathrm{i}}\pi \delta(x)\ .\label{eq_replicas:sokh}
\end{equation}

This equation provides an interesting identity for the delta function, which we already used in Chapter \ref{chap:resolvent}. We can therefore write
\begin{equation}
\rho(\llambda)=\frac{1}{\pi N} \lim_{\varepsilon\to 0^+}\mathrm{Im} \Big\langle
\sum_{i=1}^N \frac{1}{\llambda-\mathrm{i}\varepsilon
-\llambda_i}\Big\rangle=\frac{-1}{\pi N} \lim_{\varepsilon\to 0^+}\mathrm{Im} \Big\langle
\sum_{i=1}^N \frac{1}{\llambda_i+\mathrm{i}\varepsilon
-\llambda}\Big\rangle\ ,\label{eq_replicas:rhofromGreen}
\end{equation}
where $\mathrm{Im}$ stands for the imaginary part, and we changed a sign for later convenience.\\

Next, we write the denominator in the sum as the derivative of a logarithm. But the denominator is a \emph{complex} number: and the logarithms of complex numbers are nasty beasts\footnote{For example, $\mathrm{Log}(z_1 z_2)$ may not be equal to $\mathrm{Log}(z_1)+\mathrm{Log}(z_2)$!}. Anyway, we can choose the principal branch of the logarithm - and denote it by $\mathrm{Log}$ - to write
\begin{equation}
\rho(\llambda)=\frac{1}{\pi N} \lim_{\varepsilon\to 0^+}\mathrm{Im}
\frac{\partial}{\partial\llambda}\Big\langle \sum_{i=1}^N \mathrm{Log}
(\llambda_i+\mathrm{i}\varepsilon -\llambda)\Big\rangle\ .\label{eq_replicas:rhofromGreen2}
\end{equation}

Next, we use the identity
\begin{equation}
Z(\llambda)=(2\pi)^{N/2}\exp\left[-\frac{1}{2}\sum_{i=1}^N\mathrm{Log}(\llambda_i+\mathrm{i}\epsilon-\llambda)+\mathrm{i}\frac{N\pi}{4}\right]\ ,\label{eq_replicas:identityFresnel}
\end{equation}
where $Z(\llambda)$ is given by the multiple integral in \eqref{eq_replicas:Zllambda1v1}. You can check this identity with the code [$\spadesuit$ \verb"Zmultiple.m"] \\

Now, compare the last two equations. Clearly, the final formula would be easily established if we could replace $\sum_{i=1}^N\mathrm{Log}(\llambda_i+\mathrm{i}\epsilon-\llambda)$ in \eqref{eq_replicas:rhofromGreen2} with something related to $Z(\llambda)$, using \eqref{eq_replicas:identityFresnel}.\\

To extract $\sum_{i=1}^N\mathrm{Log}(\llambda_i+\mathrm{i}\epsilon-\llambda)$ from \eqref{eq_replicas:identityFresnel}, we should take the logarithm on both sides. There is a small glitch, though, due to another mind-boggling feature of complex logarithms. Namely, $\mathrm{Log}(\exp(z))$ may not just be equal to $z$, for $z\in\mathbb{C}$!\footnote{For instance, if $z=0.2-4.4\mathrm{i}$, then $\mathrm{Log}(\exp(z))=0.2+1.88319\mathrm{i}$.}\\

However, we can still write
\be
\sum_{i=1}^N\mathrm{Log}(\llambda_i+\mathrm{i}\epsilon-\llambda)=-2 \ \mathrm{Log}\ Z(\llambda) +\text{terms that are killed by }\frac{\partial}{\partial\llambda}\ .\label{eq_replicas:sumwiped}
\ee

Inserting \eqref{eq_replicas:sumwiped} into \eqref{eq_replicas:rhofromGreen2}, we establish the final formula \eqref{eq_replicas:rhoEJ}.

\section{Averaging the logarithm}

The Edwards-Jones formula \eqref{eq_replicas:rhoEJ} thus requires computing $\Big\langle \mathrm{Log}\ Z(\llambda)\Big\rangle$, where the average is taken over several realizations of the matrix $H$.\\

This means that we should compute
\begin{equation}
\Big\langle \mathrm{Log}\ Z(\llambda)\Big\rangle=\int dH_{11}\cdots dH_{NN}\rho[H]\mathrm{Log}\left[
\int_{\mathbb{R}^N}d\bm y \exp\left[-\frac{\mathrm{i}}{2}\bm{y}^T \left(\llambda_\epsilon \mathbbm{1}-H\right)\bm{y}\right]\right]\ ,\label{eq_Replicas:logintheway}
\end{equation}
which is very annoying: the logarithm is right in the way!\\

We would really need to exchange the order of integrals to perform the average over $H$ \emph{before} the average over $\bm y$ - otherwise we would be running the Edwards-Jones formula backwards and gain nothing!\\

There are two strategies to circumvent this obstacle, each with their own subtleties. To know more about the replica method and its applications to spin glass theory see \cite{ref_replicas:Castellani,ref_replicas:Zamponi}.

\section{Quenched \emph{vs.} Annealed}

Calling the quantity in \eqref{eq_replicas:Zllambda1v1} $Z(\llambda)$ is intentional: we wish to interpret it as the \emph{partition function} of an associated stat-mech model in the canonical ensemble. The logarithm of $Z$ will then be the \emph{free energy} of this model.\\

Looking again at the multiple integral defining $Z(\llambda)$, $Z(\llambda)=\int_{\mathbb{R}^N}d\bm y \exp\left[-\frac{\mathrm{i}}{2}\bm{y}^T \left(\llambda_\epsilon \mathbbm{1}-H\right)\bm{y}\right]$, we see that it encodes two different 'levels' of randomness: i) the random matrix $H$ - the so called \emph{disorder} - and ii) the dynamical variables $\bm y$, which morally\footnote{``Morally", since the ``Hamiltonian" $\mathcal{H}$ is actually \emph{complex}, so $P(y_1,\ldots,y_N)$ is not a proper distribution.} follow a Gibbs-Boltzmann distribution $P(y_1,\ldots,y_N)=\frac{1}{Z(\llambda)}\exp\left(-\mathcal{H}(\bm y; H,\llambda)\right)\ .$

Computing now $\Big\langle \mathrm{Log}\ Z(\llambda)\Big\rangle$ - as we should - assumes that the two levels of randomness are unfolding on different timescales: \emph{first}, the dynamical variables $\bm y$ need to equilibrate according to the Gibbs-Boltzmann distribution \emph{for a fixed instance of the random matrix} $H$ - and only afterwards the free energy is averaged over the disorder (different realizations of $H$).\\

For these reasons, the disorder is called \emph{quenched}\footnote{Quenched \emph{adj.} made less severe or intense; subdued or overcome; allayed; squelched.}: it is there, but it acts slowly. It only kicks in after the $\bm y$'s have thermalized.\\

Computing a quenched disorder average is difficult, but can be attempted - in the limit $N\to\infty$ - using the so called \emph{replica trick}, which gets rid of the logarithm inside the integral in \eqref{eq_Replicas:logintheway} and allows the integrations over $H$ and $\bm y$ to be interchanged. More on this later.\\

A second strategy - which simplifies the calculations considerably - is to cheat a bit and treat the disorder as \emph{annealed} instead.\\

This means that the associated stat-mech model is described in terms of the \emph{joint} set of dynamical variables $\{\bm y, H\}$, leading to a partition function $Z^{(ann)}(\llambda)=\int dH d\bm y (\cdots)$.\\

The dynamical variables $\bm y$ are no longer integrated over at fixed value of the disorder $H$, but rather $H$ and $\bm y$ fluctuate and thermalize \emph{together}. A questionable but widespread way to describe in words an annealed average is: instead of computing the quenched average $\langle\mathrm{Log}\ Z(\llambda)\rangle$ - as we should - move the average inside the logarithm \footnote{For the annealed average, we should more properly write $\mathrm{Log}\ Z^{(ann)}(\llambda)$ - with no further average over $H$.}, $\mathrm{Log}\langle Z(\llambda)\rangle$.\\

Clearly, this slick maneuver forces the logarithm out of the integrals, and allows for a much quicker - even though not entirely justifiable - computation.\\

In the following section, we present the annealed calculation to obtain the semicircle law for the GOE\footnote{This is only for training purposes. There is no need to use Edwards-Jones when the jpdf of eigenvalues is known!}.

\chapter{Replicas for GOE}\label{chap:replicas2}

In this Chapter, we apply the Edwards-Jones formula to compute the average spectral density of the GOE ensemble.

\section{Wigner's semicircle for GOE: annealed calculation}
The jpdf of entries in the upper triangle of a GOE is
\begin{equation}
\rho[H]=\prod_{i=1}^N \left[\exp\left(-NH_{ii}^2/2\right)/\sqrt{2\pi/N}\right]\prod_{i<j}\left[\exp\left(-NH_{ij}^2\right)/\sqrt{\pi/N}\right]\ ,\label{eq_replicas:1:jpdfentriesgaussian}
\end{equation}
where we have already rescaled the unit variance by a factor $1/N$. This has the net effect of rescaling the eigenvalues by $1/\sqrt{N}$ (why?), so the corresponding
spectral density will have edges between $-\sqrt{2}$ and $\sqrt{2}$ - not growing with $N$.\\

For the annealed calculation, we need to compute 
\be
Z^{(ann)}(\llambda)=\int_{\mathbb{R}^N}d\bm y \int\prod_{i\leq j}dH_{ij}\rho[H] \exp\left[-\frac{\mathrm{i}}{2}\bm{y}^T \left(\llambda_\epsilon \mathbbm{1}-H\right)\bm{y}\right]\ .\label{eq_replicas:Zllambda}
\ee

Separating diagonal and off-diagonal elements, and using the notation $\langle(\cdot)\rangle=\int\prod_{i\leq j}dH_{ij}\rho[H](\cdot)$, we can write
\be
Z^{(ann)}(\llambda)\propto\int_{\mathbb{R}^N}d\bm y \exp\left[-\frac{\mathrm{i}}{2}\llambda_\epsilon\sum_{i=1}^N y_i^2\right]
\Big\langle \exp\left[\frac{\mathrm{i}}{2}\sum_{i=1}^N H_{ii}y_i^2\right]\Big\rangle
\Big\langle \exp\left[\mathrm{i}\sum_{i<j}^N H_{ij}y_i y_j\right]\Big\rangle\ ,
\ee
where we neglect some overall constant terms.\\

Expanding $e^z\approx 1+z+z^2/2+\ldots$ and using the fact that the entries of $H$ are independent with $\langle H_{ij}\rangle=0$ and $\langle H_{ij}^2\rangle=1/(N(2-\delta_{ij}))$,
we can write 
\begin{align}
&\Big\langle \exp\left[\frac{\mathrm{i}}{2}\sum_{i=1}^N H_{ii}y_i^2\right]\Big\rangle =\prod_{i=1}^N \Big\langle 1+\frac{\mathrm{i}}{2}H_{ii}y_i^2-\frac{1}{8}H_{ii}^2 y_i^4+\ldots\Big\rangle=\prod_{i=1}^N \left(1-\frac{1}{8N}y_i^4+\ldots\right)\ ,\\
&\Big\langle \exp\left[\mathrm{i}\sum_{i<j}^N H_{ij}y_i y_j\right]\Big\rangle =\prod_{i<j}\Big\langle 1+\mathrm{i}H_{ij}y_i y_j-\frac{1}{2}H_{ij}^2 y_i^2 y_j^2+\ldots\Big\rangle=\prod_{i<j} \left(1-\frac{1}{4N}y_i^2 y_j^2+\ldots\right)\ .
\end{align}
Re-exponentiating, we can write
\be
\Big\langle \exp\left[\frac{\mathrm{i}}{2}\sum_{i=1}^N H_{ii}y_i^2\right]\Big\rangle\Big\langle \exp\left[\mathrm{i}\sum_{i<j}^N H_{ij}y_i y_j\right]\Big\rangle \simeq
\exp\left[-\frac{1}{8N}\sum_{i, j=1}^N y_i^2 y_j^2\right]=\exp\left[-\frac{1}{8N}\left(\sum_{i=1}^N y_i^2\right)^2\right]\ .
\ee

Introducing a Gaussian identity 
\be
\int_{-\infty}^\infty dq\ \exp\left[-\alpha q^2+\mathrm{i}\gamma q\right]\propto \exp\left(-\gamma^2/4\alpha\right)\label{eq_replicas:GaussianIdentity}
\ee
with $\gamma=\sum_{i=1}^N y_i^2$ and $\alpha=2N$ yields
\begin{align}
\nonumber Z^{(ann)}(\llambda) &\propto \int_{-\infty}^\infty dq\ e^{-2N q^2}\int_{\mathbb{R}^N}d\bm y  \exp\left[-\frac{\mathrm{i}}{2}\llambda_\epsilon\sum_{i=1}^N y_i^2+\mathrm{i}q\sum_{i=1}^N y_i^2\right]\\
&=\int_{-\infty}^\infty dq\ e^{-2N q^2}\left[\int_{\mathbb{R}}dy\ \exp\left[-\frac{1}{2}\epsilon y^2 -\mathrm{i}\left(\frac{1}{2}\llambda -q\right)y^2\right]\right]^N\ ,
\end{align}
where the $y$-integral is convergent as $\epsilon>0$. Writing $X^N=\exp\left[N\mathrm{Log} X\right]$, we have
\begin{equation}
Z^{(ann)}(\llambda) \propto\int_{-\infty}^\infty dq\ \exp\left[-N \underbrace{\left(2q^2-\frac{1}{2}\mathrm{Log}\left(\frac{2\pi}{\epsilon+\mathrm{i}(\llambda-2 q)}\right)\right)}_{\varphi_\llambda(q)}\right]\ .
\end{equation}

This integral lends itself to a nice Laplace's approximation, from which
\be
Z^{(ann)}(\llambda)\approx \exp(-N \varphi_\llambda (q^\star))\ .
\ee

The stationary point $q^\star$ is computed as
\begin{equation}
\varphi_\llambda'(q^\star)=0\Rightarrow 4 q^\star+\frac{1}{2 q^\star-\llambda_\epsilon}=0\Rightarrow q^\star = \frac{1}{4}\left(\llambda_\epsilon\pm\sqrt{\llambda_\epsilon^2-2}\right)\ ,\label{eq_Replicas:qstardetermined}
\end{equation}
where again $\llambda_\epsilon=\llambda-\mathrm{i}\epsilon$.\\

Applying now the Edwards-Jones formula - in the annealed version and for $N\to\infty$
\begin{equation}
\rho(\llambda)=\frac{-2}{\pi N} \lim_{\varepsilon\to 0^+}\mathrm{Im}
\frac{\partial}{\partial\llambda} \mathrm{Log}\ Z^{(ann)}(\llambda)
\approx \frac{-2}{\pi N} \lim_{\varepsilon\to 0^+}\mathrm{Im}
\frac{\partial}{\partial\llambda}\left[-N \varphi_\llambda (q^\star)\right] \ ,\label{eq_replicas:rhoEJannealed}
\end{equation}

Using now the chain rule
\be
\frac{\partial}{\partial\llambda}\varphi_\llambda (q^\star)=\underbrace{q^{\star '}\frac{\partial}{\partial q} \varphi_\llambda (q)\Big|_{q=q^\star}}_{=0}+\partial_\llambda \varphi_\llambda (q)\Big|_{q=q^\star}=\frac{1}{2\llambda_\epsilon -4 q^\star}\ ,
\ee
and substituting $q^\star$ with \eqref{eq_Replicas:qstardetermined}, we obtain
\begin{equation}
\rho(\llambda)=\frac{2}{\pi} \lim_{\varepsilon\to 0^+}\mathrm{Im}\frac{1}{\llambda_\epsilon\mp\sqrt{\llambda_\epsilon^2-2}}=\frac{1}{\pi} \lim_{\varepsilon\to 0^+}\mathrm{Im} \left[\llambda_\epsilon \pm \sqrt{\llambda_\epsilon^2-2}\right]\ ,
\end{equation}
after rationalizing the denominator.\\

Next, we use again the following short lemma. If $\sqrt{a+\mathrm{i}b}=p+q\mathrm{i}$, then
\begin{equation}
p =\frac{1}{\sqrt{2}}\sqrt{\sqrt{a^2+b^2}+a}\ ,\qquad
q =\frac{\mathrm{sign}\ b}{\sqrt{2}}\sqrt{\sqrt{a^2+b^2}-a}\ .\label{eq_replicas:shortlemma}
\end{equation}
Using this with $a=\llambda^2-\epsilon^2-2$ and $b=-2\epsilon\llambda$, and choosing the sign in order to get a physical solution, we obtain
\begin{equation}
\rho(\llambda) =\frac{1}{\pi}\frac{1}{\sqrt{2}}\sqrt{|\llambda^2-2|-\llambda^2+2}\ ,
\end{equation}
which is indeed zero outside $[-\sqrt{2},\sqrt{2}]$ and equal to Wigner's semicircle $\rho(\llambda) =\frac{1}{\pi}\sqrt{2-\llambda^2}$ inside, as it should.\\

In the next section, we embark in the tougher task of using Edwards-Jones in the correct (quenched) version (without shortcuts). This will require the use of the celebrated \emph{replica trick}.

\section{Wigner's semicircle: quenched calculation}
We use now Edwards-Jones in the full-fledged form
\begin{equation}
\rho(\llambda)=\frac{-2}{\pi N}\lim_{\epsilon\to 0^+}\mathrm{Im}\frac{\partial}{\partial\llambda}\Big\langle \mathrm{Log} \int_{\mathbb{R}^N}d\bm y \exp\left[-\frac{\mathrm{i}}{2}\bm{y}^T \left(\llambda_\epsilon \mathbbm{1}-H\right)\bm{y}\right]\Big\rangle\ ,\label{EJinitial}\end{equation}

where the average $\langle\cdot\rangle$ is taken again with respect to $\rho[H]$, i.e. $\langle\cdot\rangle=\int dH_{11}\cdots dH_{NN}\rho[H](\cdot)$
and $\llambda_\epsilon=\llambda-\mathrm{i}\epsilon$.\\

Recall that we \emph{cannot} perform the $\bm y$-integral \emph{before} taking the average over $H$, otherwise we would be running the Edwards-Jones formula backwards! On the other hand, we cannot exchange the two integrations as they stand, due to the logarithm standing right in the middle. How to proceed then?\\

Using the \emph{replica identity} in the form
\begin{equation}
\langle\mathrm{Log}\ Z(\llambda)\rangle=\lim_{n\to 0}\frac{1}{n}\mathrm{Log}\langle Z(\llambda)^n\rangle\ ,
\end{equation}
we replicate the $\bm y$-integral $n$ (integer) times, and we blindly hope that the analytical continuation to $n$ in the vicinity of zero makes sense. The formalism and notation we shall use in the following are similar to those introduced first in \cite{ref_replicas2:rodgers}. \\

Using again
\begin{equation}
\rho[H]=\prod_{i=1}^N \left[\exp\left(-NH_{ii}^2/2\right)/\sqrt{2\pi/N}\right]\prod_{i<j}\left[\exp\left(-NH_{ij}^2\right)/\sqrt{\pi/N}\right]\ ,\label{eq_replicas:1:jpdfentriesgaussian2}
\end{equation}
we want to compute the replicated partition function
\begin{align}
\nonumber \langle Z(\llambda)^n\rangle &=\int \left(\prod_{i\leq j}dH_{ij}\right)\prod_{i=1}^N \frac{e^{-N H_{ii}^2/2}}{\sqrt{2\pi/N}}\prod_{i<j} \frac{e^{-N H_{ij}^2}}
{\sqrt{\pi/N}}\times\\
&\times \int_{\mathbb{R}^{Nn}}\left(\prod_{a=1}^n d\bm y_a\right) \exp\left[-\frac{\mathrm{i}}{2}\sum_{i,j=1}^N\sum_{a=1}^n y_{ia}\left(\llambda_\epsilon\delta_{ij}-H_{ij}\right) y_{ja}\right]\ .
\end{align}

Now that the innermost integral has been ``replicated" $n$-times, we can exchange the order of integrations to get
\begin{align}
\nonumber\langle Z(\llambda)^n\rangle &=\int_{\mathbb{R}^{Nn}}\left(\prod_{a=1}^n d\bm y_a\right) e^{-\mathrm{i}\frac{\llambda_\epsilon}{2}\sum_{i=1}^N\sum_{a=1}^n y_{ia}^2}\int \left(\prod_{i=1}^N\frac{dH_{ii}}{\sqrt{2\pi/N}}\right) e^{-N \sum_{i=1}^N H_{ii}^2/2+\frac{\mathrm{i}}{2}\sum_{i=1}^N H_{ii}\sum_{a=1}^n y_{ia}^2}\times \\
&\times \int \left(\prod_{i< j}\frac{dH_{ij}}{\sqrt{\pi/N}}\right) e^{-N \sum_{i<j}^N H_{ij}^2+\mathrm{i}\sum_{i<j}\sum_{a=1}^n y_{ia}H_{ij} y_{ja}}\ .
\end{align}
Neglecting constants, we can perform the two multiple Gaussian integrals involving $H$ using \eqref{eq_replicas:GaussianIdentity} repeatedly, with $\alpha=N/2$ (or $N$) and $\gamma=(1/2)\sum_{a=1}^n y_{ia}^2$ (or $\gamma=\sum_{a=1}^n y_{ia}y_{ja}$)
to get
\begin{equation}
\langle Z(\llambda)^n\rangle =\int_{\mathbb{R}^{Nn}}\left(\prod_{a=1}^n d\bm y_a\right) \exp\left[-\mathrm{i}\frac{\llambda_\epsilon}{2}\sum_{i=1}^N\sum_{a=1}^n y_{ia}^2 -\frac{1}{8N}\sum_{i=1}^N\left(\sum_a y_{ia}^2\right)^2-\frac{1}{4N}\sum_{i<j}\left(\sum_{a=1}^n y_{ia}y_{ja}\right)^2\right]\ ,
\end{equation}
which can be more compactly rewritten as
\begin{equation}
\langle Z(\llambda)^n\rangle =\int_{\mathbb{R}^{Nn}}\left(\prod_{a=1}^n d\bm y_a\right) \exp\left[-\mathrm{i}\frac{\llambda_\epsilon}{2}\sum_{i=1}^N\sum_{a=1}^n y_{ia}^2-\frac{1}{8N}\sum_{i,j=1}^N\left(\sum_{a=1}^n y_{ia}y_{ja}\right)^2\right]\ .\label{eq_replicas:compact}
\end{equation}

In order to proceed further, we introduce the following normalized density
\begin{equation}
\mu(\overrightarrow y)=\frac{1}{N}\sum_{i=1}^N \prod_{a=1}^n \delta(y_a-y_{ia})\ ,\label{eq_replicas:definitiondensity}
\end{equation}
where the $n$-dimensional vector $\overrightarrow y = (y_1,\ldots,y_n)$.\\

You can now check by direct substitution that the second term in the exponential in \eqref{eq_replicas:compact} can be rewritten as
\begin{equation}
-\frac{1}{8N}\sum_{i,j=1}^N\left(\sum_{a=1}^n y_{ia}y_{ja}\right)^2=-\frac{N}{8}\int d\overrightarrow y\ d \overrightarrow w\ \mu(\overrightarrow y)\mu(\overrightarrow w)\left(\sum_{a=1}^n y_a w_a\right)^2\ ,
\end{equation}
where $d\overrightarrow y=\prod_{a=1}^n dy_a$.\\

We can enforce the definition \eqref{eq_replicas:definitiondensity} using the following functional-integral representation of the identity
\begin{equation}
1=\int \mathcal{D}\mu\mathcal{D}\hat\mu\exp\left[-\mathrm{i}\int d\overrightarrow y \hat{\mu}(\overrightarrow y)\left(N\mu(\overrightarrow y)-\sum_i\prod_a \delta(y_a-y_{ia})\right)\right]\ ,
\end{equation}
which leads to
\begin{align}
\nonumber \langle Z(\llambda)^n\rangle &=\int \mathcal{D}\mu\mathcal{D}\hat\mu\exp\left[-\mathrm{i}N\int d\overrightarrow y \mu(\overrightarrow y)\hat\mu(\overrightarrow y)-\frac{N}{8}\int d\overrightarrow y\ d \overrightarrow w\ \mu(\overrightarrow y)\ \mu(\overrightarrow w)\ \left(\sum_{a=1}^n y_a w_a\right)^2\right]\times \\
&\times \int_{\mathbb{R}^{Nn}}\left(\prod_{a=1}^n d\bm y_a\right)  \exp\left[-\mathrm{i}\frac{\llambda_\epsilon}{2}\sum_{i=1}^N\sum_{a=1}^n y_{ia}^2+\mathrm{i}\sum_i\int d\overrightarrow y\hat\mu(\overrightarrow y)\prod_a \delta(y_a-y_{ia})\right]\ .
\end{align}

In the above equations $\mathcal{D}\mu \mathcal{D}\hat\mu$ denotes again functional integration, which was already used in Chapter \ref{chap:semicircle}. If you want to know more on this, see \cite{ref_semicircle:MacKenzie}. \\

The multiple integral $\int_{\mathbb{R}^{Nn}}\left(\prod_{a=1}^n d\bm y_a\right) (\cdots)$ is just a collection of $N$-identical copies of a single integral, hence
\begin{align}
\nonumber &\int_{\mathbb{R}^{Nn}}\left(\prod_{a=1}^n d\bm y_a\right)  \exp\left[-\mathrm{i}\frac{\llambda_\epsilon}{2}\sum_{i=1}^N\sum_{a=1}^n y_{ia}^2+\mathrm{i}\sum_i\int d\overrightarrow y\hat\mu(\overrightarrow y)\prod_a \delta(y_a-y_{ia})\right]\\
\nonumber &=\left\{\int_{\mathbb{R}^{n}} d\overrightarrow y \exp\left[-\mathrm{i}\frac{\llambda_\epsilon}{2}\sum_{a=1}^n y_{a}^2+\mathrm{i}\int d\overrightarrow y\hat\mu(\overrightarrow y)\prod_a \delta(y_a-y_{1a})\right]\right\}^N\\
&=\left\{\int_{\mathbb{R}^{n}}  d\overrightarrow y \exp\left[-\mathrm{i}\frac{\llambda_\epsilon}{2}\sum_{a=1}^n y_{a}^2+\mathrm{i}\hat\mu(\overrightarrow y)\right]\right\}^N\ ,\label{eq_replicas: fullcalculation}
\end{align}
where in the last line we used the $n$ delta functions to kill the multiple integral.\\

Exponentiating the last line of \eqref{eq_replicas: fullcalculation}, we can eventually write
\begin{equation}
 \langle Z(\llambda)^n\rangle =\int \mathcal{D}\mu\mathcal{D}\hat\mu\exp\left\{N\mathcal{S}_n[\mu,\hat\mu;\llambda]\right\}\ ,\label{eq_replicas2:functionalintegralfinal}
\end{equation}
where the action is given by
\begin{align}
\nonumber \mathcal{S}_n[\mu,\hat\mu;\llambda] &=-\mathrm{i}\int d\overrightarrow y \mu(\overrightarrow y)\hat\mu(\overrightarrow y)-\frac{1}{8}\int d\overrightarrow y d \overrightarrow w \mu(\overrightarrow y)\mu(\overrightarrow w)\left(\sum_{a=1}^n y_a w_a\right)^2\\
&+\mathrm{Log}\left[\int_{\mathbb{R}^{n}}d\overrightarrow y \exp\left[-\mathrm{i}\frac{\llambda_\epsilon}{2}\sum_{a=1}^n y_{a}^2+\mathrm{i}\hat\mu(\overrightarrow y)\right]\right]\ .\label{eq_replicas2:actionvectorialfirst}
\end{align}

The expression \eqref{eq_replicas2:functionalintegralfinal} lends itself to a nice saddle-point evaluation for $N\to\infty$. The only catch is that in doing so we would reverse the right order of limits: instead of taking $n\to 0$ first, and $N\to\infty$ afterwards, we are going to do the opposite! This procedure is not mathematically justified, but we will proceed as if it were.

\subsection{Critical points}

Finding the critical points of this action yields the two equations
\begin{align}
\label{eq_replicas2:SPequations1}\frac{\delta \mathcal{S}}{\delta\mu} = 0 &\Rightarrow -\mathrm{i}\hat\mu^\star(\overrightarrow y)=\frac{1}{4}\int d\overrightarrow w\mu^\star(\overrightarrow w)\left(\sum_{a=1}^n y_a w_a\right)^2\ ,\\
\frac{\delta \mathcal{S}}{\delta\hat\mu} = 0 &\Rightarrow \mu^\star(\overrightarrow y)=\frac{ \exp\left[-\mathrm{i}\frac{\llambda_\epsilon}{2}\sum_{a=1}^n y_{a}^2+\mathrm{i}\hat\mu^\star(\overrightarrow y)\right]}{\int_{\mathbb{R}^{n}}d\overrightarrow y' \exp\left[-\mathrm{i}\frac{\llambda_\epsilon}{2}\sum_{a=1}^n {y'}_{a}^2+\mathrm{i}\hat\mu^\star(\overrightarrow y')\right]}\ .\label{eq_replicas2:SPequations2}
\end{align}

Inserting \eqref{eq_replicas2:SPequations2} into \eqref{eq_replicas2:SPequations1}, we get
\begin{equation}
-\mathrm{i}\hat\mu^\star (\overrightarrow y)=\frac{\frac{1}{4}\int d\overrightarrow w \exp\left[-\mathrm{i}\frac{\llambda_\epsilon}{2}\sum_{a=1}^n w_{a}^2+\mathrm{i}\hat\mu^\star(\overrightarrow w)\right]\left(\overrightarrow y \cdot \overrightarrow w\right)^2}{\int d\overrightarrow w \exp\left[-\mathrm{i}\frac{\llambda_\epsilon}{2}\sum_{a=1}^n w_{a}^2+\mathrm{i}\hat\mu^\star(\overrightarrow w)\right]}\label{eq_replicas2:SPequations3}
\end{equation}
where both integrals on the r.h.s. run over $\mathbb{R}^n$.\\

In order to proceed, we have to make assumptions on the behavior of $\mu^\star$ and $\hat\mu^\star$ upon permutation of replica indices. There is a good body of research - although not yet a formal proof - pointing to the exactness of the replica-symmetric high-temperature solution, i.e. the one preserving permutation-symmetry among replicas, and rotational symmetry in the space of replicas.\\

This simply means that we should look for a solution of \eqref{eq_replicas2:SPequations1} and \eqref{eq_replicas2:SPequations2} in the form $\mu^\star(\overrightarrow y)=\mu^\star(y)$, with $y=|\overrightarrow y |$, and similarly for $\hat\mu^\star$.\\

Introducing $n$-dimensional spherical coordinates, we can rewrite \eqref{eq_replicas2:SPequations3} under the replica-symmetric assumption as
\begin{equation}
\label{eq_replicas2:SPequationls3}
 -\mathrm{i}\hat\mu^\star(y)=\frac{\frac{y^2}{4}\int_0^\infty d\omega\ \omega^{n-1} \exp[-\frac{\mathrm{i}}{2}\llambda_\epsilon \omega^2 +\mathrm{i}\hat \mu^\star (\omega)]\omega^2\int_0^\pi d\phi \left(\sin\phi\right)^{n-2} (\cos\phi)^2}{\int_0^\infty d\omega\ \omega^{n-1} \exp[-\frac{\mathrm{i}}{2}\llambda_\epsilon \omega^2 +\mathrm{i}\hat \mu^\star (\omega)]\int_0^\pi d\phi \left(\sin\phi\right)^{n-2}}\ ,
\end{equation}
where $\phi$ is taken as the angle between $\overrightarrow y$ and $\overrightarrow w$, and the other angular integrals cancel out between numerator and denominator.\\

Performing the remaining angular integrals, and after an integration by parts in the denominator, we get
\begin{equation}
\mathrm{i}\hat\mu^\star (y)=\frac{\Gamma (n/2)n}{2\Gamma(1+n/2)}\frac{y^2}{4}\frac{\int_0^\infty d\omega\ \omega^{n+1} G(\omega)}{\int_0^\infty d\omega\ \omega^{n} G'(\omega)}\ ,
\end{equation}
where $G(\omega):=\exp[-\frac{\mathrm{i}}{2}\llambda_\epsilon \omega^2 +\mathrm{i}\hat \mu^\star (\omega)]$. In the replica limit $n\to 0$, we obtain
\begin{equation}
\mathrm{i}\hat\mu^\star (y)=\frac{y^2}{4}\frac{\int_0^\infty d\omega\ \omega G(\omega)}{\int_0^\infty d\omega\ G'(\omega)}=C(\llambda)y^2\ ,
\end{equation}
where $C(\llambda)$ can be determined self-consistently using
\begin{equation}
\frac{\int_0^\infty d\omega\ \omega G(\omega)}{\int_0^\infty d\omega\ G'(\omega)}=\frac{\int_0^\infty d\omega\ \omega \exp\left[-\frac{\mathrm{i}}{2}\llambda_\epsilon \omega^2 +C(\llambda)\omega^2\right]}{\int_0^\infty d\omega\  \exp\left[-\frac{\mathrm{i}}{2}\llambda_\epsilon \omega^2 +C(\llambda)\omega^2\right]2\omega \left[-\frac{\mathrm{i}}{2}\llambda_\epsilon +C(\llambda)\right]}=\frac{1}{2\left[-\frac{\mathrm{i}}{2}\llambda_\epsilon +C(\llambda)\right]}\ ,
\end{equation}
so that
\begin{equation}
C(\llambda) =\frac{1}{8\left[-\frac{\mathrm{i}}{2}\llambda_\epsilon +C(\llambda)\right]}\Rightarrow C(\llambda)=\frac{1}{4}\left(\mathrm{i}\llambda_\epsilon\pm \sqrt{2-\llambda_\epsilon^2}\right)\ .\label{eq_replicas:Cx}
\end{equation}

\subsection{One step back: summarize and continue}

Let us now pause for a second and recap what we are doing. We started from the Edwards-Jones identity
\begin{equation}
\boxed{\rho(\llambda)=\frac{-2}{\pi N} \lim_{\varepsilon\to 0^+}\mathrm{Im}
\frac{\partial}{\partial\llambda}\Big\langle \mathrm{Log}\ Z(\llambda)\Big\rangle}\ ,\label{eq_replicas:rhoEJ2}
\end{equation}
where
\be
Z(\llambda)=\int_{\mathbb{R}^N}d\bm y \exp\left[-\frac{\mathrm{i}}{2}\bm{y}^T \left(\llambda_\epsilon \mathbbm{1}-H\right)\bm{y}\right]\ ,\label{eq_replicas:Zllambda1}
\ee
and $\llambda_\epsilon=\llambda-\mathrm{i}\epsilon$.\\

The average of the logarithm is performed by using the replica identity 
\begin{equation}
\langle\mathrm{Log}\ Z(\llambda)\rangle=\lim_{n\to 0}\frac{1}{n}\mathrm{Log}\langle Z(\llambda)^n\rangle\ ,\label{eq_replicas:replicaidentity2}
\end{equation}
which in turn (for large $N$) can be approximated via a saddle-point evaluation from \eqref{eq_replicas2:functionalintegralfinal} as
\begin{equation}
 \langle Z(\llambda)^n\rangle =\int \mathcal{D}\mu\mathcal{D}\hat\mu\exp\left\{N\mathcal{S}_n[\mu,\hat\mu;\llambda]\right\}\sim\exp\left[N\mathcal{S}_n[\mu^\star,\hat\mu^\star;\llambda]\right]\ .\label{eq_replicas2:functionalintegralfinalapprox}
\end{equation}

Combining \eqref{eq_replicas:rhoEJ2}, \eqref{eq_replicas:replicaidentity2} and \eqref{eq_replicas2:functionalintegralfinalapprox}, we obtain
\begin{equation}
\rho(\llambda)=\frac{-2}{\pi}\lim_{\epsilon\to 0^+}\mathrm{Im} \lim_{n\to 0}\frac{1}{n}\frac{\partial}{\partial\llambda}\mathcal{S}_n[\mu^\star,\hat\mu^\star;\llambda]\ .
\end{equation}
The derivative with respect to $\llambda$ only acts over the last term in the action \eqref{eq_replicas2:actionvectorialfirst}, because $\llambda$ appears explicitly (not through $\mu^\star$ or $\hat\mu^\star$) only there, and the action is stationary at the saddle point. Taking the derivative and writing the integral in spherical $n$-dimensional coordinates, we obtain
\begin{equation}
\rho(\llambda)=\frac{-2}{\pi}\lim_{\epsilon\to 0^+}\mathrm{Im} \lim_{n\to 0}\frac{1}{n}\frac{-\frac{\mathrm{i}}{2}\int_0^\infty dy\ y^{n+1}\exp\left[-\mathrm{i}\frac{\llambda_\epsilon}{2}y^2+C(\llambda)y^2\right]}{\int_0^\infty dy\ y^{n-1}\exp\left[-\mathrm{i}\frac{\llambda_\epsilon}{2}y^2+C(\llambda)y^2\right]}\ .
\end{equation}
Performing the integrals and simplifying 
\begin{equation}
\rho(\llambda)=\frac{1}{\pi}\lim_{\epsilon\to 0^+}\mathrm{Re}\frac{1}{-2 C(\llambda)+\mathrm{i}x_\epsilon}\ .
\end{equation}
Recalling that $C(\llambda)=\frac{1}{4}\left(\mathrm{i}\llambda_\epsilon\pm \sqrt{2-\llambda_\epsilon^2}\right)$ and $\llambda_\epsilon=\llambda-\mathrm{i}\epsilon$, we can first extract the real and imaginary part of $C(\llambda)$ using the lemma in \eqref{eq_replicas:shortlemma}. Therefore we can write
\begin{equation}
C(\llambda)=P_\epsilon(\llambda)+\mathrm{i}Q_\epsilon(\llambda)\ ,
\end{equation}
with 
\begin{align}
P_\epsilon(\llambda) &=\frac{1}{\sqrt{2}}\sqrt{2-\llambda^2+\epsilon^2+\sqrt{(2-\llambda^2+\epsilon^2)^2+(2\epsilon x)^2}}\\
Q_\epsilon(\llambda) &=\frac{\mathrm{sign}(2\epsilon x)}{\sqrt{2}}\sqrt{\sqrt{(2-\llambda^2+\epsilon^2)^2+(2\epsilon x)^2}-(2-\llambda^2+\epsilon^2)}\ .
\end{align}
Hence
\begin{equation}
\mathrm{Re}\frac{1}{-2 C(\llambda)+\mathrm{i}x_\epsilon}=\frac{-2P_\epsilon(\llambda)}{4 P_\epsilon(\llambda)^2+(\llambda-2Q_\epsilon(\llambda))^2}\ 
\end{equation}
In the limit $\epsilon\to 0^+$ and for $-\sqrt{2}<\llambda<\sqrt{2}$, $P_\epsilon$ and $Q_\epsilon$ converge to
\begin{align}
P_0(\llambda) &=\pm \frac{\sqrt{2-\llambda^2}}{4}\\
Q_0(\llambda) &=\frac{\llambda}{4}\ ,
\end{align}
from which 
\begin{equation}
\rho(\llambda)=\frac{1}{\pi}\sqrt{2-\llambda^2}\ ,
\end{equation}
i.e. Wigner's semicircle law as expected.

\chapter{Born to be free}\label{chap:free}

We have so far dealt with the spectral properties of individual random matrix ensembles. You may have been wondering (or not) what happens when you sum or multiply random matrices belonging to \emph{different} ensembles. In this Chapter we present an overview of the rather complicated tool you will need to tackle this problem: \emph{free probability theory} \cite{ref_free:voiculescu,ref_free:speicher}. 

\section{Things about probability you probably already know}
Two random variables $X_1$ and $X_2$, with pdfs $\rho_1$ and $\rho_2$, are said to be \emph{statistically independent} when the combined random variable $(X_1,X_2)$ has a factorized jpdf of the form
\be \label{eq_free:stat_ind} \rho_{1,2}(x_1,x_2) = \rho_1(x_1) \rho_2(x_2) \ . \ee

Statistical independence means that averages factorize as well ($\langle X_1 X_2 \rangle = \langle X_1 \rangle \langle X_2 \rangle$), which in turn means that their covariance is zero, and is key to finding the distribution of the sum of random variables. Let us consider a random variable $X$ with pdf $\rho(x)$. Its \emph{characteristic function} $\varphi(t)$ is defined as
\be \varphi(t) = \langle e^{\mathrm{i} t X} \rangle = \int d x \ \rho(x) \ e^{\mathrm{i} t x} \ , \ee
i.e. it is the Fourier transform of its pdf. \\

You should easily realize that the factorized jpdf in equation \eqref{eq_free:stat_ind} implies that characteristic functions are multiplicative upon the addition of statistically independent random variables, i.e. $\varphi_{1,2}(t_1,t_2) = \varphi_1(t_1) \varphi_2(t_2)$. Even more simply, we can introduce the logarithm of the characteristic function $h(t) = \log \varphi(t)$, the so called \emph{cumulant generating function}, which is obviously additive upon the addition of random variables:
\be \label{eq_free:cgf_sum} h_{1,2}(t_1,t_2) = h_1(t_1) + h_2(t_2) \ . \ee

Therefore, the problem of finding the pdf of the sum of two independent random variables $X_1$ and $X_2$ reduces to a simple ``algorithm'': compute the characteristic functions of $X_1$ and $X_2$ from their pdfs, form the the cumulant generating function of the sum $X_1 + X_2$ via the additive law \eqref{eq_free:cgf_sum}, compute the corresponding characteristic function via exponentiation, and eventually compute the pdf of the sum $X_1 + X_2$ via inverse Fourier transform.

\section{Freeness}
So, is there a generalization of statistical independence that will allow us to compute the eigenvalue spectrum of sums of random matrices? At first it might be tempting to guess that the statistical independence of two scalar random variables could be straightforwardly generalized to the case of two random matrices $X_1$ and $X_2$ by merely requiring the mutual independence of all entries. Unfortunately, this is not the case, as independent entries are not enough to destroy all possible angular correlations between the eigenbases of two matrices. \\

The property that generalizes statistical independence to random matrices is that of \emph{freeness}. The theory of free probability was initiated a few years ago by the pioneering works by Voiculescu and Speicher as an abstract approach to Von Neumann algebras, and only later it was shown to have a concrete realization in terms of random matrices. \\

Here is how freeness works. Let us consider two $N \times N$ random matrices $X_1$ and $X_2$, and let us introduce the following operator
\be \tau(X) = \lim_{N \rightarrow \infty} \frac{1}{N} \mathrm{Tr}(X) \ . \ee 
The two matrices $X_1$ and $X_2$ are said to be \emph{free} if for all integers $n_1, m_1, n_2, m_2, \ldots \geq 1$ we have
\begin{align} 
\label{eq_free:freeness_def} 
\nonumber \tau \left ( \left(X_1^{n_1} - \tau \left(X_1^{n_1} \right) \right) \left( X_2^{m_1} - \tau \left(X_2^{m_1}\right) \right) \left( X_1^{n_2} - \tau \left(X_1^{n_2} \right) \right) \left( X_2^{m_2} - \tau \left(X_2^{m_2} \right) \right) \ldots \right ) &=& \\ 
\tau \left ( \left(X_2^{n_1} - \tau \left(X_2^{n_1} \right) \right) \left( X_1^{m_1} - \tau \left(X_1^{m_1}\right) \right) \left( X_2^{n_2} - \tau \left(X_2^{n_2} \right) \right) \left( X_1^{m_2} - \tau \left(X_1^{m_2} \right) \right) \ldots \right ) &=& 0 \ .
 \end{align}
Not so straightforward, is it?\\

It might help to put the above definition into words. Two random matrices are free if the traces of all non-commutative products of matrix polynomials, whose traces are zero, are zero. Still not very intuitive, right? Well, unfortunately it does not get much better than that, but some intuition can be gained by exploring some concrete examples of the above definition. For example, equation \eqref{eq_free:freeness_def} reduces to $\tau(X_1 X_2) = \tau(X_1)\tau(X_2)$ when $n_1 = m_1$, and it reduces to $\tau(X_1^2 X_2^2) = \tau(X_1^2)\tau(X_2^2)$ when $n_1 = m_1 = 2$. As you should quickly realize, these equations generalize the moment factorization rules for statistically independent variables, and you can verify that all such relations for higher order moments can be obtained from equation \eqref{eq_free:freeness_def}.\\

However, the interesting part comes into play when we explore cases in which matrix non-commutativity kicks in. For example, you can easily work out the following result from \eqref{eq_free:freeness_def} for $n_1 = n_2 = m_1 = m_2 = 1$:
\be \tau(X_1 X_2 X_1 X_2) = \tau^2(X_1) \tau(X_2^2) + \tau(X_1^2) \tau^2(X_2) - \tau(X_1^2) \tau(X_2^2) \ . \ee
This result has no counterpart in ``conventional'' probability theory. Hopefully, this will convince you that freeness essentially represents a generalization of moment factorization.

\section{Free addition} 
Let us now put freeness to work.\\

Suppose we want to compute the average spectral density of the sum of large (i.e. $N \rightarrow \infty$) random matrices belonging to two different ensembles.\\

The first ingredient we need is our old friend the resolvent, which we introduced in chapter \ref{chap:resolvent}. Now, given the resolvent $G_\infty^{(av)}(z)$ of a given ensemble, let us introduce its functional inverse $B(z)$:
\be \label{eq_free:blue} G_\infty^{(av)} \left(B(z) \right) = B \left(G_\infty^{(av)}(z) \right) = z \ . \ee
The above function is known as the \emph{Blue} function \cite{ref_free:zee}. In case you are wondering: yes, it is called Blue because it is the inverse of the Green's function. \\

The last ingredient we need is the so called $R$-transform. Blue functions usually display a singular behavior at the origin, and the $R$-transform is just defined as a Blue function minus its singular part:
\be \label{eq_free:R_transform} R(z) = B(z) - \frac{1}{z} \ . \ee

We are all set now. Let us consider random matrices $X_1$ and $X_2$ belonging to ensembles characterized by resolvents $G_{\infty,1}^{av}(z)$ and $G_{\infty,2}^{av}(z)$, respectively. Let us form, through equations \eqref{eq_free:blue} and \eqref{eq_free:R_transform}, the corresponding $R$-transforms $R_1$ and $R_2$. The $R$-transform of the sum $X = X_1 + X_2$ is then simply given by the sum of the two $R$-transforms:
\be \label{eq_free:R_addition} R(z) = R_1(z) + R_2(z) \ . \ee

The above addition rule is the free counterpart of \eqref{eq_free:cgf_sum} for the moment generating functions of statistically independent random variables. Just like in that case, this rule provides a simple addition ``algorithm'' for free random matrices, whose first part has been outlined above. Once the $R$-transform of the sum has been computed, the corresponding Blue function and resolvent can be obtained through equations \eqref{eq_free:R_transform} and \eqref{eq_free:blue}. Once that has been done, the eigenvalue density can be derived from the resolvent via equation \eqref{eq_resolvent:rhofromG}.

\section{Do it yourself}

Enough with theory now: let us see free calculus at work on a concrete example.\\

All we need is the spectral density of large hermitian random matrix ensembles. So, how about the eigenvalue density of the free sum of some of our usual suspects? For example, let us consider a mixture of a GOE matrix $H$ and a Wishart matrix $W$
\be \label{eq_free:GOE_plus_W} S = p H + (1-p) W \ , \ee
where $p \in [0,1]$. \\

\begin{figure}[h]
\centering
\includegraphics[width=.75\columnwidth]{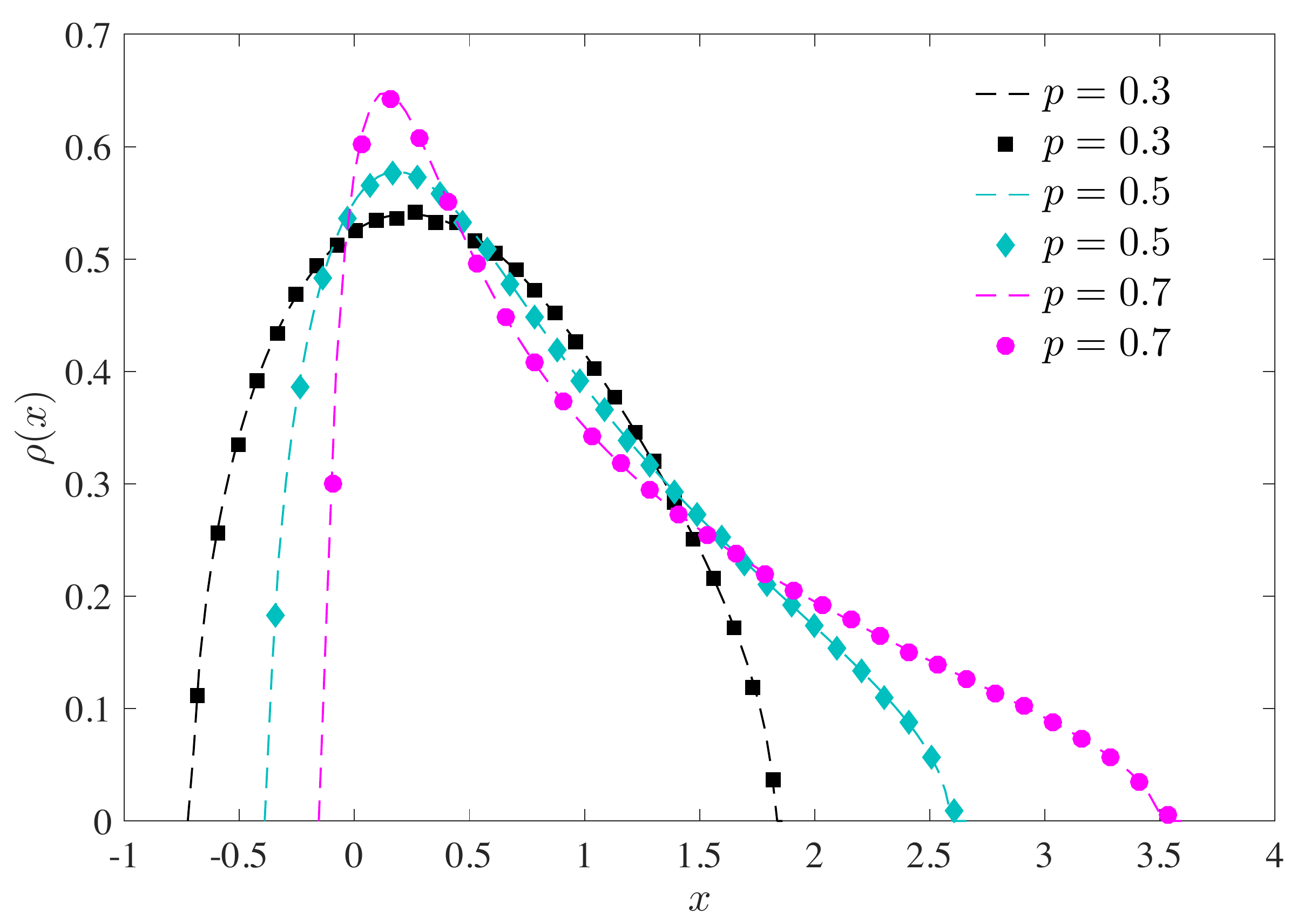}
\caption{Numerical check of the density obtained for the free addition of GOE and Wishart random matrices from the solution of Eq. (\ref{eq_free:3rddeg}). The examples shown refer to different values of the parameter $p$ that quantifies the relative weight between the two ensembles (see Eq. (\ref{eq_free:GOE_plus_W})).}
\label{fig:free_GOEplusWishart}
\end{figure}

For both ensembles we already have computed the resolvents (equations \eqref{eq_resolvent:Gsol} and \eqref{eq_Wishart:Gplusminus} with $K = 1/\alpha$). The functional inverse of those functions yield the Blue functions via equation \eqref{eq_free:blue}, and the $R$-transforms are immediately obtained via equation \eqref{eq_free:R_transform}. Please verify that they are given by the following functions for the GOE and Wishart ensembles respectively:
\begin{align} \label{eq_free:Blue_GOE_W}
& R_\mathrm{GOE}(z) = \frac{z}{2} \\ \nonumber
& R_W(z) = \frac{\alpha + 1}{1-z} \ .
\end{align}
Using the $R$-transform's scaling property $R_{cH}(z) = c R_H(c z)$ (see the box below), we can adapt the addition rule \eqref{eq_free:R_addition} to the present problem as follows:
\be R_S(z) = p \ R_\mathrm{GOE}(p z) + (1-p) R_W \left ((1-p)z \right ) \ . \ee
Plugging the functions in \eqref{eq_free:Blue_GOE_W} into the equation above gives 
\be R_S(z) = \frac{p^2}{2} z + \frac{(1-p)(1-\alpha)}{1-(1-p)z} \ , \ee
and the corresponding resolvent is obtained as $B_S(G_S(z)) = z$, where $B_S(z) = R_S(z)+1/z$ is the Blue function. The equation for the resolvent reads
\be z = \frac{p^2}{s} G_S(z) + \frac{(1-p)(1-\alpha)}{1-(1-p) G_S(z)} + \frac{1}{G_S(z)} \ . \label{eq_free:3rddeg} \ee
This is a third degree equation yielding, in general, one real solution and two complex conjugate ones for a given fixed $z$. The relationship between the eigenvalue density and the resolvent is the one in equation \eqref{eq_resolvent:rhofromG}, and that informs us that we will need to select the solution with a positive imaginary part. All this is done, and numerically verified, in the code [$\spadesuit$ \verb"GOE_Wishart_Sum.m"]. An example of the output that can be obtained is shown in Fig. \ref{fig:free_GOEplusWishart}.\\

Our goal in this Chapter was just to provide you with a short overview of the powerful tools free probability has to offer. There are plenty of papers out there if you'd like to know more. For example, you might have a look at the nice review article in \cite{ref_free:burda_free}, which also details some of the many applications that free random matrices have in quantitative finance. \\

\begin{center}
\fbox{\begin{minipage}{33em}
{\it Question.} Where does the scaling property of the $R$-transform come from? \\ \\
$\blacktriangleright$ It is inherited from the scaling properties of our good old friend the resolvent. Indeed, multiplying a matrix $H$ by a constant $c$ rescales the eigenvalues by the same factor $c$. Hence, from Eqs (\ref{eq_resolvent:StieltjesdefBis}) and (\ref{eq_resolvent:convergenceavaragedStieltjes}) it is easy to prove that the two corresponding resolvents are related to each other through this simple relationship: $G_{\infty,cH}^{(av)} = G_{\infty,H}^{(av)}(z/c) / c$. We can then write the equation for the Blue function $B_{cH}$
\be z = G_{\infty, cH}^{(av)} (B_{cH} (z)) = \frac{1}{c} G_{\infty,H}^{(av)} \left ( \frac{1}{c} B_{cH}(z) \right ) \ , \ee 
which shows that $B_{cH}(z) = cB_H(cz)$. We then have the following for the corresponding $R$ functions: 
\be c R_H(cz) = c B_H(cz) - \frac{1}{z} = B_{cH}(z) - \frac{1}{z} = R_{cH}(z) \ . \ee
\end{minipage}}
\end{center}

\begin{center}
\fbox{\begin{minipage}{33em}
{\it Question.} We know that the sum of two Gaussian scalar random variables is again Gaussian distributed. Is there an equivalent statement for the free addition of Gaussian random matrices? \\ \\
$\blacktriangleright$ Given the tools provided in this Chapter you should be able to show that the semicircle distribution is stable under free addition, i.e. if you free sum $M$ matrices each having the semicircle as spectral density, you still end up with a matrix whose spectral density is a semicircle.  
\end{minipage}}
\end{center}

\end{document}